\documentclass[aos]{imsart}

\RequirePackage{amsthm,amsmath,amsfonts,amssymb}
\RequirePackage[numbers]{natbib}
\RequirePackage[colorlinks,citecolor=blue,urlcolor=blue]{hyperref}
\RequirePackage{graphicx,float}

\usepackage{natbib}

\usepackage{algorithm}
\usepackage[noend]{algpseudocode}
\usepackage{enumitem}

\usepackage{tikz}
\usetikzlibrary{positioning}
\usetikzlibrary{decorations.pathreplacing}

\usepackage{pgfplots}

\usepackage{xcolor}

\usetikzlibrary{shapes,arrows}

\usepackage{multirow}

\usepackage{booktabs}

\startlocaldefs

\newtheorem{theorem}{Theorem}[section]
\newtheorem{lemma}[theorem]{Lemma}
\newtheorem{cor}{Corollary}[section]
\theoremstyle{remark}
\newtheorem{definition}[theorem]{Definition}
\newtheorem{example}{Example}

\newtheorem{remark}{Remark}

\newcommand{\cC}{{\cal C}}

\newcommand{\cG}{{\cal G}}

\newcommand{\cL}{{\cal L}}
\newcommand{\cM}{{\cal M}}
\newcommand{\cN}{{\cal N}}

\newcommand{\cP}{{\cal P}}

\newcommand{\cU}{{\cal U}}

\newcommand{\Np}{{\mathbb{N}}^{+}}
\newcommand{\Nn}{{\mathbb{N}}^{-}}

\def\bc{\begin{center}}
	\def\ec{\end{center}}
\def\benr{\begin{eqnarray}}
\def\eenr{\end{eqnarray}}
\def\benrr{\begin{eqnarray*}}
	\def\eenrr{\end{eqnarray*}}

\def\al{\alpha}

\def\edt{\end{document}}
\def\ep{\epsilon}
\def\g{\gamma}

\def\h{\hat}
\def\ka{\kappa}

\def\iny{\infty}
\def\ka{\kappa}

\def\la{\lambda}

\def\noi{\noindent}

\def\nn{\nonumber}

\def\Om{\Omega}
\def\om{\omega}

\def\e{\mathrm e}

\def\si{\sigma}

\def\Si{\Sigma}

\def\vep{\varepsilon}

\def\R{{\mathbb R}}
\def\Z{{\mathbb Z}}
\def\z{\zeta}

\DeclareMathOperator*{\argmin}{arg\,min}

\DeclareMathOperator*{\argmax}{arg\,max}

\endlocaldefs

\tikzstyle{block} = [rectangle, draw, fill=blue!20,
text width=5em, text badly centered, rounded corners, minimum height=4em,node distance=3cm]
\tikzstyle{line} = [draw, -latex']
\tikzstyle{linena} = [draw]
\tikzstyle{cloud} = [draw, ellipse,fill=red!20, minimum height=2em]

\usetikzlibrary{arrows,calc,angles,positioning,intersections,quotes,decorations.markings,backgrounds,patterns}
\usetikzlibrary{decorations.pathreplacing}
\pgfplotsset{compat=1.11}
\tikzset{
mynode/.style={fill,circle,inner sep=1pt,outer sep=0pt}
}

\numberwithin{equation}{section}

\begin{document}

\begin{frontmatter}
	\title{Segmentation of high dimensional means over multi-dimensional change points and connections to regression trees}
	\runtitle{HD means over 2D changes}
	
	\begin{aug}
		\author[A]{\fnms{Abhishek} \snm{Kaul}\ead[label=e1]{abhishek.kaul@wsu.edu}},
		\address[A]{Department of Mathematics and Statistics\\
			Washington State University\\
			Pullman, WA 99164, USA\\
			\printead{e1}}
	\end{aug}
	
	\begin{abstract}
		This article is motivated by the objective of providing a new analytically tractable and fully frequentist framework to characterize and implement regression trees while also allowing a multivariate (potentially high dimensional) response. The connection to regression trees is made by a high dimensional model with dynamic mean vectors over multi-dimensional change axes. Our theoretical analysis is carried out under a single two dimensional change point setting. An optimal rate of convergence of the proposed estimator is obtained, which in turn allows existence of limiting distributions. Distributional behavior of change point estimates are split into two distinct regimes, the limiting distributions under each regime is then characterized, in turn allowing construction of asymptotically valid confidence intervals for $2d$-location of change.  All results are obtained under a high dimensional scaling $s\log^2 p=o(T_wT_h),$ where $p$ is the response dimension, $s$ is a sparsity parameter, and $T_w,T_h$ are sampling periods along change axes. We characterize full regression trees by defining a multiple multi-dimensional change point model. Natural extensions of the single $2d$-change point estimation methodology are provided. Two applications, first on segmentation of {\it Infra-red astronomy satellite (IRAS)} data and second to segmentation of digital images are provided. Methodology and theoretical results are supported with monte-carlo simulations.
	\end{abstract}
	
	\begin{keyword}
		\kwd{Multi-dimensional change points}
		\kwd{Regression trees}
		\kwd{High dimensions}
		\kwd{Rate of convergence}
		\kwd{Limiting distributions}
		\kwd{Image processing}
	\end{keyword}
	
\end{frontmatter}

\section{Introduction}\label{sec:intro}

Consider the following model that describes high dimensional realizations with dynamic mean vectors observed on a two-dimensional space,

\begin{minipage}{0.6\textwidth}
	\benr\label{model:rvmcp}
	x_{(w,h)}&=&\begin{cases}\theta_{(1)}^0+\vep_{(w,h)}   & w> \tau^0_w,\,\, \&\,\, h>\tau^0_h,\\
		\theta_{(2)}^0+\vep_{(w,h)}   & w\le \tau^0_w,\,\, \&\,\, h>\tau^0_h,\\
		\theta_{(3)}^0+\vep_{(w,h)} & w\le \tau^0_w,\,\, \&\,\, h\le\tau^0_h,\\
		\theta_{(4)}^0+\vep_{(w,h)} & w> \tau^0_w,\,\, \&\,\, h\le \tau^0_h.
	\end{cases}\nn\\
	&=&\sum_{j=1}^4\theta_{(j)}^0{\bf 1}\big[(w,h)\in Q_j(\tau^0)\big]+\vep_{(w,h)},\\
	&&\hspace{2.25cm} w=1,...,T_w,\,\,h=1,...,T_h,\nn
	\eenr
\end{minipage}
\hspace{2mm}
\begin{minipage}{0.4\textwidth}
	\begin{tikzpicture}[scale=0.6]
	\draw (0,0) -- (0,6);\draw (0,0) -- (6,0);
	\draw (2,0) -- (2,6); \draw (0,2) -- (6,2);
	\draw (-0.5,6)  node{$T_h$};\draw (6,-0.5)  node{$T_w$};
	\draw[->] (4,-0.5) node[anchor=east]{Width ($w$)} -- (4.5,-0.5) node{};
	\draw[->] (-0.5,3.5) node[anchor=east, rotate=90]{Height ($h$)} -- (-0.5,4) node{};
	\node (v1) at (2,2) [ball color=black, circle, draw=black, inner sep=0.1cm] {};
	\draw (3.25,2.5) node{\footnotesize{$(\tau^0_w,\tau^0_h)$}};
	\draw (5,5) node{\footnotesize{$Q_1(\tau^0)$}}; \draw (1,5) node{\footnotesize{$Q_2(\tau^0)$}};
	\draw (1,1) node{\footnotesize{$Q_3(\tau^0)$}}; \draw (5,1) node{\footnotesize{$Q_4(\tau^0)$}};
	\end{tikzpicture}
\end{minipage}

\vspace{2mm}
\noi where $Q_j(\tau^0)$ represents the collection of indices in the $j^{th}$ quadrant with the origin shifted to  $\tau^0=(\tau^0_w,\tau^0_h)\in\{1,...,T_w\}\times\{1,...,T_h\}$\footnote{The ordering of quadrants is per usual convention, i.e.,  $Q_1(\tau^0)=\{(w,h);\, w> \tau^0_w,\,\, \&\,\, h>\tau^0_h\},$ $Q_2(\tau^0)=\{(w,h);\, w\le \tau^0_w,\,\, \&\,\, h>\tau^0_h\},$ $Q_3(\tau^0)=\{(w,h);\,  w\le \tau^0_w,\,\, \&\,\, h\le\tau^0_h\},$ and $Q_4(\tau^0)=\{(w,h);\,  w> \tau^0_w,\,\, \&\,\, h\le \tau^0_h\}.$}. The notation ${\bf 1}[\cdotp]$ represents an indicator function. Here the observed variable is $x_{(w,h)}\in\R^{p},$ $1\le w\le T_w,$ $1\le h\le T_h.$ The variables $\vep_{(w,h)}\in\R^p$ are unobserved zero mean random variables. The unknown parameters of interest are the change point $\tau^0=(\tau^0_w,\tau_h^0)^T$ and the mean vectors $\theta_{(j)}^0\in\R^p,$ $j=1,...,4,$ with $p$ being potentially high dimensional, i.e., where $p$ may diverge exponentially with respect to the number of realizations $T_wT_h,$ under a sparsity assumption to be specified later.

\begin{figure}[]
	\centering
	\begin{tikzpicture}[scale=0.6]
	\draw (5,5) circle [radius=0.85] node (A) {\footnotesize{$(w,h)$}};
	\draw (2,3) circle [radius=0.85] node (B) {\footnotesize{$w\le \tau^0_w$}};
	\draw (8,3) circle [radius=0.85] node (C) {\footnotesize{$w> \tau^0_w$}};
	\draw (0,1) circle [radius=0.85] node (D) {\footnotesize{$h\le \tau^0_h$}};
	\draw (4,1) circle [radius=0.85] node (E) {\footnotesize{$h> \tau^0_h$}};
	\draw (6,1) circle [radius=0.85] node (F) {\footnotesize{$h\le \tau^0_h$}};
	\draw (10,1) circle [radius=0.85] node (G) {\footnotesize{$h> \tau^0_h$}};
	\draw (A) -- (B);\draw (A) -- (C);
	\draw (B) -- (D);\draw (B) -- (E);
	\draw (C) -- (F);\draw (C) -- (G);
	\draw (0,0.25) node[anchor=north]{$\theta_3^0$};
	\draw (4,0.25) node[anchor=north]{$\theta_2^0$};
	\draw (6,0.25) node[anchor=north]{$\theta_4^0$};
	\draw (10,0.25) node[anchor=north]{$\theta_1^0$};
	\draw (5,5.75) node[anchor=south]{$Ex_{(w,h)}\in\R^p$};
	\end{tikzpicture}
	\caption{\footnotesize{2d change point model with HD means (\ref{model:rvmcp}) expressed equivalently as a decision tree}}
	\label{fig:dec.tree}
\end{figure}
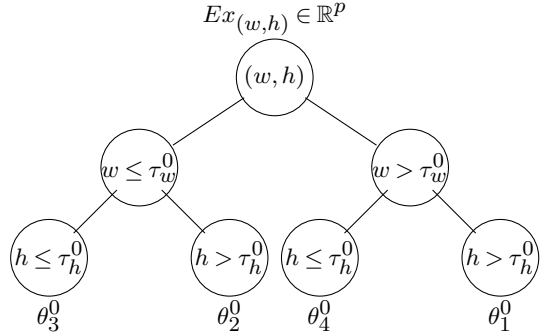

A main motivation to study model (\ref{model:rvmcp}), associated estimators and its generalizations, is that it provides a parametric framework for a frequentist analysis of regression trees.  Figure \ref{fig:dec.tree} provides a visualization of the model (\ref{model:rvmcp}) expressed as a decision tree and illustrates this connection. The equivalence of a full regression tree to generalizations of the model (\ref{model:rvmcp}) shall be illustrated later in Section \ref{sec:extensions}. The usefulness of regression trees is comprehensively established in the machine learning literature where it forms arguably one of the most empirically successful and heavily utilized tool, see, e.g. the recent review article \cite{hill2020bayesian}. On the other hand, these models also form one of the least analytically understood learning methods, wherein, to our knowledge there does not exist a frequentist parametric framework that allows analysis of statistical properties algorithms proposed for their recovery. The only analytically tractable methodology available in the literature is for Bayesian variants (BART) of \cite{chipman2010bart}, which has also only recently been recently studied in \cite{rockova2020posterior}. Furthermore, existing constructions of regression trees are typically limited to a single dimensional response ($p=1$), the model (\ref{model:rvmcp}) on the other hand allows for potential high dimensionality which is of significant interest given the nature of the current data rich landscape. We mention here that our objectives in this article shall be to make first analytical inroads to this problem, we do not attempt to address the rich body of additional problems that shall arise when viewed in its full generality, in this case we shall only attempt to provide feasible extensions that have clear analytical paths forward.

A second standalone motivation to study the model (\ref{model:rvmcp}) arises from the field of machine vision, in particular that of image recovery and segmentation. The underlying objective in this context is of segmenting a digital image into distinct clusters of pixels, each of which can be described as a vector of primary colors channels (or features) $(r,b,g)^T,$ with additional channels describing finer features of each pixel such as that of texture (or gradient in each direction) along with interactions of primary and secondary colors. Several other machine tasks are often built upon this segmentation layer, such as that of image denoising, object identification and classification amongst several others. While there are more than one heuristic techniques available in the computer science literature whose implementations are often based on computational relaxations which are not guaranteed. Analytically tractable statistical approaches to these problems are very limited. One such technique for specific the purposes of image denoising is that of total variation denoising which has been studied in the recent article of \cite{ortelli2020adaptive}.

Change point models have an extensive literature since their inception in \cite{page1955test}, with considerable effort in the recent past being devoted to allowing potential high dimensionality in these models. However, to our knowledge, all of this literature has been in the context of change on a one dimensional segmenting axis (usually {\it time}). To our knowledge, the model (\ref{model:rvmcp}) has not been described in the literature, consequently, analytically studied estimators for its parameters are unavailable in either the traditional fixed $p$ framework, diverging $p$ (with $p/T \to 0$)  under dense alternatives framework, or high dimensional $p$ (with $\log p=o(T)$) under sparse alternatives framework.  Furthermore, even under a one-dimensional segmenting axis and in the presence of potential high dimensionality, a large proportion of the existing literature is oriented towards obtaining near-optimal rates of estimation of suitably developed estimators. The question of inference under this high dimensional framework is far less understood in the literature in part owing to the until recently unavailable estimates with an optimal rate of convergence.

Following is a brief review of the recent literature on change point models under a one-dimensional segmenting axis. Estimation properties at rates slower than optimal by at least logarithmic factors are by far the most extensively studied facet of the above discussed problems. For e.g.  under a  fixed $p$ setting, the results of \cite{harchaoui2010multiple} consider a least squares estimator together with a total variation regularization. In a sparse high dimensional multiple change framework the article of \cite{wang2018high} provides a projected cusum estimator with a rate of convergence slower than optimal by a factor of $\log\log T.$ The article of \cite{cho2015multiple} yields a similar near optimal rate of estimation of the location parameters. While near optimal rates of the approximation are informative from an estimation perspective, however from an inference perspective one requires a change point estimator to obey an optimal rate of convergence in order to allow the existence of limiting distributions and in turn allow inference on change point location. In a large $p$ setting this fundamental aspect of post-estimation inference on the change location is in fairly nascent stages in the literature, and are only available under single change points. In a diverging $p,$ $p<<T$\footnote{$p<<T$ represents $p/T\to 0.$} (under dense alternatives) framework, the articles \cite{bhattacharjee2018change} and \cite{bhattacharjee2019change} develop limiting distributions for the estimator of the location of the change point, which in turn  allows inference on the unknown location.  Under stronger dimensional assumptions, the article \cite{bai2010common} provides a similar limiting distribution result. The article of \cite{wang2020dating} extends these inferential results to $p<<T^2/\log T,$ but require $p$ to be necessarily diverging. Under high dimensionality, the article of \cite{kaul2020inference} provide an plug-in estimator that yields this optimal rate of convergence and develops limiting distributions in this framework. The boundary problem of detection of existence of dynamicity in a diverging or high dimensional setting has been studied by several authors and different methods have been proposed. This problem has been addressed by several approaches, for e.g.  \cite{jirak2015uniform}, \cite{wang2019inference}, \cite{enikeeva2013high}, \cite{cho2016change} and \cite{steland2018inference} amongst others. This boundary problem  has also been approached in a selection sense by means of appropriately tuned $\ell_0$ regularization, see e.g. \cite{kaul2019efficient} in a single change point framework, \cite{kaul2019detection}, \cite{wang2019statistically} in a multiple change point framework. This $\ell_0$ regularization is also in a sense equivalent to that carried out in the regularization step of Wild Binary Segmentation Algorithm of \cite{fryzlewicz2014wild}, which is also utilized in \cite{wang2018high} and several other multiple change point methodologies in the literature. High dimensional change point models have also been studied with several other data generating processes besides mean shifts, graphical models in \cite{kaul2021graphical}, \cite{wang2017optimal}, \cite{atchade2017scalable}, stochastic block models in \cite{wang2018optimal}, \cite{bhattacharjee2018change}, markov random fields \cite{roy2017change} amongst other settings, all available article by construction assuming a one-dimensional change axis.

The main analytical contributions of this article shall be to develop algorithmic estimators for the change point parameter $\tau^0$ under the model (\ref{model:rvmcp}), so that it retains sufficient regularity despite potential high dimensionality in order to yield an optimal rate\footnote{Our use of the word {\it optimal} is a slight over-reach. This is made as a natural extension in view of the minimax optimal rate of estimation in a single change axis framework, which is known in the literature. However, under the considered setting, the optimal rate is not explicitly known. Based solely on the results of this article, the best we can claim is instead that we obtain a {\it sharp} rate of convergence.} of estimation. This in turn shall allow existence of limiting distributions under both vanishing and non-vanishing jump size regimes whose forms are then derived under high dimensional asymptotics. These results enable one to perform inference on the change point, or equivalently on the branching transitions in context of regression trees, by allowing construction of asymptotically valid confidence intervals under the assumed high dimensional and sparse framework.
Further considerations in this development shall be the following, (a) sufficient flexibility in the methodology to be directly extendable to multiple change points and regression trees in their full generality, and (b) scalability of the methodology in both dimension size of the mean parameters as well as sampling periods $T_w,T_h$ in order to allow applicability towards  applications such as analysis of large images (fixed $p,$ large $T$) as well as data sets such as those arising in regression tree contexts such as genetic sequencing studies (large $p$, small $T$).

To describe our proposed methodology first consider the squared loss,
\benr\label{def:sq.loss}
\cL(\tau_w,\tau_h,\theta)=\frac{1}{T_wT_h}\sum_{j=1}^4\,\,\sum_{(w,h)\in Q_j(\tau)}\|x_{(w,h)}-\theta_{(j)}\|_2^2,\quad{\rm where}\,\,\tau=(\tau_w,\tau_h)^T,
\eenr
and define a component-wise plug-in estimator for $\tau^0=(\tau^0_w,\tau^0_h)^T$ as follows,
\benr\label{est:optimal}
\tilde\tau_w(\h\tau_h,\h\theta)=\argmin_{1\le\tau_w< T_w} \cL(\tau_w,\h\tau_h,\h\theta),\quad{\rm and}\quad \tilde\tau_h(\h\tau_w,\h\theta)=\argmin_{1\le\tau_h< T_h} \cL(\h\tau_w,\tau_h,\h\theta),
\eenr
where $\h\tau=(\h\tau_w,\h\tau_h)$ and $\h\theta$ represent some preliminary estimates that are for the time being assumed to be available. It is evident that the behavior of the estimator $\tilde\tau=(\tilde\tau_w,\tilde\tau_h)^T$ shall be intertwined with the quality of plug in estimates $\h\tau=(\h\tau_w,\h\tau_h)$ and $\h\theta$ utilized in its construction.

To build a feasible and sufficiently regular estimator for $\tau^0,$ our strategy going forward shall be somewhat reverse of traditional. Where one usually builds an algorithm and then attempts to study its properties, instead we shall begin by obtaining statistical properties of the plug-in estimates $\tilde\tau_w,\tilde\tau_h$ of (\ref{est:optimal}) with respect to assumed properties in estimation of the preliminary estimates $\h\tau$ and $\h\theta$ used in its construction. These results shall then be aggregated in order to provide an asymptotically valid and feasible in practice, twice iterative algorithmic procedure, where the iterations are between the change point parameters and the mean parameters with an additional internal iteration in the components of the change point.

To study $\tilde\tau=(\tilde\tau_w,\tilde\tau_h)^T$ we require some more definitions and additional control parameters. Define the jump vectors that constitute the change across quadrants of model (\ref{model:rvmcp}), \benr\label{def:jump.vec.quadrants}
\eta_{(1)}^0=\theta_{(2)}^0-\theta_{(1)}^0,\,\,\eta_{(2)}^0=\theta_{(3)}^0-\theta_{(2)}^0,\,\,\eta_{(3)}^0=\theta_{(3)}^0-\theta_{(4)}^0,\,\,{\rm and}\,\,\eta_{(4)}^0=\theta_{(1)}^0-\theta_{(4)}^0.
\eenr
The direction of $\eta^0_{(j)}$'s is inconsequential for the analysis of $\tilde\tau,$ i.e., one may instead define $\eta^0_{(1)}=\theta_{(1)}^0-\theta_{(2)}^0,$ and similar for $\eta^0_{(j)},$ $j=2,3,4.$ The parameters that provide control are instead the $\ell_2$ magnitude of these vectors which represent jump sizes across quadrants, i.e.,
\benr\label{def:jump.size.quad}
\xi_j=\|\eta^0_{(j)}\|_2,\quad j=1,...,4,\quad \overline\xi=\max_{j}\{\xi_j\},\quad{\rm}\quad \underline\xi=\min_{j}\{\xi_j\} .
\eenr

\begin{figure}[]
	\centering
	\begin{minipage}{0.45\textwidth}
		\begin{tikzpicture}[scale=0.75]
		\draw (0,0) -- (0,6);\draw (0,0) -- (6,0);
		\draw (2,0) -- (2,6); \draw (0,2) -- (6,2);
		\draw (1.5,6)  node{$T_h$};\draw (6,1.5)  node{$T_w$};
		\draw (-0.5,-0.5)  node{$(0,0)$};\draw (-0.5,6)  node{$T_h$};\draw (6,-0.5)  node{$T_w$};
		\draw[->] (3.5,-0.5) node[anchor=east]{Width ($w$)} -- (4.5,-0.5) node{};
		\draw[->] (-0.5,3) node[anchor=east, rotate=90]{Height ($h$)} -- (-0.5,4) node{};
		\node (v1) at (2,2) [ball color=black, circle, draw=black, inner sep=0.1cm] {};
		\draw (1.2,2.35) node{$(\tau^0_w,\tau^0_h)$};
		\draw (5,5) node{$T_wT_h\om_1$}; \draw (1,5) node{$T_wT_h\om_2$};
		\draw (1,1) node{$T_wT_h\om_3$}; \draw (5,1) node{$T_wT_h\om_4$};
		\draw[decorate,decoration={brace,amplitude=10pt},xshift=4pt,yshift=0pt] (2,6) -- (2,2.1) node [black,midway,xshift=25pt] {$T_h\om_h$};
		\draw[decorate,decoration={brace,amplitude=10pt},xshift=0pt,yshift=4pt] (2.1,2) -- (6,2) node [black,midway,yshift=18pt] {$T_w\om_w$};
		\end{tikzpicture}
	\end{minipage}
	\hspace{4mm}
	\begin{minipage}{0.45\textwidth}
		\begin{tikzpicture}[scale=0.75]
		\draw (0,0) -- (0,6);\draw (0,0) -- (6,0);
		\draw (2,0) -- (2,6);\draw (0,2) -- (6,2);
		\draw[dashed] (4,4) node[circle,draw,anchor=west]{$\theta_{(1)}^0$} --
		(1.5,4) node[circle,draw,anchor=east]{$\theta_{(2)}^0$} --
		(1.5,1.5) node[circle,draw,anchor=north east]{$\theta_{(3)}^0$} --
		(4,1.5) node[circle,draw,anchor=north west]{$\theta_{(4)}^0$} -- cycle;
		\draw (4.75,2.75) node{$\eta_{(4)}^0,\,\xi_4$};\draw (0.75,2.75) node{$\eta_{(2)}^0,\,\xi_2$};
		\draw (2.75,4.25) node{$\eta_{(1)}^0,\,\xi_1$};\draw (2.75,1.1) node{$\eta_{(3)}^0,\,\xi_3$};
		\draw[dashed] (2.5,4) -- (2.5,1.5); \draw (2.8,3.35) node{$\xi_w$};
		\draw[dashed] (1.5,2.75) -- (4,2.75); \draw (3.5,2.5) node{$\xi_h$};
		\draw (1.5,6)  node{$T_h$};\draw (6,1.5)  node{$T_w$};
		\draw (-0.5,-0.5)  node{$(0,0)$};\draw (-0.5,6)  node{$T_h$};\draw (6,-0.5)  node{$T_w$};
		\draw[->] (3.5,-0.5) node[anchor=east]{Width ($w$)} -- (4.5,-0.5) node{};
		\draw[->] (-0.5,3) node[anchor=east, rotate=90]{Height ($h$)} -- (-0.5,4) node{};
		\node (v1) at (2,2) [ball color=black, circle, draw=black, inner sep=0.1cm] {};
		\end{tikzpicture}
	\end{minipage}
	\caption{\footnotesize{Illustration of control parameters. {\it Left panel: weight parameters $\om_j,$ $j=1,2,3,4$ and $\om_w$ and $\om_h,$ measuring proportion of available observations in each segment} {\it Right panel: underlying mean parameters $\theta_{(j)}^0,$ and change parameters $\eta^0_{(j)},$ $j=1,2,3,4,$ as well as the jump sizes $\xi_j,$ $j=1,2,3,4,$ and $\xi_w,$ and $\xi_h.$}}}
	\label{fig:parameters}
\end{figure}
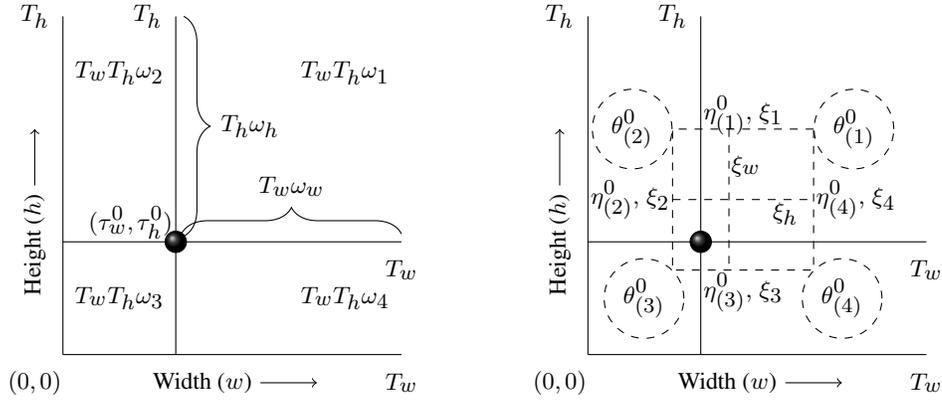

Additionally define weight parameters that measure the proportion of observations in each quadrant, as well as those that measure proportions along individual change axes,
\benr\label{def:weights.quad}
&\om_j=|Q_j(\tau^0)|\big/{T_wT_h},\quad j=1,2,3,4,\quad{\rm and}\quad \underline\om=\min_{j}\{\om_j\}\nn\\
&\om_w=(T_w-\tau^0_w)\big/{T_w},\quad{\rm and}\quad\om_h=(T_h-\tau^0_h)\big/{T_h}.
\eenr
Next define width and height-wise proportion weighted jump sizes which shall play a critical role in our analysis,
\benr\label{def:weight.jump.horiz.vert}
\,\,\,\xi_w^2= \om_h\xi_1^2+(1-\om_h)\xi_3^2,\quad \xi_h^2=\om_w\xi_4^2+(1-\om_w)\xi_2^2.\quad{\rm and }\quad\xi_{\min}=\xi_{w}\wedge\xi_h.
\eenr
We remind the reader here that the above mean, change and weight parameters are allowed to depend on $T_w,T_h,$ either directly or via the dimension $p.$ However, this dependence is notationally suppressed throughout for clarity of exposition. A visual description of these control parameters is also provided in Figure \ref{fig:parameters}.

Our first task shall be to examine the properties of $\tilde\tau$ and its relationship to assumed properties of preliminary estimates $\h\tau,\h\theta$ used in its construction. In accordance to the structure of this approach, first assume the availability of $\h\tau=(\h\tau_w,\h\tau_h)$ and $\h\theta$ satisfying mainly the following requirements,
\benr\label{eq:40}
\max_{1\le j\le 4}\|\h\theta_{(j)}-\theta_{(j)}^0\|_2\le c_{u1}\xi_{\min},\quad |\h\tau_w-\tau^0_w|\le c_{u1}T_{w},\,\,{\rm and}\,\,\,\, |\h\tau_h-\tau^0_h|\le c_{u1}T_{h}
\eenr
with probability at least $1-o(1),$ where $c_{u1}>0$ is a suitably chosen small enough constant. Then our first estimation result shall show that $\tilde\tau$ implemented with estimates satisfying (\ref{eq:40}) yields a near optimal rates of convergence,
\benr\label{eq:near.opt}
&&(\tilde\tau_w-\tau^0_w)=O_p\big(T_h^{-1}\xi_w^{-2}s\log^2(p\vee T_wT_h)\big)\quad{\rm and}\nn\\
&&(\tilde\tau_h-\tau^0_h)=O_p\big(T_w^{-1}\xi_h^{-2}s\log^2(p\vee T_wT_h)\big).
\eenr
Here $s$ is a sparsity parameter that can equivalently be thought of as either sparsity of the jump vectors $\eta^0_{(j)},$ $j=1,2,3,4,$ or that of individual mean vectors $\theta_{(j)}^0,$ $j=1,2,3,4$ (see, discussion after (\ref{def:setS}) for further details on this equivalence).

While near optimal rates of (\ref{eq:near.opt}) are of independent interest from an estimation perspective and are comparable to a large proportion of literature in a high dimensional framework under a single change axis. However, due to the absence of an optimal rate of convergence, it does not permit existence of limiting distributions and consequently disallows one to perform inference on the underlying change parameters. The next and more important estimation result shall show that when quality of preliminary estimates is improved, the corresponding plug in estimate of the change point can be upgraded from near optimality to optimality. Specifically, upon assuming the availability of preliminary estimates tightened to,
\benr\label{eq:41}
\max_{1\le j\le 4}\|\h\theta_{(j)}-\theta_{(j)}^0\|_2\le \xi_{\min}r_{T},\quad {\rm where}\quad{r_{T}}=\frac{c_{u1}}{s^{1/2}\log(p\vee T_wT_h)}\footnotemark\quad{\rm and,}\nn\\
|\h\tau_w-\tau^0_w|\le T_{w}r_{T}^2,\quad{\rm and}\quad |\h\tau_h-\tau^0_h|\le T_{h}r_{T}^2\hspace{1in}
\eenr
with probability at least $1-o(1).$ Then we shall obtain the optimal estimation result,
\benr\label{eq:opt.rate}
(\tilde\tau_w-\tau_w^0)=O_p(T_h^{-1}\xi_w^{-2})\quad{\rm and}\quad (\tilde\tau_h-\tau_h^0)=O_p(T_w^{-1}\xi_h^{-2}).
\eenr
\footnotetext{This is a sequence in both $T_w$ and $T_h,$ however to ease notation we present it in shorthand as $r_T$}
The order of each of the two relations in (\ref{eq:near.opt}) and (\ref{eq:opt.rate}) are with respect to (w.r.t.) $T_w$ and $T_h,$ respectively. We mention this subtlety to inform that these results in both the near optimal and optimal case are valid under the asymptotics $T_w\to\iny,$ $T_h\to\iny,$ additionally the first relation is also valid when $T_w\to\iny,$ $T_h<\iny,$ and symmetrically for the second relation.

The result (\ref{eq:opt.rate}) has important consequences. It characterizes the parameters ($T_h\xi_w^2$ and $T_w\xi_h^2$) that control the statistical behavior of $\tilde\tau.$ In context of regression trees, it provides the order of magnitude of change in means that is sufficient to be detectable by the proposed approach. It also highlights two important distinctions of model (\ref{model:rvmcp}) in comparison to mean shift models under a single change axis. First, in a single change axis framework, the single jump size across a change point is the estimation controlling parameter. On the other hand, in model (\ref{model:rvmcp}) there are instead four individual jump sizes $\xi_j$ $j=1,2,3,4$ characterizing each change point, and the estimation controlling parameters are instead weighted combinations of these individual jumps. Second, note that in each direction (say $\tilde\tau_w$) one is able leverage the observations in the alternate direction to yield a rate of convergence $O_p(T_h^{-1}\xi_w^{-2}).$ In comparison, a model with a one-dimensional change axis yields an optimal rate of $O_p(\xi^{-2}),$ (see, e.g. \cite{kaul2020inference}). As a direct consequence, this leveraging of observations of alternating directions allows detect jumps which may be smaller by an order of $\surd{T_h}.$ Further discussions on this and other aspects of this comparison are provided in Section \ref{sec:main}.

The most important consequence of the optimal rate (\ref{eq:opt.rate}) along with other peripheral results is that allows for the existence of limiting distributions of the estimates $\tilde\tau_w,\tilde\tau_h,$ thereby enabling inference on the corresponding parameters, despite potential high dimensionality. As for single change axis settings (see, e.g., \cite{kaul2020inference},and \cite{bhattacharjee2017common}), the distributional behavior of $\tilde\tau$ is split into two distinct regimes which are that of the vanishing and non-vanishing jump sizes. Under the former regime of $\surd{T_h}\xi_w\to 0,$ $\surd{T_w}\xi_{h}\to 0,$ we shall obtain,
\benr\label{eq:limiting.d.vanishing}
&&T_h\xi_w^2\si_{(w,\iny)}^{-2}(\tilde\tau_w-\tau^0_w)\Rightarrow \argmax_{\z\in\R}\big(2W_w(\z)-|\z|\big),\quad T_w\to \iny\nn\\
&&T_w\xi_h^2\si^{-2}_{(h,\iny)}(\tilde\tau_h-\tau^0_h)\Rightarrow \argmax_{\z\in\R}\big(2W_h(\z)-|\z|\big),\qquad T_h\to \iny
\eenr
where $\si^2_{(w,\iny)}$ and $\si^2_{(h,\iny)}$ are estimable variance parameters of these limiting processes. Here $W_w(\cdot)$ and $W_h(\cdotp)$ are both two-sided Brownian motions on $\R.$ The distribution $\argmax_{\z\in\R}\big(2 W(\z)-|\z|\big)$ is well studied in the literature and its cdf and thus its quantiles are readily available, (\cite{yao1987approximating}).

For the non-vanishing case $\surd{T_h}\xi_w\to\xi_{(w,\iny)},$ and $\surd{T_w}\xi_h\to\xi_{(h,\iny)},$  where $0<\xi_{(w,\iny)},\,\,\xi_{(h,\iny)}<\iny,$ define a negative drift two sided random walk initializing at the origin.
\benr\label{eq:neg.drift.rw}
\cC_{\iny}(\z,\xi,\si^2)=
\begin{cases}\sum_{t=1}^{\z} z_t, & \z\in \Np=\{1,2,3,...\} \\ 	
	0,				  &	\z=0 \\
	\sum_{t=1}^{-\z}z_t^*,		  &	\z\in \Nn=\{-1,-2,-3,...\},
\end{cases}
\eenr
where $z_t,z_t^*$ are independent copies of a $\cP\big(-\xi^2,4\xi^2\si^2\big)$ distribution, which are also independent over all $t,$ for a distribution law $\cP$\footnote{If one assumes $\vep_{(w,h)}\sim^{i.i.d} \cN(0,\Si),$ then $\cP$ shall also be a normal distribution.} that shall be determined by the form of the underlying distribution in model (\ref{model:rvmcp}) (see, Condition A$'$). The notation in the arguments of $\cP(\cdotp,\cdotp)$ is representative of the mean and variance of this distribution. Finally, let,
\benr\label{eq:42}
\,\,\,\,\cC_{(w,\iny)}(\z)=\cC_{\iny}\big(\z,\xi_{(w,\iny)},\si_{(w,\iny)}^2\big)\,\, {\rm and}\,\, \cC_{(h,\iny)}(\z)=\cC_{\iny}\big(\z,\xi_{(h,\iny)},\si_{(h,\iny)}^2\big),
\eenr
where $\si_{(w,\iny)}^2$ and $\si_{(h,\iny)}^2$ are variance parameters as defined earlier in context of the vanishing regime.
Then, we shall obtain the following results,
\benr\label{eq:limiting.d.non.vanishing}
&&(\tilde\tau_w-\tau^0_w)\Rightarrow \argmax_{\z\in \Z}\cC_{(w,\iny)}(\z),\quad T_w\to \iny\nn\\
&&(\tilde\tau_h-\tau^0_h)\Rightarrow \argmax_{\z\in \Z}\cC_{(h,\iny)}(\z),\,\,\quad T_h\to \iny,
\eenr
where $\Z$ is the collection of integers. Quantiles of these limiting distribution can be approximated numerically upon availability of law $\cP$ of increments of the processes in (\ref{eq:42}), thereby enabling the construction of asymptotically valid confidence intervals.

The above discussion shall establish the statistical behavior of the plug-in estimator $\tilde\tau,$ and provides sufficient properties required of the preliminary estimates used in its construction. However, two important questions remain as yet unanswered. First, since these theoretical results are established under assumed conditions on the preliminary estimates $\h\tau$ and $\h\theta,$ thus without this availability, these results remain infeasible to implement. Second, all presented results require no explicit restrictions on growth of dimensionality or the behavior of jump sizes which may appear highly suspect. Both these questions are inter-related and the following additional notation is necessary for their discussion. For any $\tau=(\tau_w,\tau_h)^T\in\{1,...,(T_w-1)\}\times\{1,...,(T_h-1)\},$ let,
\benr\label{def:empmeans}
\bar x_{(j)}(\tau)=\frac{1}{|Q_j(\tau)|}\sum_{(w,h)\in Q_j(\tau)}x_{(w,h)},\quad j=1,2,3,4.
\eenr
be the quadrant-wise sample means. Now consider the soft-thresholding operator, $k_{\la}(x)={\rm sign}(x)(|x|-\la)_{+},$ $\la>0,$ $x\in\R^p,$ where ${\rm sign}(\cdotp),$ $|\cdotp|,$ and $(\cdotp)_{+}$\footnote{For $x\in \R,$ $(x)_{+}=x,$ if $x\ge 0,$ and $x=0$ if $x<0.$} are applied component-wise. Then for any $\la_1,\la_2>0,$ define $\ell_1$ regularized quadrant-wise mean estimates,
\benr\label{est:softthresh}
\h\theta_{(j)}(\tau)=k_{\la_j}\big(x_{(j)}(\tau)\big),\quad j=1,2,3,4.
\eenr
It is well known in the literature (\cite{donoho1995noising}, \cite{donoho1995wavelet}) that the soft-thresholding operation in (\ref{est:softthresh}) is equivalent to the following $\ell_1$ regularization.
\benr\label{est:softL1construction}
\h\theta_{(j)}(\tau)&=&\argmin_{\theta\in\R^p}\big\|\bar x_{(j)}(\tau)-\theta\big\|^2_2+\la_j\|\theta\|_1,\quad \la_j>0,\quad j=1,2,3,4.
\eenr

In view of earlier discussion, the missing links required for feasibility of $\tilde\tau$ are, (a) construction of preliminary mean estimates $\h\theta,$ and (b) both preliminary estimates $\h\tau$ and $\h\theta$ requiring either the condition (\ref{eq:40}) (milder) to obtain near optimal estimates, or (\ref{eq:41}) (stronger) to obtain an optimal estimate of $\tau^0.$ We shall fulfil (a) by utilizing the soft thresholded means (\ref{est:softthresh}). In order to fulfil (b), we shall exploit the distinctions between the rate conditions (\ref{eq:40}) and (\ref{eq:41}), by building an twice iterated algorithmic estimator that shall improves a nearly arbitrarily chosen $\check\tau,$ first to a near optimal estimate $\h\tau$ in a first iteration, and then to an optimal estimate $\tilde\tau$ in a second iteration. This construction is described as Algorithm \ref{alg:single} below and is presented visually as Figure \ref{fig:schematic} in Sub-section \ref{subsec:alg1}.

\begin{algorithm}\label{alg:single}
	\caption{Optimal estimation of $\tau^0=(\tau^0_w,\tau^0_h)^T.$}
	\begin{algorithmic}[1]
		
		\vspace{1mm}
		\Statex Initialize change point $\check\tau=(\check\tau_w,\check\tau_h),$
		
		\vspace{1mm}
		\State Compute mean estimates $\check\theta_{(j)}=\h\theta_{(j)}(\check\tau),$ $j=1,2,3,4.$ and update change point estimate componentwise,
		\benr
		\h\tau_w=\argmin_{1\le \tau_w<T_w} \cL(\tau_w,\check\tau_h,\check\theta)\quad{\rm and}\quad
		\h\tau_h=\argmin_{1\le \tau_h<T_h} \cL(\check\tau_w,\tau_h,\check\theta)\nn
		\eenr
		\State Update mean estimates to $\h\theta_{(j)}=\h\theta_{(j)}(\h\tau),$ $j=1,2,3,4,$ and update,
		\benr
		\tilde\tau_w=\argmin_{1\le \tau_w<T_w} \cL(\tau_w,\h\tau_h,\h\theta)\quad{\rm and}\quad
		\tilde\tau_h=\argmin_{1\le \tau_h<T_h} \cL(\h\tau_w,\tau_h,\h\theta)\nn
		\eenr
		\Statex Output: $\tilde\tau=(\tilde\tau_w,\tilde\tau_h).$
	\end{algorithmic}
\end{algorithm}

For the validity of Algorithm \ref{alg:single} we shall assume the rate assumption,
\benr\label{eq:33}
\frac{c_u\si}{\xi_{\min}}\Big\{\frac{s\log^2 (p\vee T_wT_h)}{\surd(T_wT_h\underline\om)}\Big\}\le c_{u1}.
\eenr
Then we shall show that performing the two successive iterations between the change point and mean parameters of Algorithm \ref{alg:single}, shall in each iteration also provide estimates that satisfy (\ref{eq:40}) and (\ref{eq:41}), respectively, which then feed into the following iteration. This process is designed so as to be able to aggregate the prior developed statistical results which in turn shall guarantee the analytical validity of the proposed algorithm. In particular, the output $\tilde\tau$ of Algorithm \ref{alg:single} shall be shown to satisfy the optimal rate estimation  (\ref{eq:opt.rate}), and the limiting distribution results of (\ref{eq:limiting.d.vanishing}) and (\ref{eq:limiting.d.non.vanishing}), these shall follow as a corollary, by the construction of the iterative mechanism. Further details of this argument are postponed to Sub-section \ref{subsec:alg1}. We note that the specific choice of soft-thresholding as a regularization mechanism in (\ref{est:softL1construction}) is superficial, the eventual objective is only to obtain mean estimates that are well behaved in the high dimensional setting in the $\ell_2$ norm. Alternatively, our results are derived in sufficient generality to allow one to consider using any other suitable choice of the regularization mechanism that may also be problem specific, e.g. group $\ell_1$ regularization which assumes a partially known sparsity structure, or non-convex regularizations such as {\it scad} and {\it mcp}.

While the above results and Algorithm \ref{alg:single} have been developed independently and purely from a change point perspective, however, fortuitously there are overlapping and interesting conceptual elements to that of regression trees with four assumed partitions. In regression trees, the traditional frequentist algorithm, see, Page 308 of \cite{friedman2001elements} proceeds via a greedy iterative process with branching transitions estimated in alternating directions via half planar splits. It may be observed that Algorithm 1 is infact doing something very similar but with two key refinements. First, it performs quadrant-wise splits instead of half planar splits. Second, while Algorithm \ref{alg:single} can also be viewed as being greedy, however, we additionally show that optimality is achieved in two successive iterations and thus further iterations shall be statistically redundant, i.e., further iterations will only serve to yield data specific improvements. Further discussions and generalizations of the model (\ref{model:rvmcp}) allowing multiple hierarchical change points that provide an alternative characterization of full and fully-frequentist regression tress are provided in Section \ref{sec:extensions}.

The remainder of this article is organized as follows. Section \ref{sec:main} provides a rigorous description of the estimation and inference results discussed above as well as the analytical behavior of the proposed Algorithm 1. Section \ref{sec:extensions} shall then provide extensions of model (\ref{model:rvmcp}) and corresponding methodology to multiple hierarchical changes/full regression trees. Section \ref{sec:application} illustrates the proposed methodologies on two distinct real data application, first on performing a segmentation of the {Infrared astronomy satellite} data and second on digital image segmentation and denoising. Section \ref{sec:numerical} provides numerical support to our methodology and results via monte-carlo simulations. We conclude this section with a short note on the notation used throughout the article.

\vspace{1.5mm}
\noi{\it Notation}: $\R$ represents the real line. For any vector $\delta\in\R^p,$ $\|\delta\|_1,$ $\|\delta\|_2,$ $\|\delta\|_{\iny}$ represent the usual 1-norm, Euclidean norm, and sup-norm respectively. For any set of indices $U\subseteq\{1,2,...,p\},$ let $\delta_U=(\delta_j)_{j\in U}$ represent the subvector of $\delta$ containing the components corresponding to the indices in $U.$ Let $|U|$ and $U^c$ represent the cardinality and complement of $U.$ We denote by $a\wedge b=\min\{a,b\},$ and $a\vee b=\max\{a,b\},$ for any $a,b\in\R.$ We use a generic notation $c_u>0$ to represent universal constants that do not depend on $T_w,T_h$ or any other model parameter. All limits are with respect to the sampling periods $T_w,$ and $T_h$ simultaneously or individually. The mean and the change parameters are assumed as sequences in these sampling periods $T_w,$ and $T_h,$ however this is notationally suppressed in all to follow. The notation $\Rightarrow$ represents convergence in distribution.

\section{Theoretical Analysis}\label{sec:main}\hfill

This section is divided into two sub-sections. Sub-section \ref{subsec:rate.limiting.dist} provides sufficient conditions and main theoretical results regarding the plugin least squares estimator $\tilde\tau$ of (\ref{est:optimal}). Specifically, near optimal and optimal rates of convergence of the estimators $\tilde\tau_w$ and $\tilde\tau_h,$ together with their limiting distributions in the two regimes described earlier. Sub-section \ref{subsec:alg1} aggregates these results recursively to establish the validity of the proposed Algorithm \ref{alg:single}.

\subsection{Rate of convergence and limiting distributions of $\tilde\tau(\h\tau,\h\theta)$}\label{subsec:rate.limiting.dist}\hfill

\vspace{1.5mm}
{\it {{\noi{\bf Condition A (on underlying distributions):}} The vectors $\vep_{(w,h)}=(\vep_{(w,h,1)},...,\vep_{(w,h,p)})^T,$ $w=1,..,T_w,$ $h=1,...,T_{h}$ are independent and identically distributed (i.i.d.) subexponential random vectors with variance proxy $\si^2<\iny$ (see, Definition \ref{def:sube} and \ref{def:submult})}}

\vspace{1.5mm}

The class of subexponential distribution is well known in the literature. The distributions included in this class are the Gaussian, Laplace, mean centered Exponential, mean centered Chi-square, centered mixtures of these distributions amongst several other well known distributions. We also note that Condition A does not exclude discrete distributions, such as mean centered Bernoulli, mean centered Poisson, or any centered and bounded distribution. The monograph \cite{vershynin2019high} provides a detailed study of this large class of distributions. As is apparent, this assumption is significantly weaker than assuming a Gaussian distribution which has commonly been assumed in the change point literature.

\vspace{1.5mm}
{\it {{\noi{\bf Condition B (on model parameters):}} (i) Covariance $\Sigma:=E\vep_{(w,h)}\vep_{(w,h)}^T$ has bounded eigenvalues, i.e., $0<\ka^2\le\rm{min eigen}(\Si) <\rm{max eigen}(\Si)\le\phi^2<\iny,$ with constants $\ka^2,$ $\phi^2.$\\
		(ii) Assume a change point exists and is separated from the parametric boundary on both axes, i.e., for some positive sequence $\underline\om\to 0,$ we have $\min_{j}\{|Q_j(\tau^0)|\}\ge T_wT_h\underline\om\to\iny,$\\~
		(iii) Let $\overline\xi$  and  $\xi_j,$ $j=1,...,4$ be as defined in (\ref{def:jump.size.quad}) and let $\xi_w,\xi_h,\xi_{\min}$ be as defined in (\ref{def:weight.jump.horiz.vert}). Then we assume that
		$\overline\xi\le c_u\xi_{\min},$ for some constant $c_u>0.$}}

\vspace{1.5mm}
Condition B(i) assumes a positive definite spatial dependence structure over components $1,...,p.$ We require the assumption of bounded eigenvalues only from an inference (limiting distributions) perspective. If the objective is only that of estimation, then these assumptions can be relaxed. In this case, $\ka^2$ may be allowed to converge to zero (or identically zero), i.e., potentially rank deficient. The upper bound $\phi^2$ may be allowed to diverge with $T_w,T_h.$ The bounds for the localization error of $\tilde\tau$ and thereby its rate of convergence provided later in this section are obtained upto universal constants. Consequently the effect of this relaxation will be directly observable in these bounds.

Condition B(ii) and B(iii) are both separation conditions that ensure the {\it jump signal} is not dominated by noise in order for the estimator to catch this signal. B(ii) ensures that there are a diverging number of observations in each induced quadrant of the model (\ref{model:rvmcp}). An analogous condition is also typical in a regression tree framework, see, e.g., Definition 3.1 of \cite{rockova2020posterior}. Condition B(iii) is slightly more technical, although it is still serving a similar purpose as B(i). This condition can be interpreted as the jump metrics in the horizontal and vertical direction ($\xi_w,\xi_h$) are not dominated by either individual half planes, for e.g. if one of $\xi_2$ or $\xi_4$ dominates $\xi_w$ in its rate of divergence, the method may be unable to detect the change in the horizontal direction, and symmetrically for the vertical direction.

Next consider the following sets of non-zero indices corresponding to the $p$-dimensional mean vectors $\theta_{(j)}^0,$ $j=1,2,3,4,$
\benr\label{def:setS}
S_{j}=\big\{k\in\{1,2,...,p\};\,\,\theta^0_{(j,k)}\ne 0\big\},\quad j=1,2,3,4,
\eenr
and let $S_j^c$ $j=1,2,3,4,$ be the complement sets. Define the maximum cardinality $\max_{1\le j\le  4}|S_{j}|=s\ge 1.$ The parameter $s$ measures sparsity in the model (\ref{model:rvmcp}).  This sparsity assumption is typically made on the jump vector,  as done in \cite{wang2018high} and \cite{enikeeva2013high} under a single change axis framework. In contrast we make this assumption directly on the mean vectors $\theta_{(j)}^0$s. This version of sparsity holds with no loss of generality with respect to the former version. We refer to Appendix C of \cite{kaul2020inference} for a discussion on this aspect. To allow the viability of this assumption one may center the observed data with component-wise empirical means, i.e., consider $x_{(w,h)}$ of model (\ref{model:rvmcp}) where instead of the means $\theta_{(j)}^0,$ the jump $\eta_{(j)}^0$ are $s$-sparse, i.e., there are mean changes in at most $s$ components. Upon centering $x_{(w,h)}$ with  empirical means, $x_{(w,h)}^*=x_{(w,h)}-\bar x,$ with $\bar x=\sum_{w,h} x_{(w,h)}\big/(T_wT_h),$ the $s$-sparsity of $\eta^0_{(j)}$ is transferred onto the new mean vectors $\theta^*=Ex_{(w,h)}^*,$ as $4s$-sparsity. Heuristically, this centering operation is same as that carried out in linear regression models to get rid of the intercept parameter, which is implicitly assumed in the high dimensional linear regression literature and is known not to impact rates of estimation.

In keeping with the discussion of Section \ref{sec:intro}, for our results on the plug-in estimator $\tilde\tau$ we are agnostic about the specific choice of the estimators used to obtain the preliminary estimates and instead rely on combinations of the following condition that describes these assumed preliminary estimate properties.

\vspace{1.5mm}
{\it {{\noi{\bf Condition C (on preliminary estimates):}} Let $c_{u1}>0$ be a suitably chosen small enough constant and let $\pi_T\to 0$ be a positive sequence. Then we assume either of the combinations of  \big[(i)(a), (ii)(a,b)\big] or \big[(i)(b), (ii)(a,c)\big] below hold with probability at least $1-\pi_T.$\\~
		(i)\,\, ({\bf on preliminary location estimates $\h \tau$}):\\~
		(a)\,\, Assume that $\h\tau_w,$ and $\h\tau_h$ satisfy the absolute error bound of (\ref{eq:40})\\~
		(b)\,\,Assume that $\h\tau_w,$ and $\h\tau_h$ satisfy the absolute error bound of (\ref{eq:41})\\~
		(ii)\,\, ({\bf on preliminary mean estimates $\h\theta$}): Assume one of the pairs (a,b) or (a,c)  hold.\\~  	
		(a)\,\, The estimates $\h\theta_{(j)},$ $j=1,...,4$ satisfy $\|(\h\theta_{(j)})_{S_j^c}\|_1\le 3\|(\h\theta_{(j)}-\theta_{(j)}^0)_{S_j}\|_1,$ for each $j=1,...,4.$ Here $S_j,$ are sets of non-zero components as defined in (\ref{def:setS}). \\~
		(b)\,\, The estimates $\h\theta_{(j)},$ $j=1,...,4$  satisfy the $\ell_2$ bound (\ref{eq:40})\\~
		(c)\,\,The estimates $\h\theta_{(j)},$ $j=1,...,4$  satisfy the $\ell_2$ bound (\ref{eq:41})}}

\vspace{1.5mm}
Condition C has been carefully constructed while keeping in mind its feasibility. The second combination of $\big[(i)(b), (ii)(a,c)\big]$ is stronger version of the first $\big[(i)(a), (ii)(a,b)\big].$ In Subsection \ref{subsec:alg1} we shall exploit this distinction to show that preliminary estimates obtained recursively via Algorithm 1 satisfy the two considered combinations of this condition, at the two successive iterations, respectively. Condition C(i) and C(ii)(b) are exceptionally weak conditions on the quality these estimates. C(i) is satisfied by any $\h\tau_w,\h\tau_h$ in $o(T)$-neighborhood's of $\tau^0_w,\tau^0_h,$ respectively, i.e., all that is required is them to be consistent at any arbitrary rate of estimation. Condition C(ii)(b) requires mean estimates to be of order of the jump size $\xi_{\min},$ and may be weaker than assuming even ordinary consistency, i.e., an $o_p(1)$ approximation. Condition C(ii)(a) in a sense provides a restriction on the sparsity level of the estimated mean parameters and is common in the $\ell_1$ regularization literature. Further, other common regularization mechanisms, such as {\it scad}, {\it mcp} or the Dantzig selector are also known to induce this property.

We note  here that Condition C(ii) allows mean estimates $\h\theta_{(j)}$ to be irregular, in the sense that they are only required to be in the given $\ell_2$ neighborhoods of $\theta^0_{(j)}$s. They are not required to possess oracle properties, i.e., selection mistakes in the identification of the signs of these coefficient do not influence the eventual change point estimate $\tilde\tau$ in its rate of convergence and limiting distribution. Accordingly, we do not require minimum magnitude conditions of the coefficient vectors $\theta^0_{(j)},$ $j=1,2,3,4,$ which are typically made under high dimensionality to guarantee selection consistency of signs in the components of these mean vectors.

The tightening of assumptions across combinations $\big[C(i)(a), C(ii)(a,b)\big]$ and  $\big[C(i)(b), C(ii)(a,c)\big]$  has important consequences on the rate of convergence of $\tilde\tau.$ This aspect shall become apparent after the following estimation results and the discussion thereafter.

\begin{theorem}\label{thm:cp.nearoptimal} Suppose the model (\ref{model:rvmcp}) and assume Condition A, B, C(i)(a) and C(ii)(a,b) hold. Then, we have,
	\benr
	&(i)& 	|\tilde\tau_w-\tau^0_w|\le c_u\si^2T_h^{-1}\xi_{w}^{-2}s\log^2(p\vee T_wT_h)\nn\\
	&(ii)& |\tilde\tau_h-\tau^0_h|\le c_u\si^2T_w^{-1}\xi_{h}^{-2}s\log^2(p\vee T_wT_h)\nn
	\eenr
	with probability at least $1-2\exp\{-c_{1}\log (p\vee T_wT_h)\}-\pi_T,$ for constant $c_1>0$ that does not depend on any model parameters. In other words, we have, $(\tilde\tau_w-\tau_w^0)=O\big(T_h^{-1}\xi^{-2}_{w}s\log^2(p\vee T_wT_h)\big)$ and $(\tilde\tau_h-\tau_h^0)=O\big(T_w^{-1}\xi^{-2}_{h}s\log^2(p\vee T_wT_h)\big),$ with probability at least $1-o(1),$ where the orders are w.r.t. $T_w$ and $T_h,$ respectively.
\end{theorem}

This result provides finite sample localization error bounds that are near-optimal and are obtained upto universal constants. This is obtained under the weaker condition  $\big[C(i)(a), C(ii)(a,b)\big]$ on the preliminary estimates. While this result is informative in itself from an estimation perspective, however it does not possess the optimality necessary for the existence of limiting distributions.  Next we show the more important result that the rate of convergence can be improved to optimality by making the sole change of tightening the preliminary estimates to the combination $\big[C(i)(b), C(ii)(a,c)\big].$

\begin{theorem}\label{thm:cpoptimal} Suppose the model (\ref{model:rvmcp}) and assume Condition A, B, C(i)(b) and C(ii)(a,c) hold. Then, for any $0<a<1$ and $c_a\ge \surd(1/a),$ we have,
	\benr
	&(i)& 	|\tilde\tau_w-\tau^0_w|\le c_uc_a^2\si^2T_h^{-1}\xi_{w}^{-2}\nn\\
	&(ii)& |\tilde\tau_h-\tau^0_h|\le c_uc_a^2\si^2T_w^{-1}\xi_{h}^{-2}\nn
	\eenr
	with probability at least $1-a-o(1)-\pi_T.$ Equivalently, we have, $(\tilde\tau_w-\tau_w^0)=O_p\big(T_h^{-1}\xi^{-2}_{w}\big)$ and $(\tilde\tau_h-\tau_h^0)=O_p\big(T_w^{-1}\xi^{-2}_{h}\big),$ where the orders are w.r.t. $T_w$ and $T_h,$ respectively.
\end{theorem}

Theorem \ref{thm:cpoptimal} provides the optimal rate of convergence of $\tilde\tau.$ This is the same rate of convergence one would have obtained for $\tilde\tau_w$ if the parameters $\tau_h^0$ and $\theta_{(j)}^0,$$j=1,2,3,4$ used in its construction were known, and analogous for $\tilde\tau_h.$ This observation provides a key insight, it allows one to conclude that $\tilde\tau$ statistically behaves as if these preliminary estimates are known. This property in turn shall allow limiting distributions to exist and be characterized. This is effectively the adaptation property as described in \cite{bickel1982adaptive} but is observed here despite potential high dimensionality of plug-in estimates and in context of change point parameters.

We can now proceed to establishing the limiting distributions of $\tilde\tau.$ To this end, begin by noting that as a direct consequence of Theorem \ref{thm:cpoptimal} one may observe that when $\surd{T_h}\xi_w\to \iny,$ and $\surd{T_w}\xi_h\to \iny,$ then the estimates $\tilde\tau_w$ and $\tilde\tau_h,$ perfectly identify the corresponding change point parameters,  in probability, i.e., the limiting distribution of $\tilde\tau$ in these cases are degenerate. As a result in the following we shall only be concerned with two regimes, first of a vanishing jump $\surd{T_h}\xi_w\to 0,$ $\surd{T_w}\xi_h\to 0,$ or that of a non-vanishing jump $\surd{T_h}\xi_w\to \xi_{(w,\iny)},$ $\surd{T_w}\xi_h\to \xi_{(h,\iny)}$ where $0<\xi_{(w,\iny)},\,\,\xi_{(h,\iny)}<\iny.$ We begin here with a mild condition that shall ensure stability of asymptotic variances of the limiting processes to be characterized.

\vspace{1.5mm}
{\it {{\noi{\bf Condition D (stability of asymptotic variances):}} Let $\Si,$ $\eta^0_{(j)},$ $j=1,2,3,4,$ and $\xi_w,\xi_h$ be as defined in Condition B, (\ref{def:jump.vec.quadrants}) and (\ref{def:weight.jump.horiz.vert}) respectively. Then, assume the following limits exist,
		\benr
		&(i)&\,\,\, \frac{1}{\xi^2_w}\Big[\om_h\eta^{0T}_{(1)}\Si\eta^0_{(1)}+(1-\om_h)\eta^{0T}_{(3)}\Si\eta^0_{(3)}\Big]\to \si^2_{(w,\iny)},\quad{\rm and}\nn\\
		&(ii)&\,\,\, \frac{1}{\xi^2_h}\Big[\om_w\eta^{0T}_{(4)}\Si\eta^0_{(4)}+(1-\om_w)\eta^{0T}_{(2)}\Si\eta^0_{(2)}\Big]\to \si^2_{(h,\iny)},\nn
		\eenr
		with $0<\si_{(w,\iny)},\,\,\si_{(h,\iny)}<\iny.$ Here the limit of (i) is with respect to $T_w\to\iny$ (with $T_h<\iny$ or $T_h\to\iny$), and symmetrically (ii) is with respect to  $T_h\to\iny$ (with $T_w<\iny$ or $T_w\to\iny$).}}		

\vspace{1.5mm}
Recall that all limits in the article are w.r.t. sampling periods $T_w$ or $T_h$ (either individually of simultaneously). The limits of Condition D are acting in $T_w,T_h$ via the dimension $p$ and the jump sizes $\xi_w,\xi_h.$
The quantities $\si^2_{(w,\iny)}$ and $\si^2_{(h,\iny)}$ shall serve as variance parameters of the limiting processes described in (\ref{eq:limiting.d.vanishing}) and (\ref{eq:limiting.d.non.vanishing}), thus the need for their stability.
Note that finiteness of the limits appearing in Condition D are already guaranteed by prior assumptions, and this condition only assumes their stability. To see this,  consider Part (i) and note that the assumed convergence is on a sequence that is guaranteed to be bounded, i.e.,
\benr
&&\frac{1}{\xi^2_w}\Big[\om_h\eta^{0T}_{(1)}\Si\eta^0_{(1)}+(1-\om_h)\eta^{0T}_{(3)}\Si\eta^0_{(3)}\Big]\ge \frac{\ka^2}{\xi_{w}^2}(\om_h\xi_1^2+(1-\om_h)\xi_3^2)\ge  \ka^2>0,\,\,{\rm and}\nn\\
&&\frac{1}{\xi^2_w}\Big[\om_h\eta^{0T}_{(1)}\Si\eta^0_{(1)}+(1-\om_h)\eta^{0T}_{(3)}\Si\eta^0_{(3)}\Big]\le
\frac{\phi^2}{\xi_{w}^2}(\om_h\xi_1^2+(1-\om_h)\xi_3^2)\le  \phi^2<\iny,\nn
\eenr
and similar for Part (ii). Here the inequalities follow from the bounded eigenvalues assumption on $\Si$ \big(Condition B(i)\big). An easier to interpret, but stronger sufficient condition for the finiteness for the limits with respect to the covariance matrix $\Si$ is by assuming absolute summability of each row or column of $\Si.$ This condition is satisfied by large classes of covariances such as banded and toeplitz type matrices. We refer to Condition D of \cite{kaul2021graphical} for further details on this argument.

\begin{theorem}[Limiting distribution under vanishing jump regime]\label{thm:wc.vanishing} Suppose Conditions A, B and D hold. Assume that the jump sizes are vanishing $\surd{(T_h)}\xi_w\to 0,$ and $\surd{(T_w)}\xi_h\to 0.$ Let the parameters $\tau^0_w,\tau^0_h$ and $\theta_{(j)}^0,$ $j=1,2,3,4,$ be known and let $\tilde\tau^*_w=\tilde\tau_w(\tau_h^0,\theta^0),$ $\tilde\tau^*_h=\tilde\tau_h(\tau_w^0,\theta^0).$ Then, we have,
	\benr\label{eq:wc.vanishing}
	&&T_h^{-1}\xi^{2}_w(\tilde\tau^*_w-\tau^0_w)\Rightarrow \argmax_{\z\in\R}\big\{2\si_{(w,\iny)}W_w(\z)-|\z|\},\qquad T_w\to\iny,\nn\\
	&&T_w^{-1}\xi^{2}_h(\tilde\tau^*_h-\tau^0_h)\Rightarrow \argmax_{\z\in\R}\big\{2\si_{(h,\iny)}W_h(\z)-|\z|\},\hspace{2mm}\qquad T_h\to \iny,
	\eenr
	where $W_w(\z),$ and $W_h(\z)$ are both two sided Brownian motions\footnote{A two-sided Brownian motion $W(\z)$ is defined as $W(0) = 0,$ $W(\z) = W_1(\z),$ $\z > 0$ and $W(\z) = W_2(-\z),$ $\z < 0,$ where $W_1(\z),$ $W_2(\z)$ are two independent Brownian motions defined on the non-negative half real line}. Alternatively, when $\tau^0_w,$ $\tau^0_h$ and $\theta_{(j)}^0,$ $j=1,2,3,4,$ are unknown, suppose $\tilde\tau_w=\tilde\tau_w(\h\tau_h,\h\theta),$ $\tilde\tau_h=\tilde\tau_h(\h\tau_w,\h\theta).$ assume Condition C(i)(b) and C(ii)(a,c) are satisfied with $	r_T=o(1)\big/\{s^{1/2}\log(p\vee T)\}.$ Then, the convergence (\ref{eq:wc.vanishing}) also holds when $\tilde\tau^*_w,\tilde\tau_h^*$ are replaced with $\tilde\tau_w,\tilde\tau_h,$ respectively.
\end{theorem}

The limiting distributions of $\tilde\tau_w$ and $\tilde\tau_h$ can be utilized to construct asymptotically valid component-wise confidence intervals for the change parameters in the horizontal and vertical directions under the assumed vanishing jump regime. It can be observed that a change of variable to $\z=\si_{\iny}^2\z',$ yields that $\argmax_{\z\in\R}\big\{2\si_{\iny}W(\z)-|\z|\}=^d\si_{\iny}^2\argmax_{\z'\in\R}\big\{2W(\z')-|\z'|\},$ which in turn yields the relations in (\ref{eq:limiting.d.vanishing}) provided in Section \ref{sec:intro}. This distribution is well studied in the literature and its cdf is available in \cite{yao1987approximating}.

Next we consider non-vanishing regime of $\surd{(T_h)}\xi_w\to \xi_{(w,\iny)},$ and $\surd{(T_w)}\xi_h\to \xi_{(h,\iny)}.$ The literature on distributional properties of $\tilde\tau$ in this case is quite sparse. The only articles that provide an examination of this regime under high dimensional asymptotics are \cite{kaul2020inference} and \cite{kaul2021graphical}, under a single change axis framework. In the same framework, the articles of \cite{bhattacharjee2017common} and \cite{bhattacharjee2019change} consider the diverging $p$ case with $p<<T.$ We require an additional distributional assumption for the analysis of this regime provided below.

\vspace{1.5mm}
{\it {{\noi{\bf Condition A$'$ (additional distributional assumptions):}} Suppose Condition A, B and D hold. Assume the non-vanishing jump size regime, i.e.,  $\surd{(T_h)}\xi_w\to\xi_{(w,\iny)},$ and $\surd{(T_w)}\xi_h\to\xi_{(h,\iny)},$ with $0<\xi_{(w,\iny)},\xi_{(h,\iny)}<\iny.$ For each $w=1,...,T_w$ and $h=1,...,T_h$ define,
		\benr
		&&\psi_{w,T_w}=\Big[\sum_{h=\tau^0_h+1}^{T_h}\vep_{(w,h)}^T\eta^0_{(1)}+\sum_{h=1}^{\tau_h^0}\vep_{(w,h)}^T\eta^0_{(3)}\Big],\quad{\rm and}\nn\\
		&&\psi_{h,T_h}=\Big[\sum_{w=\tau^0_w+1}^{T_w}\vep_{(w,h)}^T\eta^0_{(4)}+\sum_{w=1}^{\tau_w^0}\vep_{(w,h)}^T\eta^0_{(2)}\Big]\nn
		\eenr
		The we assume that for any constants $c_1,c_2\in\R,$ and for some  distribution $\cP,$ which is continuous and supported in $\R,$ we have,
		\benr
		&(i)& c_1+c_2\psi_{w,T_w}\Rightarrow \cP\big(c_1,c_2^2\xi_{(w,\iny)}^2\si^2_{(w,\iny)}\big),\quad{\rm and,}\nn\\
		&(ii)& c_1+c_2\psi_{h,T_h}\Rightarrow \cP\big(c_1,c_2^2\xi_{(h,\iny)}^2\si^2_{(h,\iny)}\big).\nn
		\eenr
		Here $\si^2_{(w,\iny)}$ and $\si^2_{(h,\iny)}$ are as defined in Condition D.}}

Note that the only additional requirement imposed by Condition B$'$, in comparison to Conditions B and D, is that the random variables under consideration are continuously distributed, which is trivially true in the typically assumed Gaussian framework. To see this, note that the mean $E\psi_{w,T_w}=c_1$ follows directly from definition of $\psi_{w,T_w}$ and by the zero mean assumption of $\vep_{(w,h)}.$ The variance ${\rm var}(\psi_{w,T_w})\to c_2^2\xi_{(w,\iny)}\si^2_{(w,\iny)}$ follows from Condition D together with the considered non-vanishing jump regime. Consequently, the limiting distribution of the sequence $\psi_{w,T_w}$ (in $T_w$ via $p$ and $\xi_w$) is well defined, i.e. i.e., supported in $\R.$ Consequently, Condition B$'$ simply provides a notation $\cP$ to whatever distribution this may be. Analogously for the sequence $\psi_{h,T_h}.$ We also note there that the arguments in the notation $\cP(\mu,\si^2)$ are used to represent the mean and variance of the distribution $\cP,$ i.e,  $E\cP(\mu,\si^2)=\mu,$ and ${\rm var}\big(\cP(\mu,\si^2)\big)=\si^2.$  Further note that the representation $\cP(\mu,\si^2)$ is only for ease of presentation and does not imply that $\cP$ is characterized by only its mean and variance. If one assumed $\vep_{(w,h)}\sim \cN(0,\Si)$ on the data generating process of (\ref{model:rvmcp}), then it is straightforward to observe that $\cP=^d\cN.$

The two-sided random walk defined in (\ref{eq:neg.drift.rw}) can now be utilized in order to characterize the limiting distribution of the change point estimators $\tilde\tau_w$ and $\tilde\tau_h,$ in the current non-vanishing jump size regime. For the width change parameter this stochastic process shall have the increments as $z_t\sim^{i.i.d}\cP\big(-\xi_{(w,\iny)}^2, \,\,4\xi_{(w,\iny)}^2\si^2_{(w,\iny)}\big)$ and $z_t^*\sim^{i.i.d}\cP\big(-\xi_{(w,\iny)}^2, \,\,4\xi_{(w,\iny)}^2\si^2_{(w,\iny)}\big),$ and $z_t$ and $z_t^*$ are also independent of each other over all $t,$ with an analogous construction for the height change parameter. The only additional assumption of Condition B$',$ of continuity of the distribution law $\cP$ is assumed for the regularity of the \textit{argmax} of this two sided random walk.

\begin{theorem}[Limiting distribution under non-vanishing jump regime]\label{thm:wc.non.vanishing} Suppose Conditions A$'$, B, D hold. Assume that the jump sizes are non-vanishing $\surd{(T_h)}\xi_w\to \xi_{(w,\iny)},$ and $\surd{(T_w)}\xi_h\to \xi_{(h,\iny)}.$ Let the parameters $\tau^0_w,\tau^0_h$ and $\theta_{(j)}^0,$ $j=1,2,3,4,$ be known and let $\tilde\tau^*_w=\tilde\tau_w(\tau_h^0,\theta^0),$ $\tilde\tau^*_h=\tilde\tau_h(\tau_w^0,\theta^0).$ Then, we have,
	\benr\label{eq:wc.non.vanishing}
	&&(\tilde\tau^*_w-\tau^0_w)\Rightarrow \argmax_{\z\in \Z}\cC_{(w,\iny)}(\z),\qquad T_w\to\iny,\nn\\
	&&(\tilde\tau^*_h-\tau^0_h)\Rightarrow \argmax_{\z\in \Z}\cC_{(h,\iny)}(\z),\hspace{2mm}\qquad T_h\to \iny,
	\eenr
	where $\cC_{(w,\iny)}(\z)$ and $\cC_{(h,\iny)}(\z)$ are as defined in (\ref{eq:neg.drift.rw}) and (\ref{eq:42}). Alternatively, when $\tau^0_w,$ $\tau^0_h$ and $\theta_{(j)}^0,$ $j=1,2,3,4,$ are unknown, suppose $\tilde\tau_w=\tilde\tau_w(\h\tau_h,\h\theta),$ $\tilde\tau_h=\tilde\tau_h(\h\tau_w,\h\theta).$ assume Condition C(i)(b) and C(ii)(a,c) are satisfied with $	r_T=o(1)\big/\{s^{1/2}\log(p\vee T)\}.$ Then, the convergence (\ref{eq:wc.vanishing}) also holds when $\tilde\tau^*_w,\tilde\tau_h^*$ are replaced with $\tilde\tau_w,\tilde\tau_h,$ respectively.
\end{theorem}

The only distinction between the assumptions of Theorem \ref{thm:wc.vanishing} and Theorem \ref{thm:wc.non.vanishing} is the change of regime from a vanishing jump size to the non-vanishing jump size regime, respectively. Since the analytical form of the distribution $\argmax_{\z\in\Z}\cC_{(w,\iny)}(\z)$ and $\argmax_{\z\in\Z}\cC_{(h,\iny)}(\z)$ are unavailable, one may resort to obtaining quantiles of these distributions via monte-carlo simulations, i.e., simulating the two sided random walk process and in turn obtaining realizations from the distribution under consideration.

While the above discussion provides the statistical properties of $\tilde\tau,$ however note that it is not as yet implementable in practice since its construction requires preliminary estimates that have so far not been explicitly defined. Nonetheless, one may perhaps be surprised to note that thus far we have not assumed any conditions on the rate of divergence of the model dimensions ($s,p$) or on the jump size $\xi_{\min},$ w.r.t the sampling periods $T_w,T_h.$ The reason for this observation is that, effectively, the burden of these assumptions have been pushed to the preliminary nuisance estimates through Condition C. These conditions shall materialize in the following sub-section where we provide a feasible methodology along with the construction of nuisance estimates. Some intuition into the origination of these rate assumptions and their inter-relationship to Condition C are discussed below.

In order to aggregate our theoretical results into a feasible twice iterative algorithm, we shall utilize the near optimal change point estimates of Theorem \ref{thm:cp.nearoptimal} as preliminary nuisance estimates feeding into Theorem \ref{thm:cpoptimal}. Consequently we shall require the near optimal rate yielded by Theorem \ref{thm:cp.nearoptimal} to satisfy the assumed requirement of Condition C(i)(b) for  Theorem \ref{thm:cpoptimal}.  Accordingly, for this relation to be maintained, we must have,
\benr\label{eq:32}
\frac{c_u\si}{\xi_{\min}}\Big\{\frac{s\log^2(p\vee T_wT_h)}{\surd(T_wT_h)}\Big\}\le c_{u1},
\eenr
for a suitably chosen small enough constant $c_{u1}>0.$ This aspect provides the strongest requirement on the rate of divergence of model dimensions and the jump size that we shall require for the validity of all results to follow. The preliminary mean estimates shall also be feasible to obtain under the same requirement. To see this, note that the sharpest $\ell_2$ rate of estimation of mean parameters $\h\theta_{(j)}$ is known to be,
\benr\label{eq:31}
\max_j\|\h\theta_{(j)}-\theta_{(j)}^0\|_2\le r_T=c_u\si\Big(\frac{s\log (p\vee T_wT_h)}{T_wT_h\underline\om}\Big)^{\frac{1}{2}},
\eenr
with probability $1-o(1).$ Now Comparing (\ref{eq:31}) with Condition C(ii)(c) one can observe that in order to maintain viability of Condition C(ii)(c), we must have,  $c_us\log^{3/2}(p\vee T_wT_h)\le c_{u1}\xi_{\min}\surd{(T_wT_h\underline\om)}.$ Finally observe that this rate requirement is weaker than (\ref{eq:32}) in all but the weight parameter $\underline\om.$ Thus in the interest of notational simplicity, we shall be utilizing the condition (\ref{eq:33}) described in Section \ref{sec:intro}. The following sub-section provides a detailed examination of this discussion.

\subsection{Preliminary estimates $\h\tau,$ $\h\theta$ and validity of Algorithm 1}\label{subsec:alg1}\hfill

The main purpose of this sub-section is to fill gaps that remain to allow the feasibility of $\tilde\tau,$ i.e., to develop preliminary estimates $\h\tau,\h\theta$ in a principled manner so that results of Theorem's \ref{thm:cpoptimal} - \ref{thm:wc.non.vanishing} can be appealed to in context of  the output of Algorithm 1.  To this end, recall that these preliminary estimates require either condition's $\big[C(i)(a), C(ii)(a,c)\big]$ (milder) to obtain a near optimal estimate, or condition's $\big[C(i)(b),C(ii)(a,c)\big]$ (stronger) to obtain an optimal estimate of $\tau^0.$ Algorithm \ref{alg:single} utilizes the soft thresholded means (\ref{est:softthresh}) and the distinctions between the pairs of assumed conditions to provide an estimator that improves a nearly arbitrarily chosen $\check\tau,$ to a near optimal estimate $\h\tau$ in a first iteration, and then to an optimal estimate $\tilde\tau$ in a second iteration. To further describe the idea and constraints behind the validity of Algorithm 1 we require the following two additional conditions.

\vspace{1.5mm}
{\it {{\noi{\bf Condition E (on dimensional and jump size rate restrictions):}}  Let  $\xi_{\min}$ be as defined in (\ref{def:weight.jump.horiz.vert}) and let $s,p$ be the sparsity parameter (see, (\ref{def:setS})) and dimension size, respectively. Then, for an appropriately chosen small enough constant $c_{u1}>0,$ assume the following relation holds.
		\benr
		\Big(\frac{c_u\si}{\xi_{\min}}\Big)\Big\{\frac{s\log^{2} (p\vee T_wT_h)}{\surd (T_wT_h\underline\om)}\Big\}\le c_{u1}.\nn
		\eenr
		Additionally, assume that $s\log(p\vee T_wT_h)\le c_u T_wT_h\underline\om,$ for some constant $c_u>0.$
}}

The purpose of Condition E is chiefly to ensure that the near optimal estimates obtained from the first iteration of Algorithm \ref{alg:single} satisfy the sharper conditions of \big[C(i)(b), C(ii)(c)\big]. This allows Algorithm \ref{alg:single} to proceed to step 2 with these as the preliminary estimates.

Next recall that condition C(i)(a) is very mild, in particular, all it requires are any $\tau_w,\tau_h$ in $o(T_w)$ or $o(T_h)$ neighborhood's of the change parameters $\tau^0_w,\tau^0_h,$ respectively. Following is an almost equivalent version of this condition and a discussion on the utility, requirement and viability of both of them immediately thereafter.

\vspace{1.5mm}
{\it {{\noi{\bf Condition F (initializer of Algorithm \ref{alg:single}):}} Let $\psi=\max_{1\le j\le 4}\|\eta^0_{(j)}\|_{\iny},$ and assume that the initializer $\check\tau=(\check\tau_w,\check\tau_h)^T$ of Algorithm \ref{alg:single} satisfies the relations.
		\benr
		(i)\,\, |\check\tau_w-\tau^0_w|\le \frac{c_{u1}T_w\underline\om}{\big(\surd{s\psi\big/\xi_{\min}}\big)},\quad{\rm and}\quad  |\check\tau_h-\tau^0_h|\le (ii)\,\,\frac{c_{u1}T_h\underline\om}{\big(\surd{s\psi\big/\xi_{\min}}\big)}.\nn
		\eenr		
		Additionally assume  $(iii)\,\,\min_{1\le j\le }|Q_j(\check\tau)|\ge c_uT_wT_h\underline\om.$ Here $\underline\om$ is as defined in Condition B, $c_{u}>0$ is any constant and $c_{u1}>0$ is an appropriately chosen small enough constant.}}

Requirement (iii) of Condition F is clearly innocuous, all it requires is a marginal separation of the chosen $\check\tau$ from the parametric boundary of the $2d$ change point. It is satisfied with $\check\tau=(\lfloor T_wk_w\rfloor,\,\lfloor T_hk_h\rfloor)^T=,$ with any $(k_w,k_h)^T\in [c_{u1},c_{u2}]\times[c_{u1},c_{u2}]\subset(0,1)\times(0,1).$

Requirement (i) and (ii) are symmetrical versions in the horizontal and vertical directions, respectively. Thus we only discuss the mildness of (i), first from a theoretical and then followed by a practical perspective, these arguments shall also symmetrically hold for (ii). We begin by illustrating the near equivalence of F(i) to Condition C(i)(a). Consider the case when $\underline\om\ge c_u,$ i.e., the true change point in the fractional scale is in some bounded subset of $(0,1)\times(0,1),$ and that $\big(\surd{s\psi\big/\xi}\big)=O(1),$ i.e., the entries of the change vectors $\eta^0_w,\eta^0_h$ are roughly evenly spread across its non-zero components and not with uneven diverging spikes, this is also satisfied if one assumes $\psi\le c_u,$ i.e., when all mean parameters are bounded above. Both of these restrictions are common to the change point literature. Then, requirement F(i) becomes identical to C(i)(a), moreover, both are satisfied for all $\check\tau_w$ in an $o(T_w)$ neighborhood of $\tau^0_w,$ i.e., any $\check\tau_w$ satisfying $|\check\tau_w-\tau^0_w|=o(T_w).$ The reason for these conditions to appear separately despite their near equivalence is that Condition F ensures weak regularity of the initializing mean estimates $\check\theta_{(j)},$ $j=1,2,3,4$ and Condition C(i)(a)
ensures a weak regularity of the Step 1 (of Algorithm \ref{alg:single}) change point estimate $\h\tau.$ One may interpret both Condition C(i)(a) and Condition F as ordinary consistency of the initializer $\check\tau,$ while also recalling that this property under bounded a parametric space is a very weak statement.

\begin{figure}[]
	\centering
	\begin{tikzpicture}[scale=0.7]
	\draw (-6,1)  ellipse (1cm and 0.5cm) node (A) {\footnotesize{$\check\tau_w$}};
	\draw (-6,-1)  ellipse (1cm and 0.5cm) node (B) {\footnotesize{$\check\tau_h$}};
	\node (ctheta) at (-2.5,-2) [draw,minimum width=6mm,minimum height=4mm,align=left] {\footnotesize{$\check\theta_j=\h\theta_j(\check\tau)$} \\ {\footnotesize $j=1,2,3,4.$}};
	\draw (2,1) ellipse (1.6cm and 0.7cm)  node (D) {\footnotesize{$\h\tau_w=\tilde\tau(\check\tau_h,\check\theta)$}};
	\draw (2,-1) ellipse (1.6cm and 0.7cm) node (E) {\footnotesize{$\h\tau_h==\tilde\tau(\check\tau_w,\check\theta)$}};
	\node (htheta) at (6.5,-2) [draw,minimum width=6mm,minimum height=4mm,align=left] {\footnotesize{$\h\theta_j=\h\theta_j(\h\tau)$} \\ {\footnotesize $j=1,2,3,4.$}};
	\draw (10.75,1) ellipse (1.6cm and 0.7cm)  node (F) {\footnotesize{$\tilde\tau_w=\tilde\tau(\h\tau_h,\h\theta)$}};
	\draw (10.75,-1) ellipse (1.6cm and 0.7cm) node (G) {\footnotesize{$\tilde\tau_h==\tilde\tau(\h\tau_w,\h\theta)$}};
	\node (Cond1) at (-6,-5) [draw,minimum width=6mm,minimum height=6mm,align=left] {\footnotesize{Condition} \\ \footnotesize {F satisfied}\\ \footnotesize {(nearly}\\ \footnotesize{arbitrary}\\ \footnotesize{choice)}};
	\node (Cond2) at (-2.5,-5) [draw,minimum width=6mm,minimum height=6mm,align=left] {\footnotesize{Condition} \\ {\footnotesize {C(ii)(b) satisfied}}};
	\node (Cond3) at (2,-5) [draw,minimum width=6mm,minimum height=6mm,align=left] {\footnotesize{Near optimal,} \\ \footnotesize{Condition}\\ \footnotesize{C(i)(b) satisfied}};
	\node (Cond4) at (6.5,-5) [draw,minimum width=6mm,minimum height=6mm,align=left] {\footnotesize{Condition} \\ {\footnotesize {C(ii)(c) satisfied}}};
	\node (Cond5) at (10.75,-5) [draw,minimum width=6mm,minimum height=6mm,align=left] {\footnotesize{Optimal,}\\ \scriptsize{$O_p\big(T^{-1}\xi^{-2}\big)$}};
	\path [line] (A) -- (ctheta);\path [line] (B) -- (ctheta);
	\path [line] (ctheta) -- (D);\path [line] (ctheta) -- (E);
	\path [line] (A) -- (E);\path [line] (B) -- (D);
	\path [line] (D) -- (htheta);	\path [line] (E) -- (htheta);
	\path [line] (htheta) -- (F);	\path [line] (htheta) -- (G);
	\path [line] (D) -- (G);	\path [line] (E) -- (F);
	\draw[dashed] (-6,0.5) -- (-6,-0.5); 	\draw[dashed] (-6,-1.5) -- (-6,-3.5);
	\draw[dashed] (-2.5,-2.7) -- (-2.5,-4.5); 	\draw[dashed] (2,-1.7) -- (2,-4);
	\draw[dashed] (2,0.3) -- (2,-0.35);	\draw[dashed] (6.5,-2.7) -- (6.5,-4.5);
	\draw[dashed] (10.75,-1.7) -- (10.75,-4.25); 	\draw[dashed] (10.75,0.3) -- (10.75,-0.35);
	\path [line] (Cond1) -- (Cond2);\path [line] (Cond2) -- (Cond3);
	\path [line] (Cond3) -- (Cond4);\path [line] (Cond4) -- (Cond5);
	\draw[decorate,decoration={brace,amplitude=10pt},xshift=4pt,yshift=0pt] (-4,2) -- (3.5,2) node [black,midway,yshift=14pt,xshift=-10pt] {Step 1};
	\draw[decorate,decoration={brace,amplitude=10pt},xshift=4pt,yshift=0pt] (4.5,2) -- (12.5,2) node [black,midway,yshift=14pt,xshift=-10pt] {Step 2};
	\draw[decorate,decoration={brace,amplitude=9pt},xshift=4pt,yshift=0pt] (-7.25,2) -- (-5,2) node [black,midway,yshift=14pt,xshift=-10pt] {Initialize};
	\end{tikzpicture}
	\caption{\footnotesize{A schematic of the underlying working mechanism of Algorithm \ref{alg:single}.}}
	\label{fig:schematic}
\end{figure}
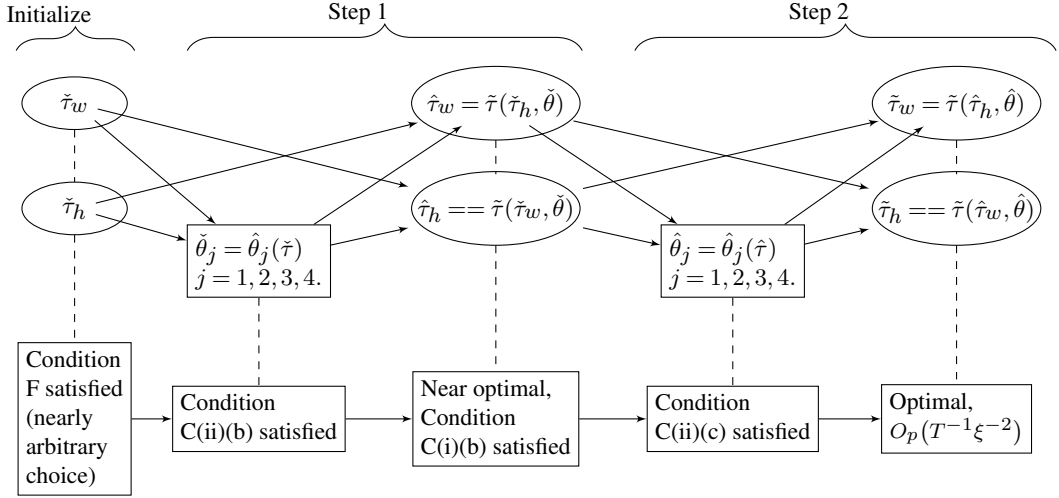

We can now describe the working mechanism of Algorithm \ref{alg:single} in its entirety. Any nearly arbitrarily (Condition F) chosen $\check\tau=(\check\tau_w,\check\tau_h)^T,$ yields Step 1 means $\check\theta_{(j)}=\h\theta_{(j)}(\check\tau),$ $j=1,2,3,4,$ of (\ref{est:softthresh}) that satisfy the weaker Condition C(ii)(a,c). Theorem \ref{thm:cp.nearoptimal} now guarantees that update $\h\tau=(\h\tau_w,\h\tau_h)^T,$ where $\h\tau_w=\tilde\tau_w\big(\check\tau_h,\check\theta),$ and similar for $\h\tau_h,$ shall be a near optimal estimate of $\tau^0.$ Under the rate assumption of Condition D, this near optimal $\h\tau$ together with the updated mean estimates $\h\theta_{(j)}=\h\theta_{(j)}(\h\tau),$ $j=1,2,3,4,$ satisfy the stronger requirements of Condition C(i)(b) and C(ii)(a,c). This allows us to perform another update $\tilde\tau=(\tilde\tau_w,\tilde\tau_h),$ with $\tilde\tau_w=\tilde\tau(\h\tau_h,\h\theta),$ and similar for $\h\tau_h.$ Theorem \ref{thm:cpoptimal} now guarantees optimality of this Step 2 updated estimate $\tilde\tau,$ moreover, its limiting distributions can also be characterized as per Theorem \ref{thm:wc.vanishing} and Theorem \ref{thm:wc.non.vanishing}. Thus, in performing these updates (two each of the change point and the mean, with internal iterations on the components of $\tau$) we have taken a $\check\tau$ from a nearly arbitrary neighborhood of $\tau^0,$ and deposited it in an optimal neighborhood of $\tau^0,$ with an intermediate $\h\tau$ that lies in a near optimal neighborhood, i.e.  $o(T_w)$-nbd.$\longrightarrow^{\rm Step 1}$ near optimal-nbd., $O_p(T_h^{-1}\xi^{-2}_ws\log^2 p)$ $\longrightarrow^{\rm Step 2}$ optimal-nbd., $O_p(T_h^{-1}\xi^{-2}_w),$ in context of the width change parameter $\tau_w$, and symmetrically for $\tau_h.$ This is the process stated as Algorithm \ref{alg:single} and described visually in Figure \ref{fig:schematic}.

From a practical perspective, robustness w.r.t choice of initializer $\check\tau$ has also been illustrated in \citep{kaul2019efficient,kaul2020inference} in other dynamic contexts through extensive numerical experiments. We note that this initializer question also arise in the traditional regression tree algorithm where the same robustness has been observed in innumerable studies. While we have characterized the initializing Condition F, similar conditions have also been implicitly assumed in the literature, e.g., \cite{atchade2017scalable}. Nevertheless, one may choose a theoretically valid initializer $\check\tau$ satisfying Condition F by utilizing a preliminary coarse grid search, e.g. one may choose any slowly diverging sequence  (say $\log (\cdotp)$ ) and choose $\log T_w$ equally separated values in $\{1,...,T_w\},$ and  $\log T_h$ equally separated values in $\{1,...,T_h\},$ forming a coarse grid of $\log T_w\log T_h$ possible $2d$-initializer values. Upon choosing the best fitting value $\check\tau=(\check\tau_w,\check\tau_h)$ for Algorithm \ref{alg:single} from this coarse initializer grid and assuming that the best fitting value is component-wise closest to $\tau^0,$ amongst the chosen grid points. Then by the pigeonhole principle the choice of $\check\tau_w$ and $\check\tau_h$ must be in an $T_w/\log T_w=o(T_w)$ and $T_h/\log T_h=o(T_h)$ neighborhoods of $\tau^0_w$ and $\tau^0_h,$ respectively. Thereby this $\check\tau$ shall form a theoretically valid initializer. A similar preliminary coarse grid search has also been heuristically utilized in \cite{roy2017change} in a different model setting, where it also points towards an implicit need for an assumption similar to Condition F. In applications of Section \ref{sec:application} and simulation experiments of Section \ref{sec:numerical} we consider a preliminary grid search of $(\check\tau_w,\check\tau_h)\in\{\lfloor 0.25\cdotp T_w\rfloor,\lfloor 0.5\cdotp T_w\rfloor,\lfloor 0.75\cdotp T_w\rfloor\}\times \{\lfloor 0.25\cdotp T_h\rfloor,\lfloor 0.5\cdotp T_h\rfloor,\lfloor 0.75\cdotp T_h\rfloor\}$ to choose the initializer.

The following corollary provides a precise description of the above discussion, in particular, it aggregates the results of Sub-section \ref{subsec:rate.limiting.dist} to provide the validity of Algorithm \ref{alg:single}, in context of its estimation and inference properties.

\begin{cor}\label{cor:alg1.validity} Suppose Condition A, B, E and F hold and assume the regularizers for mean estimates of Step 1 of Algorithm 1 are chosen as in (\ref{eq:la.step1.choice}), Then, \\~
	(a) the Step 1 estimate $\h\tau=(\h\tau_w,\h\tau_h)^T$ satisfies the near optimal bounds of Theorem \ref{thm:cp.nearoptimal}.\\~
	Additionally, suppose the regularizers for mean estimates of Step 2 are chosen as in  (\ref{eq:la.step2.choice}) and assume  $(\psi/{\xi_{\min}}\big)\le c_u\surd\{\log(p\vee T)\}.$ Then, \\~
	(b) the Step 2 estimate $\tilde\tau=(\tilde\tau_w,\tilde\tau_h)^T$ satisfies the optimal error bounds of Theorem \ref{thm:cpoptimal}.\\~
	Furthermore, suppose the rate assumption of Condition E is slightly tightened to $\{s\log^2(p\vee T_wT_h)\}=o\big(\xi_{\min}\surd (T_wT_h\underline\om)\big)$\footnote{We do not necessarily require the order $o(1)$ here to hold simultaneously w.r.t $T_w,T_h.$ If this order holds w.r.t. $T_w\to\iny$ ($T_h<\iny$ or $T_h\to\iny$) then it is sufficient in context of the width change parameter $\tau_w^0,$ and symmetrically for the height change parameter.}, and assume Condition A$'$ and D holds. Then,\\~
	(c) the Step 2 estimate $\tilde\tau=(\tilde\tau_w,\tilde\tau_h)^T$ satisfies the limiting distributions of Theorem \ref{thm:wc.vanishing} and Theorem \ref{thm:wc.non.vanishing}, in the vanishing and non-vanishing jump size regimes, respectively.
\end{cor}

The above result completes the description of Algorithm \ref{alg:single}. Following are two observations regarding minor additional assumptions that have thus far not been discussed. First on tightening the dimensional rate requirement from $O(1)$ in the r.h.s. of Condition E, to $o(1)$ for Part (c) of Corollary \ref{cor:alg1.validity}. This can be viewed as price paid to obtain existence of limiting distributions in comparison to only the availability of optimal rates of estimation. This slight tightening is also in coherence with classical results in the fixed dimensional and single change axis framework, see, e.g., \citep{bai1994, bai1997estimation}. The second additional assumption here is  $(\psi/{\xi_{\min}}\big)\le c_u\surd\{\log(p\vee T)\},$ made for Part (b). Some insight into this assumption was provided after Condition F. It can be viewed as restriction stating the jump vectors in both the horizontal and vertical directions are somewhat evenly spread across its non-zero components, with control of order $\log (p\vee T_wT_h)$ on the rate of divergence on individual spikes within these components. The sufficient condition $\psi\le c_u,$ i.e., bounded components of all mean vectors, can also be made to ensure this assumption. Such boundedness of mean vectors has also been prevalent assumption in the existing change point literature.

The remainder of this sub-section provides two important remarks, first on a comparison with change point models with a single change axis. Followed by one on weakening the assumption of existence of a change point (Condition B(ii)) and extending the methodology to allow for selection in this context.

\begin{remark}[Comparison with dynamic mean models with a one-dimensional change axis] The defining advantage of model (\ref{model:rvmcp}) over a conventional $1d$-change axis framework is the ability to recover this simultaneous change with guaranteed statistical properties. This ability from a jump size perspective can be viewed as being able to leverage observed data in a secondary direction to detect much finer changes in a either directions. More precisely, model (\ref{model:rvmcp}) and Algorithm \ref{alg:single} leads to the ability to detect a change in the horizontal direction with a jump size magnitude that may be smaller by an order of $\surd T_h,$ and vice-versa. This can be observed from the assumed Condition E, where under fixed $p$ we have $\xi\ge c_u\big(1\big/\surd(T_wT_h)\big),$ as opposed to the $1d$-change axis framework where one will require at the very least $\xi\ge c_u\big(1\big/\surd T_w\big),$ see, e.g., \cite{wang2018high} and \cite{kaul2020inference}.
\end{remark}

\begin{remark}[Boundary cases of $\tau^0=(T_w,\tau_h^0)^T,$ $\tau^0=(\tau^0_w,T_h)^T$ or $\tau^0=(T_w,T_h)^T$\footnote{While boundary values can be equivalently characterized as either, \big[$\tau^0=(T_w,\tau_h^0)^T,$ $\tau^0=(\tau^0_w,T_h)^T$ or $\tau^0=(T_w,T_h)^T$\big]  or \big[$\tau^0=(0,\tau_h^0)^T,$ $\tau^0=(\tau^0_w,0)^T$ or $\tau^0=(0,0)^T,$\big]. However, clearly both these characterizations are not simultaneously identifiable since at these boundary values, realizations from two or three of the underlying distributions are not observed.}]\label{rem:boundary} Recall that Condition B(ii) assumes that the $2d$-change point is separated from parametric boundaries. This assumption can be relaxed by utilizing a conventional $0$-norm regularization technique in order to allow selection of the change point at boundary values. This can be achieved by replacing Step 1 of Algorithm \ref{alg:single} with a regularized version,
	\benr
	\h\tau^*_w&=&\argmin_{1\le \tau_w\le T_w}\big\{\cL(\tau_w,\check\tau_h,\check\theta)+\g_w{\bf 1}[\tau_w\ne T_w]\big\},\quad \g_w>0,\nn\\
	\h\tau^*_h&=&\argmin_{1\le\tau_h\le T_h}\big\{\cL(\check\tau_w,\tau_h,\check\theta)+\g_h{\bf 1}[\tau_h\ne T_h]\big\},\quad\,\,\,\, \g_h>0.\nn
	\eenr
	Here $\g_w,\g_h$ are tuning parameters. It can be observed that the $0$-regularized $\h\tau^*$ can also be equivalently represented as,
	\benr\label{eq:45}
	\h\tau^*_w&=&\begin{cases} 	T_w			  &	 {\rm if}\,\, \{\cL(T_w,\check\tau_h,\check\theta)-\cL(\h\tau_w,\check\tau_h,\check\theta)\}<\g_w, \\
		\h\tau_w & {\rm else},
	\end{cases}\nn\\
	\h\tau^*_h&=&\begin{cases} 	T_h			  &	 {\rm if}\,\, \{\cL(\check\tau_w,T_h,\check\theta)-\cL(\check\tau_h, \h\tau_h,\check\theta)\}<\g_h, \\
		\h\tau_h & {\rm else},
	\end{cases}
	\eenr
	where $\h\tau_w$ and $\h\tau_h$ are the un-regularized versions under the restricted search space obtained  in Step 1 of Algorithm \ref{alg:single}. Representation (\ref{eq:45}) is more common to change point literature, see, e.g. \cite{fryzlewicz2014wild} and \cite{wang2018high}, where it is utilized to extend a single axis and single change point method to multiple change points via variants of binary segmentation. 	A version of Algorithm \ref{alg:single} obtained by introducing this relaxation is described as Algorithm \ref{alg:single.regularized}.

\begin{algorithm}\label{alg:single.regularized}
	\caption{Optimal estimation of $\tau^0=(\tau^0_w,\tau^0_h)^T$ with boundary selection}
	\begin{algorithmic}[1]
		
		\vspace{1mm}
		\Statex Initialize change point $\check\tau=(\check\tau_w,\check\tau_h),$
		
		\vspace{1mm}
		\State Compute mean estimates $\check\theta_{(j)}=\h\theta_{(j)}(\check\tau),$ $j=1,2,3,4.$ and change point estimate $\h\tau=(\h\tau_w,\h\tau_h)^T$ as Step 1 of Algorithm \ref{alg:single}. Additionally perform selection by obtaining $\h\tau^*=(\h\tau_w^*,\h\tau_h^*)^T$ as (\ref{eq:45}).
		\State If $\h\tau^*_w=T_w,$ then set $\tilde\tau_w=T_w.$ Else update mean estimates to $\h\theta_{(j)}=\h\theta_{(j)}(\h\tau),$ $j=1,2,3,4,$ and update,
		\benr
		\tilde\tau_w=\argmin_{1\le \tau_w<T_w} \cL(\tau_w,\h\tau_h^*,\h\theta).\nn
		\eenr
		Similarly, if $\h\tau^*_h=T_h,$ then set $\tilde\tau_h=T_h.$ Else, update,
		\benr
		\tilde\tau_h=\argmin_{1\le \tau_h<T_h} \cL(\h\tau_w^*,\tau_h,\h\theta)\nn
		\eenr
		\Statex Output: $\tilde\tau=(\tilde\tau_w,\tilde\tau_h).$
	\end{algorithmic}
\end{algorithm}

Selection consistency \big($pr(\tilde\tau_w=T_w)\to 1,$ $T_w\to\iny,$ when $\tau^0_w=T_w$ and symmetrical for $\tilde\tau_h.$ \big) yielded by this $0$-regularization can be additionally verified via fairly conventional arguments, see, e.g. \cite{kaul2019efficient}.
\end{remark}

\section{Generalizations to full regression trees}\label{sec:extensions}

Regression trees in the current literature are limited to a one-dimensional response $p=1.$ These objects are typically represented as the following model (see, e.g. Page 307 of \cite{friedman2001elements}),
\benr\label{mod.tree.generic}
x_{(w,h)}=\sum_{m=1}^{M}\theta^0_{(m)}{\bf 1}[(w,h)\in \Om_m]+\vep_{(w,h)},
\eenr
in the case of $q=2$\footnote{The case of $q>2$ generalizes to hyper-rectangles $\Om_m,$ $m=1,...,M.$} dimensional feature space (or $2d$-change axes in our terminology). Here $\Om_m,$ $m=1,...,M$ represent disjoint partitioning sub-rectangles of the sampling space $\{1,...,T_w\}\times\{1,...,T_h\},$ and $\theta^0_{(m)}$ are representative of underlying $p=1$ dimensional dynamic mean parameters. The frequentist approach (see, e.g. Page 307, 308 of \cite{friedman2001elements}) to recovery of the associated tree, i.e., estimation of rectangle $\Om_m,$ $m=1,..,M,$ proceeds via binary half-plane splits of the sampling space via a greedy algorithm. Despite its prevalent use in the machine learning landscape, to our knowledge there is no analytical support available in the current literature towards statistical properties of these estimated transition points. Trees with a multi-dimensional $p>1$ response have not been considered in the literature. Model (\ref{model:rvmcp}) views the regression tree (\ref{mod.tree.generic}) from a $2d$-change point perspective under the special case of $M=4$ partitioning sub-rectangles under a high dimensional response $p>>(T_wT_h),$ furthermore, Algorithm \ref{alg:single} and the results of the previous section provide both estimation and inferential properties of the behavior of the recovered transitioning points under this HD framework with $M=4.$ The relaxation of Remark \ref{rem:boundary} extends these results to the $M=1,2\,\,or\,\, 4$ cases. Conceptual similarities of this algorithm to the proposed Algorithm 1 as discussed in Section \ref{sec:intro} also provides alternative insight into the equivalence of the model (\ref{mod.tree.generic}) (under $M=1,2\,\,or\,\, 4$) and the change point perspective taken thus far in this article.

The remainder of this section is devoted to characterizing generalizations of the model (\ref{model:rvmcp}), so as to allow the regression tree model structure (\ref{mod.tree.generic}) under any unknown number $M$ of partitioning rectangles (or hyper-rectangles), and allowing high dimensionality of the response. Then to provide methodological extensions to recover these underlying HD trees. We mention here that the methodology extensions to follow shall be heuristic although these shall be validated numerically in this article. These extensions are motivated by classical results in the change point literature  with established statistical properties (in a $1d$-change axis and fixed $p$ framework). The important contribution here shall be in the formulation of the generic regression tree in a change point framework, thereby allowing a clear path forward toward a statistical examination of the proposed methods as future work. The reason for our inability to examine these properties is particulary owed to high dimensionality of the response, where inferential properties under multiple change points are unknown in the literature even at a more fundamental level of a $1d$-change axis framework, which first need addressal before analogous question can be pursued in this framework.

Accommodating regression trees in full generality require the following generalizations of model (\ref{model:rvmcp}), which shall in turn also require us to define additional notation and conventions.
\begin{enumerate}[label={G.\arabic*},ref=G.\arabic*]
	\item Allow boundary cases of change point parameters, $\tau^0=(T_w,\tau_h^0)^T,$ $\tau^0=(\tau^0_w,T_h)^T$ or $\tau^0=(T_w,T_h)^T$ in order to allow `no partition' and `half-plane partition' cases.\label{list:gen.boundary}
	\item Generalize model (\ref{model:rvmcp}) to allow multiple $2d$-change points that are to be introduced in a hierarchical manner, i.e., to allow each quadrant/half-plane split by a previous change point to be further divided into fourths/halfs by new change points. \label{list:gen.multiple.cp}
	\item Allow feature space to be of $q>2$ dimensions, i.e., allow framework of a $p$-dimensional response and a $q$-dimensional feature space (or $q$-dimensional change points). \label{list:gen.multaxes}
\end{enumerate}

The generalization of \ref{list:gen.boundary} has already been considered in Remark \ref{rem:boundary} and the corresponding extension described in full in Algorithm \ref{alg:single.regularized}. Next we proceed to the  generalization of \ref{list:gen.multiple.cp}. Recall the collection of indices defined as quadrants $Q_j(\tau)\subseteq\{1,...,T_w\}\times\{1,...,T_h\},$ $j=1,2,3,4.$ By convention, the ordering of these $j=1,2,3,4,$ quadrants is assumed as {\it top right}, {\it tope left}, {\it bottom left} and {\it bottom right}, respectively, with both change axes labeled in ascending order. We shall retain this ordering in all to follow. Moreover, when $\tau$ is in a boundary condition, then two or three of these quadrants are to be defined as empty sets, e.g., if $\tau=(T_w,T_h)^T,$ then $Q_j(\tau)=\phi,$ $j=1,2,4,$ and $Q_3(\tau)=\{1,...,T_w\}\times\{1,...,T_h\}.$ Similarly, if $\tau=(\tau_w,T_h)^T,$ $\tau_w<T_w,$ then $Q_j(\tau)=\phi,$ $j=1,2$ and $Q_3(\tau)=\{1,...,\tau_w\}\times\{1,...,T_h\}$ and $Q_4(\tau)=\{(\tau_w+1),...,T_w\}\times\{1,...,T_h\}.$ Symmetrically, if  $\tau=(T_w,\tau_h)^T,$ $\tau_h<T_h,$ then $Q_j(\tau)=\phi,$ $j=2,3.$

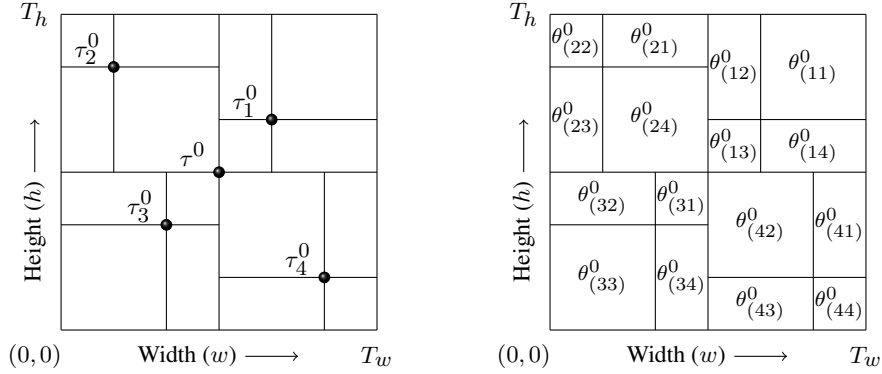
\begin{figure}[]
	\centering
	\begin{minipage}{0.45\textwidth}
		\begin{tikzpicture}[scale=0.7]
		\draw (0,0) -- (0,6);\draw (0,0) -- (6,0);
		\draw (6,0) -- (6,6);\draw (0,6) -- (6,6);
		\draw (3,0) -- (3,6); \draw (0,3) -- (6,3);
		\draw (3,4) -- (6,4); \draw (4,3) -- (4,6);
		\draw (0,5) -- (3,5); \draw (1,3) -- (1,6);
		\draw (2,0) -- (2,3); \draw (0,2) -- (3,2);
		\draw (5,0) -- (5,3); \draw (3,1) -- (6,1);
		\draw (-0.5,-0.5)  node{$(0,0)$};\draw (-0.5,6)  node{$T_h$};\draw (6,-0.5)  node{$T_w$};
		\draw[->] (3.5,-0.5) node[anchor=east]{Width ($w$)} -- (4.5,-0.5) node{};
		\draw[->] (-0.5,3) node[anchor=east, rotate=90]{Height ($h$)} -- (-0.5,4) node{};
		\node (v1) at (3,3) [ball color=black, circle, draw=black, inner sep=0.05cm] {};
		\draw (2.5,3.35) node{$\tau^0$};
		\node (v2) at (4,4) [ball color=black, circle, draw=black, inner sep=0.05cm] {};
		\draw (3.5,4.35) node{$\tau^0_{1}$};
		\node (v3) at (1,5) [ball color=black, circle, draw=black, inner sep=0.05cm] {};
		\draw (0.5,5.35) node{$\tau^0_{2}$};	
		\node (v4) at (2,2) [ball color=black, circle, draw=black, inner sep=0.05cm] {};
		\draw (1.5,2.35) node{$\tau^0_{3}$};	
		\node (v5) at (5,1) [ball color=black, circle, draw=black, inner sep=0.05cm] {};
		\draw (4.5,1.35) node{$\tau^0_{4}$};		
		
		\end{tikzpicture}
	\end{minipage}
	\begin{minipage}{0.45\textwidth}
		\begin{tikzpicture}[scale=0.7]
		\draw (0,0) -- (0,6);\draw (0,0) -- (6,0);
		\draw (6,0) -- (6,6);\draw (0,6) -- (6,6);
		\draw (3,0) -- (3,6); \draw (0,3) -- (6,3);
		\draw (3,4) -- (6,4); \draw (4,3) -- (4,6);
		\draw (0,5) -- (3,5); \draw (1,3) -- (1,6);
		\draw (2,0) -- (2,3); \draw (0,2) -- (3,2);
		\draw (5,0) -- (5,3); \draw (3,1) -- (6,1);
		\draw (-0.5,-0.5)  node{$(0,0)$};\draw (-0.5,6)  node{$T_h$};\draw (6,-0.5)  node{$T_w$};
		\draw[->] (3.5,-0.5) node[anchor=east]{Width ($w$)} -- (4.5,-0.5) node{};
		\draw[->] (-0.5,3) node[anchor=east, rotate=90]{Height ($h$)} -- (-0.5,4) node{};
		\draw (5,5) node{\scriptsize{$\theta_{(11)}^0$}};\draw (3.5,5) node{\scriptsize{$\theta_{(12)}^0$}};
		\draw (3.5,3.5) node{\scriptsize{$\theta_{(13)}^0$}};\draw (5,3.5) node{\scriptsize{$\theta_{(14)}^0$}};
		\draw (2,4) node{\scriptsize{$\theta_{(24)}^0$}};\draw (0.5,4) node{\scriptsize{$\theta_{(23)}^0$}};
		\draw (0.5,5.5) node{\scriptsize{$\theta_{(22)}^0$}};\draw (2,5.5) node{\scriptsize{$\theta_{(21)}^0$}};
		\draw (2.5,1) node{\scriptsize{$\theta_{(34)}^0$}};\draw (1,1) node{\scriptsize{$\theta_{(33)}^0$}};
		\draw (1,2.5) node{\scriptsize{$\theta_{(32)}^0$}};\draw (2.5,2.5) node{\scriptsize{$\theta_{(31)}^0$}};
		\draw (5.5,0.5) node{\scriptsize{$\theta_{(44)}^0$}};\draw (4,0.5) node{\scriptsize{$\theta_{(43)}^0$}};
		\draw (4,2) node{\scriptsize{$\theta_{(42)}^0$}};\draw (5.5,2) node{\scriptsize{$\theta_{(41)}^0$}};
		\end{tikzpicture}
	\end{minipage}
	\caption{\footnotesize{Visualization of underlying  mean and change point parameters of $\ell$-hierarchical model (\ref{mod:mult.rmvcp}) with $\ell=1.$}}
	\label{fig:mult.parameters}
\end{figure}

Next, we define a notion of hierarchical multiple change points where each induced quadrant is allowed to further sub-partition. In other words each parent change point gives rise to four child change points in a hierarchical sense. Following notation is necessary to describe this precisely. Let $\tau^0=(\tau_{w}^0,\tau_{h}^0)^T\in\{1,...,T_w\}\times\{1,...,T_h\}$ represent the zero$^{th}$ level change point as considered in model (\ref{model:rvmcp}). Then for a first hierarchical step, define four children change points spawned by $\tau^0,$ in same order as the quadrants, i.e.,
\benr\label{eq:50}
\tau^0_{i_1}\in Q_{i_1}(\tau^0), \quad i_1=1,2,3,4,
\eenr
Moreover, each new change point induces four sub-quadrants and consequently four new mean vectors, we represent the means yielded by $\tau_{i_1}$ as follows, for each $i_1=1,2,3,4,$ define,
\benr
Ex_{(w,h)}=\theta_{(i_1i_2)}^0{\bf 1}\big[(w,h)\in Q_{i_1}(\tau^0)\cap Q_{i_2}(\tau^0_{i_1})\big],\quad i_2=1,2,3,4.
\eenr
Continuing this hierarchical construction, i.e., further sub-splitting each partitioning rectangle by children change points, we have from the second to the $\ell^{th}$ hierarchical level,
\benr\label{eq:51}
&&\quad\tau_{i_1i_2}^0\in Q_{i_1}(\tau^0)\cap Q_{i_2}(\tau^0_{i_1}),\quad {i_1,i_2=1,2,3,4}\nn\\
&&\quad\tau_{i_1i_2i_3}^0\in Q_{i_1}(\tau^0)\cap Q_{i_2}(\tau^0_{i_1})\cap Q_{i_3}(\tau^0_{i_1i_2}),\quad {i_1,i_2,i_3=1,2,3,4}\nn\\
&&\quad\vdots\nn\\
&&\quad\tau_{i_1i_2\dots i_{\ell}}^0\in Q_{i_1}(\tau^0)\cap Q_{i_2}(\tau^0_{i_1})\cap Q_{i_3}(\tau^0_{i_1i_2})...\cap Q_{i_{\ell}}\big(\tau_{i_1i_2\dots i_{(\ell-1)}}\big) \quad {i_1,...,i_{\ell}=1,2,3,4}.
\eenr
In the following we represent the finest level parametric space of the change point parameters with a slight misuse of notation as,
\benr
\bigcap_{j=1}^{\ell}Q_{i_j}\big(\tau_{\Pi_{k=1}^{j-1}i_k}^0\big)=Q_{i_1}(\tau^0)\cap Q_{i_2}(\tau^0_{i_1})\cap Q_{i_3}(\tau^0_{i_1i_2})\cdots\cap Q_{i_l}\big(\tau_{i_1i_2\cdots i_{(\ell-1)}}\big).\footnotemark\nn
\eenr
\footnotetext{The notation $\Pi_{k=1}^{j}i_k$ only represents the indexing $(i_1i_2,...,i_j),$ and does not represent any actual product.}
Now, the corresponding means at the finest partitioning level ($\ell^{th}$ hierarchical level) become,
\benr
Ex_{(w,h)}=\theta_{(i_ii_2\dots i_{(\ell+1)})}^0{\bf 1}\Big[(w,h)\in \bigcap_{j=1}^{\ell+1}Q_{i_j}\big(\tau_{\Pi_{k=1}^{j-1}i_k}^0\big)\Big].\nn
\eenr
The above notations now allow a parametric description of a $\ell$-hierarchical multiple $2d$-change point model as,
\benr\label{mod:mult.rmvcp}
x_{(w,h)}&=&\underbrace{\sum_{i_1=1}^{4}\sum_{i_2=1}^{4}...\sum_{i_{(\ell+1)}=1}^4}_{\ell+1\,\,{\rm iterated\,\, sums}}\theta_{(i_1i_2...i_{(\ell+1)})}^0{\bf 1}\Big[(w,h)\in \bigcap_{j=1}^{\ell+1}Q_{i_j}\big(\tau_{\Pi_{k=1}^{j-1}i_k}^0\big)\Big]\\
&&\hspace{4cm} +\,\,\vep_{(w,h)},\qquad w=1,...,T_w,\,\,h=1,...,T_h.\nn
\eenr
Construction (\ref{mod:mult.rmvcp}) is a natural $2d$-extension of a $1d$ multiple change point model. Recall that even under $1d$ multiple change points, each change point can be viewed hierarchically w.r.t. prior and post change points, in the sense that each new change point has its parametric space restricted by the prior and post change points. The additional complications seen above arise due these parametric spaces being restricted on two axes. A visualization of this construction is provided in Figure \ref{fig:mult.parameters}. Following is a final note on notation. Recall that by construction, each change point is also allowed to be on its respective parametric boundaries, thus, we call the model (\ref{mod:mult.rmvcp}) as being $\ell$-hierarchical if there exists at least one change point at the $\ell^{th}$ hierarchy which is away from its parametric boundaries (also see Remark \ref{rem:mult.boundary}).

\begin{remark}[On the number of induced partitions]\label{rem:mult.boundary}
	Observe that when all change points at all levels of hierarchy are separated from their parametric boundaries, the maximum possible number of change points induced by the model (\ref{mod:mult.rmvcp}) is $1+4+4^2+....4^l=(4^{l+1}-1)/3.$ Similarly, at this finest level, the sampling period $\{1,...,T_w\}\times\{1,...,T_h\}$ is segmented into a maximum number of $4^{l+1}$ distinct $p$-dimensional mean parameter vectors. On the other hand, when change point are at boundaries, the model (\ref{mod:mult.rmvcp}) allows the number of sampling period partitions to be much smaller. For e.g., consider the $1$-hierarchical model visualized in Figure \ref{fig:mult.parameters}. Suppose the first of the first level change points is at a boundary, e.g, $\tau^0_{1}=(T_w,T_h)^T.$ Then the number of distinct partitions of the sampling period becomes $3*4+1=13.$ Another consequence of the boundary change $\tau^0_{1}=(T_w,T_h)^T$ is that three of the four sub-quadrants induced by this boundary change are empty sets, in particular $Q_1(\tau^0)\cap Q_j(\tau^0_1)=\phi,$ $j=1,2,4.$ Thus, no further hierarchical changes can be spawned within these sub-quadrants, in other words, further change points within these sub-quadrants are all at the same boundary point $(T_w,T_h)^T.$ Analogous patterns can be observed at half plane partitioning boundary change points, e.g., $\tau^0_{1}=(\tau^0_{1,w},T_h)^T$ or  $\tau^0_{1}=(T_w,\tau^0_{1,h})^T.$
\end{remark}

The $\ell$-hierarchical construction of model (\ref{mod:mult.rmvcp}) allows a natural extension of the methodology of Algorithm \ref{alg:single} and \ref{alg:single.regularized}. Recall that in a $1d$-multiple change point framework a well established technique that allows a similar extension is that of binary segmentation. Under our framework, the natural variant of this technique become {\it quarterly segmentation}. More specifically, one may apply the regularized version of  Algorithm \ref{alg:single.regularized} on the entire sampling period to first find the zero$^{th}$ level change point. If this change point is at the boundary $\h\tau=(T_w,T_h)^T$ then no change points are observed. Else, one may segment the observed realizations into the half plane or quarter partitions induced by the estimated change point $\h\tau,$ and again apply Algorithm \ref{alg:single.regularized} within each of these partitions to detect further hierarchical changes. One may continue these recursions until no further changes are detected in any partition induced by a change point of the prior hierarchy. This process is described as Algorithm \ref{alg:quarter.seg}.


\begin{algorithm}\label{alg:quarter.seg}
	\caption{Quarterly segmentation}
	\begin{algorithmic}[]
		
		\vspace{1mm}	
		\State {\bf Initialize}: $\tilde\tau_{{\rm st}}=\phi$ (empty matrix ($2$-rows) collecting all $2d$-change points to be estimated over columns);
		\State Implement $\tilde\tau$= Alg. \ref{alg:single.regularized} \big($\{1,...,T_w\}\times\{1,...,T_h\}$\big)
		\If {$\tilde\tau=(T_w,T_h)$ (simultaneous boundaries on both axes) }\State STOP	
		\Else \,\,$\tilde\tau_{{\rm up}}=cbind(\tau_{{\rm st}},\tilde\tau)$ (updated matrix of estimated change points)	\EndIf
		
		\vspace{-3mm}
		\State $\rule{3.5in}{0.05mm}$
		\State Let $\ell=1$ (first hierarchical level)
		\While {$ncol(\tilde\tau_{up})$> $ncol(\tilde\tau_{st})$}
		\State $\tilde\tau_{st}=\tilde\tau_{up}$
		\For {$m \in 1:4^{\ell}$}
		\State $(i_1,i_2...,i_{\ell})=indexFinder(\ell,m)$;
		\State $partition{(i_1,i_2...,i_{\ell})}=\bigcap_{j=1}^{\ell}Q_{i_j}\big(\tilde\tau_{\Pi_{k=1}^{j-1}i_k}\big)$\footnotemark
		\If {$|partition(i_1,i_2,...,i_{\ell})|>0$} \State $\tilde\tau_{i_1i_2...i_{\ell}}=$Alg.\ref{alg:single.regularized}  \big($partition(i_1,i_2,...,i_{\ell})$\big)
		\EndIf
		\If {$\tilde\tau_{i_1i_2...i_{\ell}}$ is away from simultaneous parametric boundary on both axes} \State $\tilde\tau_{{\rm up}}=cbind(\tilde\tau_{{\rm up}},\tilde\tau_{i_1i_2...i_{\ell}})$
		\EndIf
		\EndFor
		\State $\ell=\ell+1$ (next hierarchical level)
		\EndWhile
		\State {\bf Output}: All change points $\tilde\tau_{\big(\Pi_{k=1}^{j}i_k\big)},$ for each $i_1,...i_k\in\{1,2,3,4\}$ and  $k=0,1,...\ell.$
	\end{algorithmic}
\end{algorithm}

\footnotetext{Define $\tilde\tau_{\big(\Pi_{k=1}^{j}i_k\big)}=\tilde\tau$ at $j=0,$ i.e., the zero$^{th}$ level change point estimate.}

\begin{remark}[On notations used in Algorithm \ref{alg:quarter.seg}] The following shorthand notations have been utilized in the description of Algorithm \ref{alg:quarter.seg}. First, the function $ncol()$ represents the number of columns of the argument matrix. The function $cbind()$ represents a column-wise concatenation of a matrix and a vector. Finally, the function $indexFinder(\ell,m)$ represents a function providing the $1-1$ mapping between the indexing $1,...,4^{\ell}$ and the indexing $(i_1,i_2,...,i_{\ell}),$ with $i_j=1,2,3,4,$ for each $j,$ in keeping with the ordering convention assumed in the construction of model (\ref{mod:mult.rmvcp}).
\end{remark}

As briefly mentioned earlier, Algorithm \ref{alg:quarter.seg} is the natural extension of binary segmentation, which is perhaps the most widely used method for estimation of multiple change points and has well studied statistical properties at least in the fixed $p$ and $1d$-change axis framework, see, e.g., \cite{venkatraman1993consistency} and \cite{fryzlewicz2014wild}.  We conclude this section with a discussion on the final generalization  (\ref{list:gen.multaxes}) that provide an equivalence to the standard form of regression trees with the additional complexity layer of a $p$-dimensional response.

\begin{remark}[On generalization (\ref{list:gen.multaxes})]  Model (\ref{model:rvmcp}) and its $\ell$-hierarchical version (\ref{mod:mult.rmvcp}) both possess sufficient flexibility to allow extensions to any finite number of $q>2$ change axes. For e.g., in a $3d$-change point setting, all one requires is to first consider the single change point methodology and extend it to $3d$ by including another internal update in the change point vector, i.e., all three components shall be
updated component-wise while utilizing the remaining two and all preliminary mean estimates as plug-in estimates into the squared loss. Another consequence of this extension shall be to introduce {\it octants} instead of quadrants, across which the mean parameters are allowed to be dynamic. Consequently, the {\it for} loop of Algorithm \ref{alg:quarter.seg} shall now be over a total of $8^\ell$ indices. More generally, under $q$-dimensional change axis this partitioning number shall inflate exponentially as $2^q,$ which is also the case in regression trees.
\end{remark}

\section{Applications}\label{sec:application}

This section considers two distinct applications of models and methods developed in the previous sections. First we consider an application of the single $2d$-change point model (\ref{model:rvmcp}) and the corresponding Algorithm \ref{alg:single} together with Theorem \ref{thm:wc.vanishing} and Theorem \ref{thm:wc.non.vanishing}. The second considers an application of the $\ell$-hierarchical multiple change point model (\ref{mod:mult.rmvcp}) together with the proposed Algorithm \ref{alg:single.regularized} and \ref{alg:quarter.seg} (quarterly segmentation).

\subsection{Segmenting IRAS data}

The NASA mission of the Infra-Red Astronomy Satellite (IRAS) was the first attempt to map the sky at infra-red wavelengths. This could not be done from ground observatories because large portions of the infrared spectrum is absorbed by the atmosphere. This database contains observation vectors of high quality spectra, each on $44$ blue band and $49$ red band channels of usable flux measurements ($p=93$). Moreover, each observation is made at a point in the sky associated with the equatorial celestial coordinates system of {\it Right Ascension (RA)} and {\it Declination (DE)}. These coordinates can be viewed as the analogs of terrestrial longitude and latitude on the {\it celestial sphere}, respectively. The {\it DE} coordinate is the angle of view above or below the {\it celestial equator} ranging from $90^{\circ}$ to $-90^{\circ}.$ The {\it RA} coordinate is measured from the sun at the march equinox (zero point of $RA$) and varies between  $0$ to $24$ {\it hours (hr)}, see, \url{https://solarsystem.nasa.gov/basics/chapter2-2/} for further details on this coordinate system. The data set is publicly available at the UCI repository at \url{https://archive.ics.uci.edu/ml/datasets/Low+Resolution+Spectrometer} and consists of observations with $RA$ between $12\,hr$ and $24\,hr.$

It is known that observed spectral types are associated with particular stellar types and consequently the observed light spectrum may not be uniformly distributed across the sky in its components. It is thus of interest to perform an unsupervised segmentation of the {\it celestial sphere} into regions of distinct spectra. For this purpose we utilize the single $2d$-change model (\ref{model:rvmcp}), the proposed  Algorithm \ref{alg:single}\footnote{The mean regularizers $\la_j$ for Step 1 and Step 2 are chosen via a BIC-type criteria, see, (\ref{eq:bic})} for estimation, and Theorem \ref{thm:wc.vanishing} and Theorem \ref{thm:wc.non.vanishing} for inference\footnote{See, Section \ref{sec:numerical} and Appendix \ref{app:numerical.supplement} for details on all necessary additional estimations such as that of the jump size, drifts, asymptotic variances and quantiles of limiting distributions} on location of the partitioning break.

The change axes $(w,h)$ in the notation of model (\ref{model:rvmcp}) shall be representative of the coordinates {\it RA} and {\it DE}, respectively. A few data pre-processing steps are carried out prior to implementation, chiefly to allow compatibility with the assumed model. Since available observations are largely clustered around the {\it celestial equator}, we limit our analysis to {\it DE} between $-50^{\circ}$ and $50^{\circ}.$ Additionally we also restrict {\it RA} to between $14h$ and $22h.$ This sub-setting reduces the available observations to $316,$ and is done in order to yield observations that are more uniformly spread over the restricted $2d$-grid. The coordinate system is then binned into a uniform grid of size $25\times 25$ and the observation vector for each point on this uniform grid is set as the sample mean of the flux measurements of the $10$ nearest neighbors to each location on the uniform grid. The flux data is mean centered to allow the sparsity assumption as described earlier in (\ref{def:setS}). The binned uniform grid is then re-labeled w.r.t change axes $w=1,...,T_w$ and $h=1,...,T_h$  with $T_w=T_h=25$ in keeping with the construction of model (\ref{model:rvmcp}). All estimation and inference is carried out under this framework. Estimates and confidence intervals are then mapped back to the binned uniform coordinate grid to obtain corresponding values in the {\it RA} and {\it DE} coordinates.

Upon implementation of Algorithm 1 we obtain estimated change points as $(\tilde\tau_w,\tilde\tau_h)=(8,10).$ Mapping these break points to the uniform coordinate axes yields coordinates $(RA,DE)=(16.32\,hr,-12.43^{\circ})$ as the $2d$-partitioning change point. This change point segments the region under consideration of the {\it celestial sphere} into four sub-regions of distinct spectral flux, specifically,
\benr
&Q_1(\tilde\tau)&= \{22\,hr \ge RA>16.32\,hr\,\,\, \&\,\,\, 50^{\circ}\ge DE> -12.43^{\circ}\},\nn\\
&Q_2(\tilde\tau)&= \{14\,hr \le RA\le 16.32\,hr \,\,\,\&\,\,\, 50^{\circ}\ge DE> -12.43^{\circ}\},\nn\\
&Q_3(\tilde\tau)&= \{14\,hr\le RA\le 16.32\,hr\,\,\, \&\,\,\, -50^{\circ}\le DE\le -12.43^{\circ} \},\quad{\rm and}\nn\\
&Q_4(\tilde\tau)&= \{22\,hr \ge RA> 16.32\,hr\,\,\, \&\,\,\, -50^{\circ}\le DE\le -12.43^{\circ}\}.\nn
\eenr
Next, we perform inference on the change parameters in accordance with Theorem \ref{thm:wc.vanishing} and Theorem \ref{thm:wc.non.vanishing}. We set the significance level at $\al\in\{0.05,0.01\}.$ The obtained confidence intervals are provided in Table \ref{tab:cis}. It is observed that the estimated jump size $\h\xi_h$ in the $DE$ (or horizontal) direction is relatively larger w.r.t the estimated asymptotic variance $\h\si^2_{(h,\iny)},$ thus leading to the quantile of the limiting distribution under the non-vanishing regime to be zero at $95\%$ coverage, in turn causing the margin of error to be zero. Thus the corresponding interval is the single point of the estimated change.  This unsupervised analysis can potentially utilized by a researcher to pre-identify regions of interest in the sky to focus upon in order to search for specific spectral signatures. Alternatively, one may potentially utilize this analysis and build secondary supervised models to aid identification of specific stellar processes.

\begin{table}[]
	\centering
	\resizebox{1\textwidth}{!}{
		\begin{tabular}{lcccc}
			\toprule
			& \multicolumn{2}{c}{$\al=0.05$}     & \multicolumn{2}{c}{$\al=0.01$}     \\ \midrule
			& Vanishing          & Non-Vanishing      & Vanishing      & Non-vanishing \\
			\midrule
			Integer Scale ($w$)  & $[6.73, 9.26]$	  & $[7,9]$  		   & $[6.28, 9.71]$ & $[6, 10]$  \\
			Right Ascension 	 & $[15.99\,hr, 16.79\,hr]$ & $[16.00\,hr, 16.65\,hr]$ & $[15.85\,hr, 16.93\,hr]$ & $[15.67\,hr, 16.98\,hr]$  \\
			Integer Scale ($h$)  & $[9.52, 10.47]$	  & $[10,10]$  	       & $[9.35, 10.64]$ & $[9, 11]$  \\
			Declination   		 & $[-13.84^{\circ}, -10.06^{\circ}]$      & $[-12.43^{\circ}, -12.43^{\circ}]$  & $[-14.52^{\circ}, -9.38^{\circ}]$       & $[-16.58^{\circ}, -8.28^{\circ}]$  \\ \hline
	\end{tabular}}
	\vspace{1mm}
	\caption{\footnotesize{Confidence intervals under vanishing and non-vanishing regimes at $95\%$ and $99\%$ coverage. Intervals presented in integer level coordinates $(w,h)$ and corresponding $RA$ and $DE$ coordinates. CI's under the vanishing jump regime obtained by mapping integer level coordinates to $RA,\,DE$ coordinates via associated quantiles.}}
	\label{tab:cis}
\end{table}

\subsection{Image processing}\label{subsec:image}

A digital image comprises of $T_w\times T_h$ observational units (pixels) where each pixel is a three dimensional vector ($p=3$) of intensity of the three primary color channels $(r,g,b).$ These pixels can often be contaminated with noise due to a variety of reasons and the primary goal of image denoising is to recover the underlying clean image. The basic idea utilized in nearly all methods designed for this objective is to perform a local averaging of pixels, this includes the total variation estimator studied in \cite{ortelli2020adaptive}, where the total variation regularization is designed to locally average pixels driven by gradient changes. Fundamentally our approach also pursues the same idea of local averaging, however, we instead utilize model (\ref{mod:mult.rmvcp}) and the proposed quarterly segmentation methodology of Algorithm \ref{alg:quarter.seg} to perform an explicit identification of `local'  pixels as identified by underlying changes in means, in turn induced by hierarchical $2d$-change points.

The first example below are synthetic images that provide a visual proof of principle of our methodology. Specifically, we create images with atmost one $2d$ simultaneous change. Recovery of these shall illustrate the working mechanism of the regularized Algorithm \ref{alg:single.regularized}, which also performs selection of change parameters. The second example shall then consider two real images.

{Additional computations details:} Imaging applications utilizing the raw $(r,g,b)$ data is $p=3$ dimensional problem, i.e., high dimensionality of the response does not arise here. Thus, soft-thresholding of (\ref{est:softthresh}) is in this case redundant. Accordingly, we utilize sample means $\bar x(\check\tau),$ and $\bar x(\h\tau)$ to serve as mean estimates $\check\theta$ and $\h\theta,$ respectively, for Algorithm \ref{alg:single.regularized}. Doing so only leaves two tuning parameters $\g_w$ and $\g_h$ arising from Step 1 of Algorithm \ref{alg:single.regularized}. These tuning parameters are chosen via a BIC type criteria as follows.
\benr\label{eq:bic.g}
BIC\big(\tau_w(\g_w),\tau_h(\g_h)\big)&=& \sum_{j=1}^4\sum_{(w,h)\in Q_j(\tau(\g))}\big\|x_{(w,h)}-\bar x_{(j)}(\tau(\g))\big\|_2^2\nn\\
&&\hspace{1cm}+c_{bic}\big(|\h S|+{\bf 1}[\tau_w(\g_w)\ne T_w]+{\bf 1}[\tau_h(\g_h)\ne T_h]\big)\log T_wT_h.
\eenr
Here $\h S=\cup_{j=1}^4 \h S_j,$ with $\h S_j=\big\{k;\,\,\bar x_{(j)k}(\tau)\ne 0\big\}.$ When $\tau_w$ and $\tau_h$ are within parametric boundaries then $|\h S|=12$ (each quadrant with $p=3$ dim. means). The tuning parameters $\g_w,\g_h$ are chosen componentwise as maximizer values of (\ref{eq:bic.g}), i.e., $\g_w$ is chosen as the value such that $\h\tau(\g_w)$ maximizes $BIC(\h\tau_w(\g_w),\check\tau_h)$ and symmetrically for $\h\tau_h.$ The constant $c_{bic}$ is introduced to allow for additional control on the tuning process, with the classical BIC criteria requiring $c_{bic}=1.$ The finite and discrete nature of the optimization of Step 1 of Algorithm \ref{alg:single.regularized} provides a significant simplification. Here tuning with BIC criteria (\ref{eq:bic.g}) is equivalent to choosing $\g_w$ and $\g_h$ of Algorithm \ref{alg:single.regularized} as $\g_w=\g_h= 7c_{bic}(\log T_wT_h)\big/(T_wT_h).$ This can be observed by noting that when $\tau_w$ is at the parametric boundary, the quadrants $Q_1(\tau)$ and $Q_4(\tau)$ are empty sets, consequently the underlying degrees of freedom in this case is $7$ ($3+3$ for mean, $1$ for the height parameter), analogously when $\tau_w$ is within the parametric boundary the degrees of freedom is $14.$ Symmetrically for the height change parameter.

\begin{example}[Synthetic images] To provide a visual proof of principle, we construct synthetic images under the following cases. I(a) Simultaneous change in both horizontal and vertical directions, I(b) change in only the horizontal direction  I(c) change in only the vertical direction, and I(d) no change in either direction. The construction in all four cases of (I) is in direct equivalence to the model (\ref{model:rvmcp}), with the relaxation of boundary valued change points as discussed in Remark \ref{rem:boundary}. (II) Additional  images are constructed with multiple hierarchical change points as described by model (\ref{mod:mult.rmvcp}).   Algorithm \ref{alg:quarter.seg} is implemented in all cases to estimate the underlying $2d$-transition and in turn the induced disjoint partitions. A clean image is recovered by replacing each pixel with the sample mean of the partition in which it lies.
	
	The true, noisy and the recovered images are illustrated in Figure \ref{fig:ex1.single} for case I(a) and Figure \ref{fig:ex1.multiple} for case (II). All remaining cases are provided in Figure \ref{fig:ex1.single.supp} and Figure \ref{fig:ex1.multiple.supp} in Appendix \ref{app:numerical.supplement}. The results of these cases are visually self explanatory. The proposed method successfully estimates transition points as well as boundary valued cases and consequently recovers underlying images in all cases, despite a significant amount of noise.
	\begin{figure}
		\includegraphics[width=.3\textwidth]{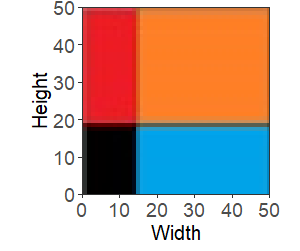}\hfill
		\includegraphics[width=.3\textwidth]{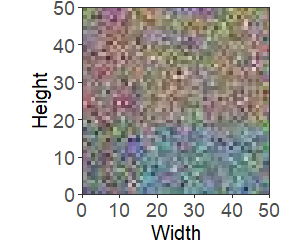}\hfill
		\includegraphics[width=.3\textwidth]{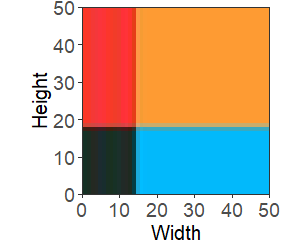}
		\caption{\footnotesize{Image denoising with Algorithm \ref{alg:quarter.seg}. Image is $50\times 50$ pixels with one $2d$-change point. {\it Left panel:} True image (unobserved), {\it Center panel:} Noisy image (observed), {\it Right panel:} Recovered image. Change point estimated as $\tilde\tau=(14,18)^T.$  Noise set to $\Si=I_{3\times 3},$ and tuning constant $c_{bic}=1.$}}\label{fig:ex1.single}
	\end{figure}
	\begin{figure}
		\includegraphics[width=.3\textwidth]{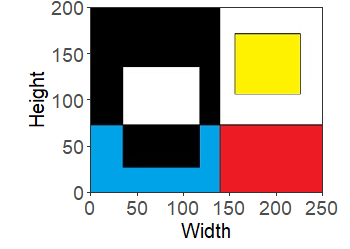}\hfill
		\includegraphics[width=.3\textwidth]{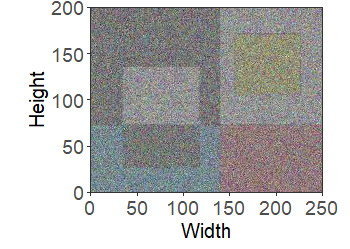}\hfill
		\includegraphics[width=.3\textwidth]{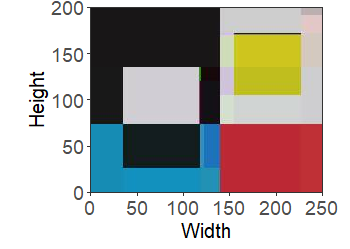}
		\caption{\footnotesize{Image denoising with Algorithm \ref{alg:quarter.seg}. Image is $250\times 200$ pixels with multiple hierarchical $2d$-change points. {\it Left panel:} True images (unobserved), {\it Center panel:} Noisy images (observed), {\it Right panel:} Recovered image. Estimated model recovered with $\ell=5$ hierarchical change points inducing a total of $32$ disjoint partitions of the sampling space. Noise set to $\Si=I_{3\times 3},$ and tuning constant $c_{bic}=1.$}}
		\label{fig:ex1.multiple}
	\end{figure}
\end{example}

\begin{example}[Real images] In this example we considered two real images. First is an image of {\it Lena} which has traditionally been utilized as a benchmark example for image denoising techniques. A second image of {\it Charlie Chaplin} has also been considered. The $\ell$-hierarchical change model (\ref{mod:mult.rmvcp}) and proposed Algorithm \ref{alg:quarter.seg} is utilized to fit the observed $p=3$ dimensional realizations (pixels) of each considered image. A large number of disjoint partitions are expected in any realistic image. Consequently, the noise added to these images is set lower in comparison to the previous example. i.e., we set $\Si=0.05\cdotp I_{3\times 3}.$ Implementation is performed for three cases of the tuning constant $c_{bic}\in\{0.25,0.5,1\}.$
	
	Figure \ref{fig:ex2.lena} provides the true, noisy and recovered images for the {\it Lena} image with $c_{bic}=0.25.$ Figure \ref{fig:ex2.chaplin}  provides these images for the {\it Charlie Chaplin} image with $c_{bic}=0.5.$ The remaining cases are provided in Figure \ref{fig:ex2.lena.supp} and Figure \ref{fig:ex2.chaplin.supp}.

	\begin{figure}
		\includegraphics[width=.33\textwidth]{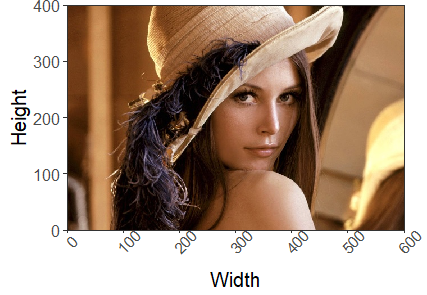}\hfill
		\includegraphics[width=.33\textwidth]{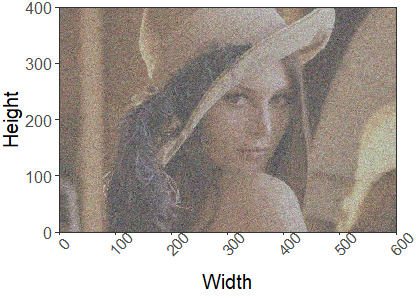}\hfill
		\includegraphics[width=.33\textwidth]{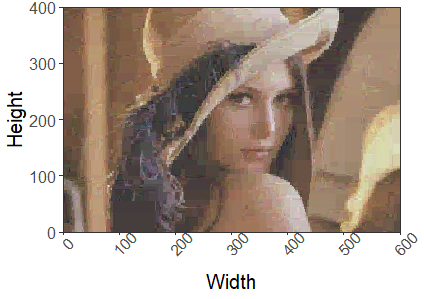}
		\caption{\footnotesize{Image denoising with Algorithm \ref{alg:quarter.seg}. Image is $600\times 400$ pixels. {\it Left panel:} True image (unobserved), {\it Center panel:} Noisy image (observed), {\it Right panel:} Recovered image. Estimated model recovered with $\ell=13$ hierarchical change points inducing a total of $7254$ disjoint partitions of the sampling space. Noise set to $\Si=0.05\cdotp I_{3\times 3},$ and tuning constant $c_{bic}=0.25.$
		}}\label{fig:ex2.lena}
	\end{figure}

	\begin{figure}
		\includegraphics[width=.33\textwidth]{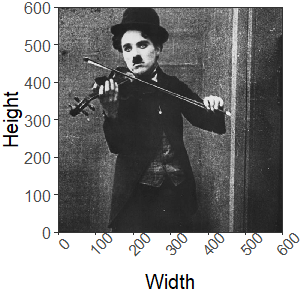}\hfill
		\includegraphics[width=.33\textwidth]{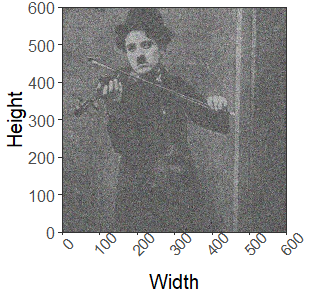}\hfill
		\includegraphics[width=.33\textwidth]{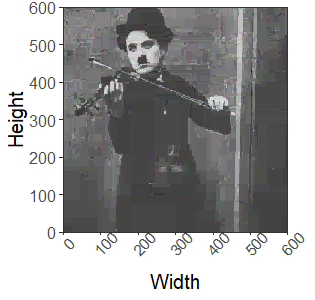}
		\caption{\footnotesize{Image denoising with Algorithm \ref{alg:quarter.seg}. Image is $600\times 600$ pixels. {\it Left panel:} True image (unobserved), {\it Center panel:} Noisy image (observed), {\it Right panel:} Recovered image. Estimated model recovered with $\ell=13$ hierarchical change points inducing a total of $4201$ disjoint partitions of the sampling space. Noise set to $\Si=0.05\cdotp I_{3\times 3},$ and tuning constant $c_{bic}=0.5.$}}
		\label{fig:ex2.chaplin}
	\end{figure}
	
\end{example}


To conclude this sub-section we mention that denoising is perhaps the simplest application of the proposed model and methods. The segmenting process being carried out prior to denoising is of a much larger consequence, in particular, this segmentation layer produced by the proposed methodology has several other potential applications such as that of object identification via a secondary layer of supervised model(s).

\section{Numerical experiments}\label{sec:numerical}

This section considers the single $2d$-change model (\ref{model:rvmcp}) and provides numerical support to the proposed estimation method of Algorithm \ref{alg:single} and the inference results of Theorem \ref{thm:wc.vanishing} and Theorem \ref{thm:wc.non.vanishing}. In all simulations to follow, no underlying parameter is assumed to be known. We begin with a description of the simulation design. In all cases considered, the mean vectors are set as $\theta_{(1)}^0=\theta_{(3)}^0=\big(\theta_{s\times 1}^T,0...,0\big)^T_{p\times 1},$ where $\theta=(0.75,...,0.25)_{s\times 1},$ contains evenly spaced $s=5$ entries. The remaining two means are set to zero, i.e.,  $\theta_{(2)}^0=\theta_{(4)}^0=0_{p\times 1}.$ The covariance matrix $\Si$ is chosen to be a toeplitz type matrix defined as $\Si_{ij}=\rho^{|i-j|},$ $i,j=1,...,p$ and $\rho=0.5.$ We consider all combinations of the sampling periods  $T_w,T_h\in\{30,35,40,45\},$ model dimension $p\in\{10,50,100,250\}.$ The $2d$-change point $\tau^0=(\tau_w^0,\tau^0_h)^T$ chosen as all combinations of $\tau_w^0\in \in\big\{\lfloor 0.2\cdotp T_w\rfloor,\lfloor 0.4\cdotp T_w\rfloor,\lfloor 0.6\cdotp T_w\rfloor,\lfloor 0.8\cdotp T_w\rfloor\big\}$ and   $\tau_h^0\in \in\big\{\lfloor 0.2\cdotp T_h\rfloor,\lfloor 0.4\cdotp T_h\rfloor,\lfloor 0.6\cdotp T_h\rfloor,\lfloor 0.8\cdotp T_h\rfloor\big\}.$ The unobserved noise variables $\vep_{(w,h)}\in\R^p$ are generated as i.i.d. Gaussian r.v.'s, more precisely we set $\vep_{(w,h)}\sim^{i.i.d} \cN(0,\Si),$ $w=1,...,T_w,\,\,h=1,...,T_h.$

We construct confidence intervals using both the limiting distributions of Theorem \ref{thm:wc.vanishing} and Theorem \ref{thm:wc.non.vanishing}. The significance level is set to $\al=0.05.$ Confidence intervals are constructed component-wise, i.e., for the width change parameter as, $\big[(\tilde\tau_w-ME_w),\, (\tilde\tau_w+ME_w)\big],$ where $\tilde\tau_w$ is the width component of the output of Algorithm 1.The margin of error ($ME_w$) is computed as  $ME_w=q_{\alpha}^v\si^2_{(w,\iny)}\big/\big(T_h\xi^2_w\big)$ or $ME=q_{\al}^{nv}$ based on the results of Theorem \ref{thm:wc.vanishing} and Theorem \ref{thm:wc.non.vanishing}, respectively. Here $q_{\al}^v$ represents the $\big(1-\alpha/2\big)^{th}$ quantile of the argmax of two sided negative drift Brownian motion of Theorem \ref{thm:wc.vanishing}. This critical value is evaluated as $c_{\alpha}=11.03$ by using its distribution function provided in \cite{yao1987approximating}. The $\big(1-\alpha/2\big)^{th}$ quantile $q_{\al}^{nv}$ of the argmax of the two sided negative drift random walk is computed as its monte carlo approximation by simulating $4000$ realizations of this distribution.
As per the assumed data generating process, the distribution $\cP$ here is Gaussian. For implementation of the confidence interval, we utilize plugin estimates of $\si^2_{(w,\iny)}$ and $\xi^2_w,$ whose computational details of which are provided in Appendix \ref{app:numerical.supplement} of the supplementary materials. Symmetrical computations are carried out for the height change parameter $\tau_h.$

Choice of tuning parameters: The regularizers $\la_j,$ $j=1,2,3,4$ used to obtain soft thresholded mean estimates in Step 1 and Step 2 of Algorithm \ref{alg:single} are tuned via a BIC type criteria. Specifically we set $\la_j=\la,$ $j=1,2,3,4,$ and evaluate $\h\theta_{(j)}(\la),$ $j=1,2,3,4$ over an equally spaced grid of twenty five values in the interval $(0,0.5).$ Upon letting $\h S=\big\{k;\quad\cup_{j=1}^{4}\h\theta_{(j)k}\ne 0\}$ we evaluate the criteria,
\benr\label{eq:bic}
BIC(\la,\tau_w,\tau_h)= \sum_{j=1}^4\sum_{(w,h)\in Q_j(\tau)}\big\|x_{(w,h)}-\h\theta_{(j)}(\la)\big\|_2^2+|\h S|\log T_wT_h.
\eenr
For Step 1 of Algorithm 1 we set $\la$ as the minimizer of $BIC(\la,\check\tau_w,\check\tau_h),$ and for Step 2 of Algorithm 1 we choose $\la$ as the minimizer of $BIC(\la,\h\tau_w,\h\tau_h).$

To present our results we report the following metrics, bias ($|E(\h\tau_w-\tau^0_w)|$) and root mean squared error ($E^{1/2}(\h\tau_w-\tau^0_w)^2$) measure estimation performance. Coverage (relative frequency of the number of times $\tau^0_w$ lies in the confidence interval), and the average margin of error (average over replications of the margin of error of each confidence interval) measure inference performance. Symmetrical metrics are also presented w.r.t the height change parameter $\tau_h.$ All approximations are made based on $500$ monte carlo replications. Partial results regarding estimation of the width change parameter are provided in Tables \ref{tab:wres1} - Table \ref{tab:wres4}. Corresponding results for the height change parameter are provided in Table \ref{tab:hres1} - Table \ref{tab:hres4} in Appendix \ref{app:numerical.supplement} of the Supplement.

Change point estimates in both directions are largely observed to exhibit little bias with an expected deterioration with larger dimension sizes $p.$ The proposed inference methodology is observed to provide good control on the nominal significance level if one provides a slight leeway given the discrete nature of the underlying problem and confidence intervals. The cases where coverage is observed to be significantly away from nominal are again the larger values of $p,$ or where change points are near the parametric boundary (see, Table \ref{tab:wres2}, case: $p=250,$ $\tau^0=0.8$). It can be observed that this deviation from nominal coverage is primarily due to bias in estimation and not the computation of margin of error (the margin of error as expected remains stable for all values of $T_w$).  Importantly, bias is observed to diminish and coverage to catchup to nominal as the effective sample size of $T_wT_h\underline\om$ increases, i.e., when the sampling periods increase or the change point move away from parametric boundaries.

\begin{table}[]
	\centering
	\resizebox{0.8\textwidth}{!}{
	\begin{tabular}{llcccccc}
		\toprule
		\multicolumn{2}{c}{\multirow{2}{*}{\begin{tabular}[c]{@{}c@{}}$T_h=30,$\\ $\tau^0_h/T_h=0.2$\end{tabular}}} & \multicolumn{3}{c}{$p=10$}                                                                                & \multicolumn{3}{c}{$p=50$}                                                                               \\ \cmidrule{3-8}
		\multicolumn{2}{c}{}                                                                                        & \multirow{2}{*}{bias (rmse)}      & \multicolumn{2}{c}{coverage (av. ME)}                                 & \multirow{2}{*}{bias (rmse)}      & \multicolumn{2}{c}{coverage (av. ME)}                                \\ \cmidrule{1-2} \cmidrule{4-5} \cmidrule{7-8}
		\multicolumn{1}{c}{$\tau^0_w/T_w$}                         & \multicolumn{1}{c}{$T_w$}                        &                                   & Vanishing                         & Non-Vanishing                     &                                   & Vanishing                         & Non-Vanishing                    \\ \midrule
		0.2                                                      & 30                                               & 0.008 (0.155)                     & 0.976 (0.468)                     & 0.978 (0.018)                     & 0.01 (0.148)                      & 0.978 (0.394)                     & 0.978 (0)                        \\
		0.2                                                      & 35                                               & 0.02 (0.2)                        & 0.96 (0.508)                      & 0.966 (0.04)                      & 0.022 (0.195)                     & 0.968 (0.401)                     & 0.968 (0)                        \\
		0.2                                                      & 40                                               & 0.016 (0.21)                      & 0.962 (0.514)                     & 0.964 (0.032)                     & 0.02 (0.2)                        & 0.96 (0.409)                      & 0.96 (0)                         \\
		0.2                                                      & 45                                               & 0.032 (0.253)                     & 0.958 (0.516)                     & 0.96 (0.02)                       & 0.006 (0.214)                     & 0.972 (0.418)                     & 0.972 (0)                        \\ \midrule
		0.4                                                      & 30                                               & 0.02 (0.228)                      & 0.954 (0.476)                     & 0.956 (0.02)                      & 0.034 (0.344)                     & 0.964 (0.471)                     & 0.964 (0.004)                    \\
		0.4                                                      & 35                                               & 0.028 (0.268)                     & 0.958 (0.522)                     & 0.958 (0.02)                      & 0.064 (0.477)                     & 0.954 (0.472)                     & 0.954 (0.002)                    \\
		0.4                                                      & 40                                               & 0.006 (0.224)                     & 0.966 (0.528)                     & 0.968 (0.018)                     & 0.05 (0.293)                      & 0.962 (0.48)                      & 0.962 (0.002)                    \\
		0.4                                                      & 45                                               & 0.02 (0.228)                      & 0.96 (0.523)                      & 0.964 (0.012)                     & 0.008 (0.141)                     & 0.98 (0.482)                      & 0.982 (0.004)                    \\ \midrule
		0.6                                                      & 30                                               & 0.06 (0.358)                      & 0.948 (0.495)                     & 0.95 (0.028)                      & 0.078 (0.361)                     & 0.936 (0.504)                     & 0.938 (0.014)                    \\
		0.6                                                      & 35                                               & 0.008 (0.245)                     & 0.97 (0.522)                      & 0.972 (0.02)                      & 0.056 (0.303)                     & 0.954 (0.501)                     & 0.954 (0.008)                    \\
		0.6                                                      & 40                                               & 0.006 (0.265)                     & 0.958 (0.533)                     & 0.96 (0.026)                      & 0.054 (0.326)                     & 0.946 (0.507)                     & 0.946 (0.004)                    \\
		0.6                                                      & 45                                               & 0.034 (0.224)                     & 0.962 (0.535)                     & 0.966 (0.02)                      & 0.048 (0.228)                     & 0.948 (0.515)                     & 0.952 (0.01)                     \\ \midrule
		0.8                                                      & 30                                               & 0.118 (0.882)                     & 0.944 (0.529)                     & 0.948 (0.064)                     & 0.212 (0.976)                     & 0.876 (0.474)                     & 0.886 (0.036)                    \\
		0.8                                                      & 35                                               & 0.036 (0.237)                     & 0.968 (0.521)                     & 0.974 (0.054)                     & 0.13 (0.65)                       & 0.916 (0.475)                     & 0.916 (0.014)                    \\
		0.8                                                      & 40                                               & 0.024 (0.179)                     & 0.97 (0.513)                      & 0.972 (0.022)                     & 0.092 (0.42)                      & 0.932 (0.488)                     & 0.932 (0.012)                    \\
		0.8                                                      & 45                                               & \multicolumn{1}{l}{0.046 (0.265)} & \multicolumn{1}{l}{0.958 (0.519)} & \multicolumn{1}{l}{0.966 (0.034)} & \multicolumn{1}{l}{0.066 (0.319)} & \multicolumn{1}{l}{0.938 (0.488)} & \multicolumn{1}{l}{0.94 (0.006)} \\ \bottomrule
	\end{tabular}}
	\vspace{1mm}
\caption{\footnotesize{Simulation results for estimation of $\tau^0_w$ based on 500 replications. All reported metrics rounded to three decimals. Other data generating parameters: $T_h=30,$ $\tau^0_h=\lfloor 0.2\cdotp T_h\rfloor$ and $p\in\{10,50\}.$}}
\label{tab:wres1}
\end{table}

\begin{table}[]
	\centering
	\resizebox{0.8\textwidth}{!}{
	\begin{tabular}{llllllll}
		\toprule
		\multicolumn{2}{c}{\multirow{2}{*}{\begin{tabular}[c]{@{}c@{}}$T_h=30,$\\ $\tau^0_h/T_h=0.2$\end{tabular}}} & \multicolumn{3}{c}{$p=100$}                                                                                          & \multicolumn{3}{c}{$p=250$}                                                                                          \\ \cmidrule{3-8}
		\multicolumn{2}{c}{}                                                                                        & \multicolumn{1}{c}{\multirow{2}{*}{bias (rmse)}} & \multicolumn{2}{c}{coverage (av. ME)}                             & \multicolumn{1}{c}{\multirow{2}{*}{bias (rmse)}} & \multicolumn{2}{c}{coverage (av. ME)}                             \\ \cmidrule{1-2} \cmidrule{4-5} \cmidrule{7-8}
		\multicolumn{1}{c}{$\tau^0_w/T_w$}                         & \multicolumn{1}{c}{$T_w$}                        & \multicolumn{1}{c}{}                             & \multicolumn{1}{c}{Vanishing} & \multicolumn{1}{c}{Non-Vanishing} & \multicolumn{1}{c}{}                             & \multicolumn{1}{c}{Vanishing} & \multicolumn{1}{c}{Non-Vanishing} \\ \midrule
		0.2                                                      & 30                                               & 0.032 (0.738)                                    & 0.966 (0.352)                 & 0.966 (0.002)                     & 0.024 (0.74)                                     & 0.968 (0.306)                 & 0.968 (0.002)                     \\
		0.2                                                      & 35                                               & 0.016 (0.167)                                    & 0.972 (0.362)                 & 0.972 (0)                         & 0.008 (0.19)                                     & 0.964 (0.325)                 & 0.964 (0)                         \\
		0.2                                                      & 40                                               & 0.008 (0.167)                                    & 0.972 (0.375)                 & 0.972 (0)                         & 0.01 (0.173)                                     & 0.97 (0.342)                  & 0.97 (0)                          \\
		0.2                                                      & 45                                               & 0.026 (0.173)                                    & 0.97 (0.381)                  & 0.97 (0)                          & 0.01 (0.615)                                     & 0.964 (0.35)                  & 0.964 (0.002)                     \\ \midrule
		0.4                                                      & 30                                               & 0.048 (0.29)                                     & 0.952 (0.457)                 & 0.952 (0.002)                     & 0.104 (0.544)                                    & 0.938 (0.438)                 & 0.938 (0)                         \\
		0.4                                                      & 35                                               & 0.04 (0.303)                                     & 0.962 (0.463)                 & 0.962 (0)                         & 0.086 (0.546)                                    & 0.932 (0.451)                 & 0.932 (0)                         \\
		0.4                                                      & 40                                               & 0.038 (0.224)                                    & 0.95 (0.47)                   & 0.95 (0)                          & 0.044 (0.261)                                    & 0.956 (0.455)                 & 0.956 (0)                         \\
		0.4                                                      & 45                                               & 0.024 (0.245)                                    & 0.962 (0.481)                 & 0.962 (0)                         & 0.028 (0.228)                                    & 0.97 (0.463)                  & 0.97 (0.002)                      \\ \midrule
		0.6                                                      & 30                                               & 0.142 (0.801)                                    & 0.922 (0.49)                  & 0.922 (0.01)                      & 0.122 (0.852)                                    & 0.928 (0.467)                 & 0.93 (0.01)                       \\
		0.6                                                      & 35                                               & 0.068 (0.346)                                    & 0.94 (0.491)                  & 0.94 (0)                          & 0.114 (0.68)                                     & 0.934 (0.481)                 & 0.934 (0.008)                     \\
		0.6                                                      & 40                                               & 0.068 (0.696)                                    & 0.954 (0.498)                 & 0.954 (0.004)                     & 0.072 (0.369)                                    & 0.942 (0.485)                 & 0.942 (0.004)                     \\
		0.6                                                      & 45                                               & 0.046 (0.272)                                    & 0.958 (0.505)                 & 0.96 (0.004)                      & 0.066 (0.387)                                    & 0.948 (0.49)                  & 0.954 (0.008)                     \\ \midrule
		0.8                                                      & 30                                               & 0.276 (1.31)                                     & 0.856 (0.444)                 & 0.858 (0.022)                     & 0.568 (2.416)                                    & 0.776 (0.424)                 & 0.78 (0.034)                      \\
		0.8                                                      & 35                                               & 0.12 (0.639)                                     & 0.928 (0.456)                 & 0.93 (0.01)                       & 0.27 (0.838)                                     & 0.82 (0.433)                  & 0.82 (0.018)                      \\
		0.8                                                      & 40                                               & 0.176 (0.699)                                    & 0.884 (0.468)                 & 0.892 (0.018)                     & 0.168 (0.593)                                    & 0.88 (0.431)                  & 0.882 (0.01)                      \\
		0.8                                                      & 45                                               & 0.094 (0.377)                                    & 0.928 (0.466)                 & 0.928 (0.006)                     & 0.294 (1.599)                                    & 0.866 (0.445)                 & 0.866 (0.014)                     \\ \bottomrule
	\end{tabular}}
\vspace{1mm}
\caption{\footnotesize{Simulation results for estimation of $\tau^0_w$ based on 500 replications. All reported metrics rounded to three decimals. Other data generating parameters: $T_h=30,$ $\tau^0_h=\lfloor 0.2\cdotp T_h\rfloor$ and $p\in\{100,250\}.$}}
\label{tab:wres2}
\end{table}

\begin{table}[]		\centering
	\resizebox{0.8\textwidth}{!}{
	\begin{tabular}{llllllll}
		\toprule
		\multicolumn{2}{c}{\multirow{2}{*}{\begin{tabular}[c]{@{}c@{}}$T_h=30,$\\ $\tau^0_h/T_h=0.4$\end{tabular}}} & \multicolumn{3}{c}{$p=10$}                                                                                           & \multicolumn{3}{c}{$p=50$}                                                                                           \\ \cmidrule{3-8}
		\multicolumn{2}{c}{}                                                                                        & \multicolumn{1}{c}{\multirow{2}{*}{bias (rmse)}} & \multicolumn{2}{c}{coverage (av. ME)}                             & \multicolumn{1}{c}{\multirow{2}{*}{bias (rmse)}} & \multicolumn{2}{c}{coverage (av. ME)}                             \\ \cmidrule{1-2} \cmidrule{4-5} \cmidrule{7-8}
		\multicolumn{1}{c}{$\tau^0_w/T_w$}                        & \multicolumn{1}{c}{$T_w$}                       & \multicolumn{1}{c}{}                             & \multicolumn{1}{c}{Vanishing} & \multicolumn{1}{c}{Non-Vanishing} & \multicolumn{1}{c}{}                             & \multicolumn{1}{c}{Vanishing} & \multicolumn{1}{c}{Non-Vanishing} \\ \midrule
		0.2                                                       & 30                                              & 0.028 (0.219)                                    & 0.958 (0.464)                 & 0.962 (0.028)                     & 0.106 (1.093)                                    & 0.954 (0.421)                 & 0.954 (0.02)                      \\
		0.2                                                       & 35                                              & 0.056 (0.303)                                    & 0.936 (0.517)                 & 0.948 (0.048)                     & 0.054 (0.88)                                     & 0.964 (0.443)                 & 0.964 (0.012)                     \\
		0.2                                                       & 40                                              & 0.02 (0.19)                                      & 0.97 (0.512)                  & 0.972 (0.03)                      & 0.008 (0.155)                                    & 0.976 (0.441)                 & 0.976 (0.004)                     \\
		0.2                                                       & 45                                              & 0.028 (0.237)                                    & 0.956 (0.519)                 & 0.958 (0.036)                     & 0.004 (0.19)                                     & 0.964 (0.442)                 & 0.964 (0)                         \\ \midrule
		0.4                                                       & 30                                              & 0.002 (0.279)                                    & 0.964 (0.467)                 & 0.964 (0.022)                     & 0.02 (0.219)                                     & 0.968 (0.443)                 & 0.968 (0.006)                     \\
		0.4                                                       & 35                                              & 0.01 (0.272)                                     & 0.96 (0.524)                  & 0.96 (0.03)                       & 0.002 (0.195)                                    & 0.968 (0.445)                 & 0.968 (0.004)                     \\
		0.4                                                       & 40                                              & 0.01 (0.214)                                     & 0.96 (0.534)                  & 0.962 (0.032)                     & 0.002 (0.161)                                    & 0.974 (0.456)                 & 0.974 (0.004)                     \\
		0.4                                                       & 45                                              & 0.014 (0.272)                                    & 0.964 (0.531)                 & 0.964 (0.016)                     & 0.014 (0.184)                                    & 0.966 (0.459)                 & 0.966 (0.002)                     \\ \midrule
		0.6                                                       & 30                                              & 0.026 (0.205)                                    & 0.964 (0.471)                 & 0.964 (0.012)                     & 0.04 (0.522)                                     & 0.968 (0.474)                 & 0.97 (0.012)                      \\
		0.6                                                       & 35                                              & 0.022 (0.326)                                    & 0.948 (0.523)                 & 0.952 (0.024)                     & 0.016 (0.179)                                    & 0.974 (0.466)                 & 0.974 (0.006)                     \\
		0.6                                                       & 40                                              & 0.022 (0.232)                                    & 0.968 (0.521)                 & 0.97 (0.01)                       & 0.018 (0.195)                                    & 0.968 (0.476)                 & 0.968 (0.004)                     \\
		0.6                                                       & 45                                              & 0.008 (0.21)                                     & 0.962 (0.529)                 & 0.966 (0.016)                     & 0.01 (0.184)                                     & 0.966 (0.49)                  & 0.968 (0.006)                     \\ \midrule
		0.8                                                       & 30                                              & 0.028 (0.261)                                    & 0.95 (0.486)                  & 0.952 (0.034)                     & 0.146 (1.15)                                     & 0.926 (0.46)                  & 0.926 (0.016)                     \\
		0.8                                                       & 35                                              & 0.028 (0.253)                                    & 0.954 (0.505)                 & 0.96 (0.034)                      & 0.076 (0.927)                                    & 0.956 (0.465)                 & 0.956 (0.008)                     \\
		0.8                                                       & 40                                              & 0.022 (0.224)                                    & 0.968 (0.519)                 & 0.972 (0.042)                     & 0.028 (0.19)                                     & 0.964 (0.472)                 & 0.966 (0.004)                     \\
		0.8                                                       & 45                                              & 0.048 (0.253)                                    & 0.948 (0.523)                 & 0.956 (0.036)                     & 0.038 (0.3)                                      & 0.962 (0.475)                 & 0.962 (0.006)                     \\ \bottomrule
	\end{tabular}}
\vspace{1mm}
\caption{\footnotesize{Simulation results for estimation of $\tau^0_w$ based on 500 replications. All reported metrics rounded to three decimals. Other data generating parameters: $T_h=30,$ $\tau^0_h=\lfloor 0.4\cdotp T_h\rfloor$ and $p\in\{10,50\}.$}}
\label{tab:wres3}
\end{table}

\begin{table}[]
		\centering
	\resizebox{0.85\textwidth}{!}{
	\begin{tabular}{llllllll}
		\toprule
		\multicolumn{2}{c}{\multirow{2}{*}{\begin{tabular}[c]{@{}c@{}}$T_h=30,$\\ $\tau^0_h/T_h=0.4$\end{tabular}}} & \multicolumn{3}{c}{$p=100$}                                                                                          & \multicolumn{3}{c}{$p=250$}                                                                                          \\ \cmidrule{3-8}
		\multicolumn{2}{c}{}                                                                                        & \multicolumn{1}{c}{\multirow{2}{*}{bias (rmse)}} & \multicolumn{2}{c}{coverage (av. ME)}                             & \multicolumn{1}{c}{\multirow{2}{*}{bias (rmse)}} & \multicolumn{2}{c}{coverage (av. ME)}                             \\ \cmidrule{1-2} \cmidrule{4-5} \cmidrule{7-8}
		\multicolumn{1}{c}{$\tau^0_w/T_w$}                        & \multicolumn{1}{c}{$T_w$}                       & \multicolumn{1}{c}{}                             & \multicolumn{1}{c}{Vanishing} & \multicolumn{1}{c}{Non-Vanishing} & \multicolumn{1}{c}{}                             & \multicolumn{1}{c}{Vanishing} & \multicolumn{1}{c}{Non-Vanishing} \\ \midrule
		0.2                                                       & 30                                              & 0.09 (1.122)                                     & 0.968 (0.4)                   & 0.968 (0.008)                     & 0.076 (0.908)                                    & 0.966 (0.357)                 & 0.966 (0.008)                     \\
		0.2                                                       & 35                                              & 0.03 (0.488)                                     & 0.966 (0.412)                 & 0.966 (0.004)                     & 0.008 (0.179)                                    & 0.968 (0.367)                 & 0.968 (0)                         \\
		0.2                                                       & 40                                              & 0.01 (0.195)                                     & 0.962 (0.418)                 & 0.962 (0)                         & 0.046 (0.866)                                    & 0.97 (0.387)                  & 0.97 (0.006)                      \\
		0.2                                                       & 45                                              & 0.006 (0.161)                                    & 0.974 (0.425)                 & 0.974 (0)                         & 0.014 (0.173)                                    & 0.976 (0.397)                 & 0.976 (0)                         \\ \midrule
		0.4                                                       & 30                                              & 0.002 (0.224)                                    & 0.956 (0.438)                 & 0.962 (0.008)                     & 0.008 (0.2)                                      & 0.966 (0.421)                 & 0.966 (0.008)                     \\
		0.4                                                       & 35                                              & 0.032 (0.316)                                    & 0.976 (0.434)                 & 0.976 (0.002)                     & 0.03 (0.392)                                     & 0.97 (0.432)                  & 0.97 (0)                          \\
		0.4                                                       & 40                                              & 0.01 (0.205)                                     & 0.97 (0.438)                  & 0.97 (0)                          & 0.002 (0.214)                                    & 0.96 (0.437)                  & 0.96 (0.002)                      \\
		0.4                                                       & 45                                              & 0.01 (0.148)                                     & 0.978 (0.441)                 & 0.978 (0)                         & 0.004 (0.11)                                     & 0.988 (0.445)                 & 0.988 (0)                         \\ \midrule
		0.6                                                       & 30                                              & 0.018 (0.173)                                    & 0.97 (0.459)                  & 0.97 (0.004)                      & 0.014 (0.173)                                    & 0.976 (0.46)                  & 0.976 (0)                         \\
		0.6                                                       & 35                                              & 0.038 (0.605)                                    & 0.97 (0.471)                  & 0.97 (0.004)                      & 0.018 (0.195)                                    & 0.968 (0.479)                 & 0.968 (0.004)                     \\
		0.6                                                       & 40                                              & 0.006 (0.118)                                    & 0.986 (0.482)                 & 0.986 (0.002)                     & 0.006 (0.148)                                    & 0.978 (0.475)                 & 0.978 (0)                         \\
		0.6                                                       & 45                                              & 0.022 (0.232)                                    & 0.958 (0.484)                 & 0.96 (0.004)                      & 0.002 (0.184)                                    & 0.978 (0.487)                 & 0.978 (0)                         \\ \midrule
		0.8                                                       & 30                                              & 0.14 (1.103)                                     & 0.94 (0.434)                  & 0.942 (0.01)                      & 0.138 (1.127)                                    & 0.94 (0.403)                  & 0.94 (0.016)                      \\
		0.8                                                       & 35                                              & 0.052 (0.486)                                    & 0.954 (0.437)                 & 0.954 (0.002)                     & 0.06 (0.54)                                      & 0.954 (0.417)                 & 0.954 (0.004)                     \\
		0.8                                                       & 40                                              & 0.034 (0.214)                                    & 0.96 (0.45)                   & 0.96 (0)                          & 0.04 (0.253)                                     & 0.954 (0.426)                 & 0.954 (0.004)                     \\
		0.8                                                       & 45                                              & 0.052 (1.083)                                    & 0.978 (0.463)                 & 0.978 (0.002)                     & 0.02 (0.2)                                       & 0.966 (0.433)                 & 0.966 (0)                         \\ \bottomrule
	\end{tabular}}
\vspace{1mm}
\caption{\footnotesize{Simulation results for estimation of $\tau^0_w$ based on 500 replications. All reported metrics rounded to three decimals. Other data generating parameters: $T_h=30,$ $\tau^0_h=\lfloor 0.4\cdotp T_h\rfloor$ and $p\in\{100,250\}.$}}
\label{tab:wres4}
\end{table}

\section{Discussion}

Regression trees are amongst the most heavily utilized and perhaps the least analytically understood modelling techniques. We illustrate that viewing regression trees via multidimensional change points provides an analytically tractable construction which allows for a multivariate response, and more importantly allows for fundamental statistical properties of rates of estimation and limiting distributions to be established, despite potential high dimensionality. The purpose of this article is only to make first inroads into this connection. Clearly there is a rich body of further questions that need analytical addressal particularly in Section \ref{sec:extensions} above, these are however beyond the scope of this article. Moreover, solutions to these questions still require considerable development of the change point literature at a more fundamental level of one dimensional change points, particularly in context of inference.


\clearpage

\setcounter{page}{1}

\section*{\hfil\Large{Supplementary Materials}}\hfil

\vspace{2mm}
{\bc \Large{ 
\sc{Segmentation of high dimensional means over multi-dimensional change points and connections to regression trees}}\ec}

\appendix

\begin{appendix}

	\section{Proofs of results in Section \ref{sec:main}}
	
	To present the arguments of this section we require additional notation. In all to follow let  $\h\eta_{(j)}$ represent estimates of the jump parameters $\eta^0_{(j)},$ $j=1,2,3,4$ respectively.  For any non-negative sequences $0\le v_{T_w}\le u_{T_w}\le 1,$ in $T_w$ and $0\le v_{T_h}\le u_{T_h}\le 1,$ in $T_h,$ define the following collections,
	\benr\label{def:setG}
	&&\cG_w(u_{T_w},v_{T_w})=\Big\{\tau_w\in\{1,2,...,T_w\};\,\,T_wv_{T_w}\le |\tau_w-\tau^0_w|\le T_wu_{T_w} \Big\},\,\,{\rm and}\nn\\
	&&\cG_h(u_{T_h},v_{T_h})=\Big\{\tau_h\in\{1,2,...,T_h\};\,\,T_hv_{T_h}\le |\tau_h-\tau^0_h|\le T_hu_{T_h} \Big\}.\nn
	\eenr
	Finally, we define,
	\benr\label{def:cU}
	&&\cU_w(\tau_w,\tau_h,\theta)=T_wT_h\Big(\cL(\tau_w,\tau_h,\theta)-\cL(\tau_w^0,\tau_h,\theta)\Big),\,\,{\rm and}\nn\\
	&&\cU_h(\tau_w,\tau_h,\theta)=T_wT_h\Big(\cL(\tau_w,\tau_h,\theta)-\cL(\tau_w,\tau_h^0,\theta)\Big),\nn
	\eenr
	where $\cL(\cdotp,\cdotp,\cdotp)$ is the squared loss defined in (\ref{def:sq.loss}). Clearly, the plug-in estimates $\tilde\tau_w(\h\tau_h,\h\theta)$ and $\tilde\tau_h(\h\tau_w,\h\theta)$ of (\ref{est:optimal}) can then equivalently be written as,
	\benr\label{est:optimal.cU}
	\tilde\tau_w(\h\tau_h,\h\theta)=\argmin_{1\le\tau_w< T_w} \cU_w(\tau_w,\h\tau_h,\h\theta),\quad{\rm and}\quad \tilde\tau_h(\h\tau_w,\h\theta)=\argmin_{1\le\tau_h< T_h} \cU_h(\h\tau_w,\tau_h,\h\theta)
	\eenr
	The change of representation of estimates to (\ref{est:optimal.cU}) is made solely for notational convenience in the proofs to follow. We begin this section with Lemma \ref{lem:unif.lb}-\ref{lem:unif.lb.optimal.height}, which are all closely related. These lemma's shall serve as critical tools that upon which our arguments for Theorem \ref{thm:cp.nearoptimal} and Theorem \ref{thm:cpoptimal} rely. Lemma \ref{lem:unif.lb} and Lemma \ref{lem:unif.lb.height} shall serve towards the argument for the near optimal rate of Theorem \ref{thm:cp.nearoptimal} for the estimates $\tilde\tau_w,$ and $\tilde\tau_h,$ respectively. Lemma \ref{lem:unif.lb.optimal} and Lemma \ref{lem:unif.lb.optimal.height} shall sever towards Theorem \ref{thm:cpoptimal} for $\tilde\tau_w,$ and $\tilde\tau_h,$ respectively. All four of lemma's provide uniform lower bounds over the collections $\cG_w$ and $\cG_h,$ under different conditions of preliminary nuisance estimates.

	\begin{lemma}\label{lem:unif.lb} Suppose the model (\ref{model:rvmcp}) and assume that Condition A, B, C(i)(a) and C(ii)(a,b) hold. Let $0\le v_{T_w}\le u_{T_w}\le 1,$ be any non-negative sequences. Then,
		\benr\label{eq:3a}
		\inf_{\tau_w\in\cG_w(u_{T_w},v_{T_w})}\cU_w(\tau_w,\h\tau_h,\h\theta)\ge \frac{T_wT_h\xi^2_w}{2}\Big[v_{T_w}-c_u\log(p\vee T_wT_h)\frac{\si}{\xi_w}\Big\{\frac{su_{T_w}}{T_wT_h}\Big\}^{\frac{1}{2}}\Big]
		\eenr
		with probability at least $1-2\exp\{-c_1\log(p\vee T)\}-\pi_T,$ for constant $c_1>0$ that does not depend on any model parameters.
	\end{lemma}

	\begin{proof}[{\bf Proof of Lemma \ref{lem:unif.lb}}] The proof of this result is fairly long and is thus broken up into the following three steps for the convenience of the reader.
		\begin{enumerate}[itemsep=0pt, labelindent=!]
			\item[{\bf Step 1}] Utilize Condition B, Condition C and results of Appendix \ref{app:deviation} to obtain upper and lower bounds on some stochastic quantities of interest.
			\item[{\bf Step 2}] Perform an algebraic decomposition of $\cU_w(\tau_w,\h\tau_h,\h\theta)$ to obtain a manageable expression in terms of jump sizes (\ref{def:jump.size.quad}) and additional noise terms.
			\item [{\bf Step 3}] Apply bounds of Step 1 to expression of Step 2 to obtain the desired uniform lower bound of the statement of this lemma.
		\end{enumerate}	
		
		\noi We begin with {\noi\bf Step 1} that provides a few observations that shall be required to obtain the desired lower bound. Using Condition C(ii)(a,b) we have the following relations,	
		\benr\label{eq:4}
		\|\h\eta_{(1)}-\eta^0_{(1)}\|_2&\le& \|\h\theta_{(1)}-\theta_{(1)}^0\|_2+\|\h\theta_{(2)}-\theta_{(2)}^0\|_2\le 2c_{u1}\xi_{\min}\quad{\rm and\,\, similarly},\nn\\
		\|\h\eta_{(1)}-\eta^0_{(1)}\|_1&\le& 4\surd{s}\|\h\theta_{(1)}-\theta_{(1)}^0\|_2+4\surd{s}\|\h\theta_{(2)}-\theta_{(2)}^0\|_2\le 8c_{u1}\surd{s}\xi_{\min}
		\eenr
		with probability at least $1-\pi_T.$ Here the third inequality follows from Condition C(ii)(a). Next, consider,
		\benr\label{eq:5}
		\|\h\eta_{(1)}\|_2&\le& \|\h\eta_{(1)}-\eta^0_{(1)}\|_2+\|\eta^0_{(1)}\|_2\le \xi_1 +2_{u1}\xi_{\min}\le c_u\xi_{\min},\quad{\rm and\,\, similarly},\nn\\
		\|\h\eta_{(1)}\|_1&\le&\|\h\eta_{(1)}-\eta^0\|_1+\|\eta^0_{(1)}\|_1\le 8c_{u1}\surd{s}\xi_{\min}+\surd{s}\xi_1\le c_u\surd{s}\xi_{\min},
		\eenr
		which holds with probability at least $1-\pi_T.$ Here the second inequality for the $\ell_2$ bound follows from (\ref{eq:4}) and the third follows from Condition B(ii). The $\ell_1$ bound follows analogously. Expression (\ref{eq:5}) provides an upper bound for $\|\h\eta_{(1)}\|_2$ that holds with probability at least $1-\pi_T,$ below we show that this quantity can also be bounded from below. Consider,
		\benr\label{eq:7}
		\|\h\eta_{(1)}\|_2^2&=&\|\eta^0_{(1)}+(\h\eta_{(1)}-\eta^0_{(1)})\|_2^2\ge \|\eta^0_{(1)}\|_2^2+2(\h\eta_{(1)}-\eta^0_{(1)})^T\eta^0_{(1)}\nn\\
		&\ge& \xi_1^2-2\|\h\eta_{(1)}-\eta^0_{(1)}\|_2\xi_1\ge \xi_1^2-2c_{u1}\xi_{\min}\overline\xi
		\eenr
		with probability at least $1-\pi_T.$ Here the second inequality follows by an application of the Cauchy-Schwarz inequality. The final inequality follows from (\ref{eq:4}). Analogous arguments can be utilized to obtain versions of the above bounds for $\|\h\eta_{(3)}\|_2,$ specifically,
		\benr\label{eq:8}
		&\|\h\eta_{(3)}-\eta^0_{(3)}\|_2\le  2c_{u1}\xi_{\min},\qquad \|\h\eta_{(3)}-\eta^0_{(3)}\|_1\le 8c_{u1}\surd{s}\xi_{\min},\\
		&\|\h\eta_{(3)}\|_2\le c_u\xi_{\min},\qquad \|\h\eta_{(3)}\|_1\le c_u\surd{s}\xi_{\min},\quad{\rm and}\quad \|\h\eta_{(3)}\|_2^2\ge \xi_3^2-2c_{u1}\xi_{\min}\overline\xi\nn
		\eenr
		with probability $1-\pi_T.$ Additional residual terms that shall require control are as follows,
		\benr\label{eq:9}
		2\om_h(\h\theta_{(1)}-\theta_{(1)}^0)^T\h\eta_{(1)}+2(1-\om_h)(\h\theta_{(4)}-\theta_{(4)}^0)^T\h\eta_{(3)}\hspace{4cm}\nn\\
		-2\frac{(\h\tau_h-\tau_h^0)}{T_h}(\h\theta_{(1)}-\theta_{(1)}^0)^T\h\eta_{(1)}
		+2\frac{(\h\tau_h-\tau_h^0)}{T_h}(\h\theta_{(4)}-\theta_{(1)}^0)^T\h\eta_{(3)}\nn\\
		+\frac{(\h\tau_h-\tau_h^0)}{T_h}\|\h\eta_{(3)}\|_2^2-\frac{(\h\tau_h-\tau_h^0)}{T_h}\|\h\eta_{(1)}\|_2^2\hspace{3.5cm}\nn\\
		\le 2\|\h\theta_{(1)}-\theta_{(1)}^0\|_2\|\h\eta_{(1)}\|_2+2\|\h\theta_{(4)}-\theta_{(4)}^0\|_2\|\h\eta_{(3)}\|_2\hspace{3cm}\nn\\
		+2\frac{|\h\tau_h-\tau_h^0|}{T_h}\|\h\theta_{(1)}-\theta_{(1)}^0\|_2\|\h\eta_{(1)}\|_2+2\frac{|\h\tau_h-\tau_h^0|}{T_h}\|\h\theta_{(4)}-\theta_{(1)}^0\|_2\|\h\eta_{(3)}\|_2\nn\\
		+\frac{|\h\tau_h-\tau_h^0|}{T_h}\|\h\eta_{(3)}\|_2^2+\frac{|\h\tau_h-\tau_h^0|}{T_h}\|\h\eta_{(1)}\|_2^2\hspace{4.65cm}\nn\\
		\le c_{u1}\xi_{\min}^2+c_{u1}\overline\xi\xi_{\min}\hspace{2.9in}
		\eenr
		with probability at least $1-\pi_T.$ Here the first inequality follows from several applications of the Cauchy-Schwarz inequality. The second inequality follows by utilizing Condition C(i)(a), C(ii)(b) as well as (\ref{eq:5}). Here we have also utilized the triangle inequality $\|\h\theta_{(4)}-\theta_{(1)}^0\|\le \|\h\theta_{(4)}-\theta_{(4)}^0\|+\|\eta^0_{(4)}\|.$ The above inequalities provide bounds on terms where the randomness is induced solely due to the plug-in preliminary estimates $\h\tau_h$ and $\h\theta.$ The following provides upper bounds on stochastic terms where randomness is induced via the noise terms $\vep_{(w,h)}'$s. These shall follow mainly as a consequence of results of Appendix \ref{app:deviation}. Consider,
		\benr\label{eq:10}
		\sup_{\substack{\tau_w\in\cG_w(u_{T_w},v_{T_w});\\ \tau_w\ge\tau_w^0}}\Big|\sum_{w=\tau_w^0+1}^{\tau_w}\sum_{h=\tau_h^0+1}^{T_h}\vep_{(w,h)}^T\h\eta_{(1)}\Big|&\le& \sup_{\substack{\tau_w\in\cG_w(u_{T_w},v_{T_w});\\ \tau_w\ge\tau_w^0}}\Big\|\sum_{w=\tau_w^0+1}^{\tau_w}\sum_{h=\tau_h^0+1}^{T_h}\vep_{(w,h)}\Big\|_{\iny}\|\h\eta_{(1)}\|_1\nn\\
		&\le& c_u\xi_{\min}\si\log(p\vee T_wT_h)\surd(T_wT_hu_{T_w})\surd{s}
		\eenr
		with probability at least $1-2\exp\{c_1\log(p\vee T_wT_h)\}-\pi_T,$ for some $c_1>0.$ Here the second inequality follows from Lemma \ref{lem:nearoptimalcross.subE.special} and (\ref{eq:5}). The same argument also yields the same uniform bounds on other similar residual terms,
		\benr\label{eq:11}
		&&\sup_{\substack{\tau_w\in\cG_w(u_{T_w},v_{T_w});\\ \tau_w\ge\tau_w^0}}\Big|\sum_{w=\tau_w^0+1}^{\tau_w}\sum_{h=1}^{\tau_h^0}\vep_{(w,h)}^T\h\eta_{(3)}\Big|\le c_u\xi_{\min}\si\log(p\vee T_wT_h)\surd(T_wT_hu_{T_w})\surd{s}\nn\\
		&&\sup_{\substack{\tau_w\in\cG_w(u_{T_w},v_{T_w});\\ \tau_w\ge\tau_w^0}}\Big|\sum_{w=\tau_w^0+1}^{\tau_w}\sum_{h=\tau^0_h+1}^{\h\tau_h}\vep_{(w,h)}^T\h\eta_{(1)}\Big|\le c_u\xi_{\min}\si\log(p\vee T_wT_h)\surd(T_wT_hu_{T_w})\surd{s}\nn\\
		&&\sup_{\substack{\tau_w\in\cG_w(u_{T_w},v_{T_w});\\ \tau_w\ge\tau_w^0}}\Big|\sum_{w=\tau_w^0+1}^{\tau_w}\sum_{h=\tau^0_h+1}^{\h\tau_h}\vep_{(w,h)}^T\h\eta_{(3)}\Big|\le c_u\xi_{\min}\si\log(p\vee T_wT_h)\surd(T_wT_hu_{T_w})\surd{s}
		\eenr
		with probability at least $1-2\exp\{c_1\log(p\vee T_wT_h)\}-\pi_T.$ Here we have also utilized Condition C(i)(a) which provides $|\h\tau_h-\tau_h^0|\le c_{u1}T_h,$ w.p. $1-\pi_T.$ These bounds complete {\bf Step 1} of our argument and provide the necessary groundwork to proceed to the next step.
		
		\vspace{2mm}
		\noi{\bf Step 2:} We shall prove the statement of this lemma for the case $\tau_w\ge \tau_w^0$ and $\h\tau_h\ge \tau_h^0.$ The remaining three permutations of the ordering of $\tau_w\le \tau_w^0$ and $\h\tau_h\ge \tau_h^0,$ or  $\tau_w\ge \tau_w^0$ and $\h\tau_h\le \tau_h^0,$ or  $\tau_w\le \tau_w^0$ and $\h\tau_h\le \tau_h^0$ shall follow symmetrical arguments. Consider,
		\benr\label{eq:1a}
		\cU_w(\tau_w,\h\tau_h,\h\theta)&=&\cL(\tau_w,\h\tau_h,\h\theta)-\cL(\tau_w^0,\h\tau_h,\h\theta)\nn\\
		&=&\sum_{w=\tau_w+1}^{T_w}\sum_{h=\h\tau_h+1}^{T_h}\|x_{(w,h)}-\h\theta_{(1)}\|_2^2+\sum_{w=1}^{\tau_w}\sum_{h=\h\tau_h+1}^{T_h}\|x_{(w,h)}-\h\theta_{(2)}\|_2^2\nn\\
		&&+\sum_{w=1}^{\tau_w}\sum_{h=1}^{\h\tau_h}\|x_{(w,h)}-\h\theta_{(3)}\|_2^2+\sum_{w=\tau_w+1}^{T_w}\sum_{h=1}^{\h\tau_h}\|x_{(w,h)}-\h\theta_{(4)}\|_2^2\nn\\
		&&-\sum_{w=\tau_w^0+1}^{T_w}\sum_{h=\h\tau_h+1}^{T_h}\|x_{(w,h)}-\h\theta_{(1)}\|_2^2-\sum_{w=1}^{\tau_w^0}\sum_{h=\h\tau_h+1}^{T}\|x_{(w,h)}-\h\theta_{(2)}\|_2^2\nn\\
		&&-\sum_{w=1}^{\tau_w^0}\sum_{h=1}^{\h\tau_h}\|x_{(w,h)}-\h\theta_{(3)}\|_2^2-\sum_{w=\tau_w^0+1}^{T}\sum_{h=1}^{\h\tau_h}\|x_{(w,h)}-\h\theta_{(4)}\|_2^2
		\eenr
		
		An algebraic manipulation of (\ref{eq:1a}) yields the following expression.
		\benr\label{eq:2}
		\cU_w(\tau_w,\h\tau_h,\h\theta)&=&(\tau_w-\tau_w^0)\Bigg[(T_h-\tau_h^0)\Big\{\|\h\eta_{(1)}\|_2^2+2(\h\theta_{(1)}-\theta_{(1)}^0)^T\h\eta_{(1)}\Big\}\nn\\
		&&+\tau_h^0\Big\{\|\h\eta_{(3)}\|_2^2+2(\h\theta_{(4)}-\theta_{(4)}^0)^T\h\eta_{(3)}\Big\}\Bigg]\nn\\
		&&+(\tau_w-\tau_w^0)(\h\tau_h-\tau_h^0)\Bigg[\|\h\eta_{(3)}\|_2^2-\|\h\eta_{(1)}\|_2^2\nn\\
		&&-2(\h\theta_{(1)}-\theta_{(1)}^0)^T\h\eta_{(1)}+2(\h\theta_{(4)}-\theta_{(1)}^0)^T\h\eta_{(3)}\Bigg]\nn\\
		&&-2\sum_{w=\tau_w^0+1}^{\tau_w}\sum_{h=\tau_h^0+1}^{T_h}\vep_{(w,h)}^T\h\eta_{(1)}-2\sum_{w=\tau_w^0+1}^{\tau_w}\sum_{h=1}^{\tau_h^0}\vep_{(w,h)}^T\h\eta_{(3)}\nn\\
		&&+2\sum_{w=\tau_w^0+1}^{\tau_w}\sum_{h=\tau_h^0+1}^{\h\tau_h}\vep_{(w,h)}^T\h\eta_{(1)}-2\sum_{w=\tau_w^0+1}^{\tau_w}\sum_{h=\tau_h^0+1}^{\h\tau_h}\vep_{(w,h)}^T\h\eta_{(3)}
		\eenr
		The calculations yielding (\ref{eq:2}) from the defining equality (\ref{eq:1a}) are fairly tedious and in order to maintain continuity of the main argument of this lemma, these algebraic manipulations are presented as Remark \ref{rem:lem1.algebra} after the proof of this lemma.  We can now proceed to the final step of the argument where we shall apply the bounds obtained in Step 1 to the expression (\ref{eq:2}) in order to obtain the desired uniform lower bound.
		
		\vspace{2mm}
		\noi{\bf Step 3:} It shall be important in this step to recall that all constants in Step 1 above represented by $c_{u1}$ arise from the Condition C, where it is assumed as chosen to be suitably small enough. Now recalling the construction of the set $\cG_w(u_{T_w},v_{T_w})$ from (\ref{def:setG}) and applying the bounds (\ref{eq:5}), (\ref{eq:8}), (\ref{eq:9}), (\ref{eq:10}) and (\ref{eq:11}) to the expression (\ref{eq:2}) we obtain,
		
		\benr\label{eq:12}
		\inf_{\substack{\tau_w\in\cG_w(u_{T_w},v_{T_w});\\ \tau_w\ge\tau_w^0}}\cU_w(\tau_w,\h\tau_h,\h\theta)&\ge& T_wT_hv_{T_w}\Big[\om_h\xi_1^2+(1-\om_h)\xi_3^2-c_{u1}\xi_{\min}\overline\xi-c_{u1}\xi_{\min}^2\Big]\nn\\
		&&-c_u\xi_{\min}\si\log(p\vee T_wT_h)\surd(T_wT_hu_{T_w})\surd{s}\nn\\
		&\ge& \frac{T_wT_h\xi_{w}^2}{2}\Big[v_{T_w}-c_u\log(p\vee T_wT_h)\frac{\si}{\xi_w}\Big(\frac{su_{T_w}}{T_wT_h}\Big)^{\frac{1}{2}}\Big]
		\eenr
		with probability at least $1-2\exp\{c_1\log(p\vee T_wT_h)\}-\pi_T.$ To obtain the second inequality  we have used that by definition $\om_h\xi_1^2+(1-\om_h)\xi_3^2=\xi_w^2,$ and $\xi_{\min}\le \xi_{w},$ additionally from Condition B(ii) we have $\overline\xi\le c_u\xi_{\min}$ and that the constant $c_{u1}$ arises from Condition C where it is chosen to be suitable small enough. Repeating symmetrical arguments for the mirroring permutations of the ordering of $\tau_w,\tau_h$ with respect to $\tau_w^0,\h\tau_h$ shall yield the same uniform lower bound (\ref{eq:12}). This completes the proof of this lemma.
	\end{proof}
	
	\bc$\rule{3.5in}{0.1mm}$\ec

	\begin{remark}\label{rem:lem1.algebra} This remark provides the decomposition (\ref{eq:2}) of (\ref{eq:1a}) described in {\bf Step 2} of the proof of Lemma \ref{lem:unif.lb}. A note of interest here is that the calculations below become much more intuitive when viewed w.r.t. to a $2d$-visualization such as that illustrated in (\ref{model:rvmcp}). Let,
		\benr
		\h\vep_{(w,h)}=\begin{cases} x_{(w,h)}-\h\theta_{(1)} & \tau_w^0< w\le \tau_w,\,\, \h\tau_h<h\le T\\
			x_{(w,h)}-\h\theta_{(4)} & \tau_w^0< w\le \tau_w,\,\, 1\le h<\h\tau_h\nn
		\end{cases}
		\eenr
		Then picking up from the expression (\ref{eq:1a}), a simplification now yields,
		\benr\label{eq:16}
		\cU(\tau_w,\h\tau_h,\h\theta)&=&-\sum_{w=\tau_w^0+1}^{\tau_w}\sum_{h=\h\tau_h+1}^{T}\|x_{(w,h)}-\h\theta_{(1)}\|_2^2
		+\sum_{w=\tau_w^0+1}^{\tau_h}\sum_{\h\tau_h+1}^{T}\|x_{(w,h)}-\h\theta_{(2)}\|_2^2\nn\\
		&&+\sum_{w=\tau_w^0+1}^{\tau_w}\sum_{h=1}^{\h\tau_h}\|x_{(w,h)}-\h\theta_{(3)}\|_2^2
		-\sum_{w=\tau_w^0+1}^{\tau_w}\sum_{h=1}^{\h\tau_h}\|x_{(w,h)}-\h\theta_{(4)}\|_2^2\nn\\
		&=&\sum_{w=\tau_w^0+1}^{\tau_w}\sum_{h=\h\tau_h+1}^{T}\|\h\theta_{(2)}-\h\theta_{(1)}\|_2^2-2\sum_{w=\tau_w^0+1}^{\tau_w}\sum_{h=\h\tau_h+1}^{T}\h\vep_{(w,h)}^T(\h\theta_{(2)}-\h\theta_{(1)})\nn\\
		&&+\sum_{w=\tau_w^0+1}^{\tau_w}\sum_{h=1}^{\h\tau_h}\|\h\theta_{(3)}-\h\theta_{(4)}\|_2^2-2\sum_{w=\tau_w^0+1}^{\tau_w}\sum_{h=1}^{\h\tau_h}\h\vep_{(w,h)}^T(\h\theta_{(3)}-\h\theta_{(4)})\nn\\
		&=&R1-R2+R3-R4.
		\eenr
		Next we consider these remainder terms $R1,$ and $R3,$ where we have,
		\benr\label{eq:13}
		R1+R3&=&\sum_{w=\tau_w^0+1}^{\tau_w}\sum_{h=\h\tau_h+1}^{T}\|\h\theta_{(2)}-\h\theta_{(1)}\|_2^2+\sum_{w=\tau_w^0+1}^{\tau_w}\sum_{h=1}^{\h\tau_h}\|\h\theta_{(3)}-\h\theta_{(4)}\|_2^2\nn\\
		&=&\sum_{w=\tau_w^0+1}^{\tau_w}\sum_{h=\tau_h^0+1}^{T}\|\h\theta_{(2)}-\h\theta_{(1)}\|_2^2+\sum_{w=\tau_w^0+1}^{\tau_w}\sum_{h=1}^{\tau_h^0}\|\h\theta_{(3)}-\h\theta_{(4)}\|_2^2\nn\\
		&&-\sum_{w=\tau_w^0+1}^{\tau_w}\sum_{h=\tau_h^0+1}^{\h\tau_h}\|\h\theta_{(2)}-\h\theta_{(1)}\|_2^2 +\sum_{w=\tau_w^0+1}^{\tau_w}\sum_{h=\tau_h^0+1}^{\h\tau_h}\|\h\theta_{(3)}-\h\theta_{(4)}\|_2^2\nn\\
		&=& (\tau_w-\tau_w^0)\Big[(T-\tau_h^0)\|\h\theta_{(2)}-\h\theta_{(1)}\|_2^2+\tau_h^0\|\h\theta_{(3)}-\h\theta_{(4)}\|_2^2\Big]\nn\\
		&&+(\tau_w-\tau_w^0)(\h\tau_h-\h\tau_h^0)\Big[\|\h\theta_{(3)}-\h\theta_{(4)}\|_2^2-\|\h\theta_{(2)}-\h\theta_{(1)}\|_2^2\Big]
		\eenr
		
		In order to simplify the terms $R2$ and $R4,$ the double sums under consideration are split at the underlying change point parameters $\tau_w^0,\tau_h^0,$ leading to the following decompositions.
		\benr\label{eq:14}
		R2&=&2\sum_{w=\tau_w^0+1}^{\tau_w}\sum_{h=\h\tau_h+1}^{T}\h\vep_{(w,h)}^T(\h\theta_{(2)}-\h\theta_{(1)})\nn\\
		&=&2\sum_{w=\tau_w^0+1}^{\tau_w}\sum_{h=\tau_h^0+1}^{T}\h\vep_{(w,h)}^T(\h\theta_{(2)}-\h\theta_{(1)})-2\sum_{w=\tau_w^0+1}^{\tau_w}\sum_{h=\tau_h^0+1}^{\h\tau_h}\h\vep_{(w,h)}^T(\h\theta_{(2)}-\h\theta_{(1)})\nn\\
		&=&2\sum_{w=\tau_w^0+1}^{\tau_w}\sum_{h=\tau_h^0+1}^{T}\vep_{(w,h)}^T(\h\theta_{(2)}-\h\theta_{(1)})-2\sum_{w=\tau_w^0+1}^{\tau_w}\sum_{h=\tau_h^0+1}^{T}(\h\theta_{(1)}-\theta_{(1)}^0)(\h\theta_{(2)}-\h\theta_{(1)})\nn\\
		&&-2\sum_{w=\tau_w^0+1}^{\tau_w}\sum_{h=\tau_h^0+1}^{\h\tau_h}\vep_{(w,h)}^T(\h\theta_{(2)}-\h\theta_{(1)})+2\sum_{w=\tau_w^0+1}^{\tau_w}\sum_{h=\tau_h^0+1}^{\h\tau_h}(\h\theta_{(1)}-\theta_{(1)}^0)(\h\theta_{(2)}-\h\theta_{(1)})
		\eenr
		Similarly, we have,
		\benr\label{eq:15}
		R4&=&2\sum_{w=\tau_w^0+1}^{\tau_w}\sum_{h=1}^{\h\tau_h}\h\vep_{(w,h)}^T(\h\theta_{(3)}-\h\theta_{(4)})\nn\\
		&=&2\sum_{w=\tau_w^0+1}^{\tau_w}\sum_{h=1}^{\tau_h^0}\h\vep_{(w,h)}^T(\h\theta_{(3)}-\h\theta_{(4)})+2\sum_{w=\tau_w^0+1}^{\tau_w}\sum_{h=\tau_h^0+1}^{\h\tau_h}\h\vep_{(w,h)}^T(\h\theta_{(3)}-\h\theta_{(4)})\nn\\
		&=&2\sum_{w=\tau_w^0+1}^{\tau_w}\sum_{h=1}^{\tau_h^0}\vep_{(w,h)}^T(\h\theta_{(3)}-\h\theta_{(4)})-2\sum_{w=\tau_w^0+1}^{\tau_w}\sum_{h=1}^{\tau_h^0}(\theta_{(4)}-\h\theta_{(4)})(\h\theta_{(3)}-\h\theta_{(4)})\nn\\
		&& +2\sum_{w=\tau_w^0+1}^{\tau_w}\sum_{h=\tau_h^0+1}^{\h\tau_h}\vep_{(w,h)}^T(\h\theta_{(3)}-\h\theta_{(4)})-2\sum_{w=\tau_w^0+1}^{\tau_w}\sum_{h=\tau_h^0+1}^{\h\tau_h}(\h\theta_{(4)}-\theta_{(1)}^0)(\h\theta_{(3)}-\h\theta_{(4)})
		\eenr
		
		Substituting (\ref{eq:13}), (\ref{eq:14}) and (\ref{eq:15}) in (\ref{eq:16}) we obtain,
		\benr\label{eq:17}
		\cU(\tau_w,\h\tau_h,\h\theta)&=&R1-R2+R3-R4\nn\\
		&=&(\tau_w-\tau_w^0)\Bigg[(T-\tau_h^0)\Big\{\|\h\theta_{(2)}-\h\theta_{(1)}\|_2^2+2(\h\theta_{(1)}-\theta_{(1)}^0)^T(\h\theta_{(2)}-\h\theta_{(1)})\Big\}\nn\\
		&&+\tau_h^0\Big\{\|\h\theta_{(3)}-\h\theta_{(4)}\|_2^2+2(\h\theta_{(4)}-\theta_{(4)}^0)^T(\h\theta_{(3)}-\h\theta_{(4)})\Big\}\Bigg]\nn\\
		&&+(\tau_w-\tau_w^0)(\h\tau_h-\h\tau_h^0)\Bigg[\|\h\theta_{(3)}-\h\theta_{(4)}\|_2^2-\|\h\theta_{(2)}-\h\theta_{(1)}\|_2^2\nn\\
		&&-2(\h\theta_{(1)}-\theta_{(1)}^0)^T(\h\theta_{(2)}-\h\theta_{(1)})+2(\h\theta_{(4)}-\theta_{(1)}^0)^T(\h\theta_{(3)}-\h\theta_{(4)})\Bigg]\nn\\
		&&-2\sum_{w=\tau_w^0+1}^{\tau_w}\sum_{h=\tau_h^0+1}^T\vep_{(w,h)}^T(\h\theta_{(2)}-\h\theta_{(1)})-2\sum_{w=\tau_w^0+1}^{\tau_w}\sum_{h=1}^{\tau_h^0}\vep_{(w,h)}^T(\h\theta_{(3)}-\h\theta_{(4)})\nn\\
		&&+2\sum_{w=\tau_w^0+1}^{\tau_w}\sum_{h=\tau_h^0+1}^{\h\tau_h}\vep_{(w,h)}^T(\h\theta_{(2)}-\h\theta_{(1)})-2\sum_{w=\tau_w^0+1}^{\tau_w}\sum_{h=\tau_h^0+1}^{\h\tau_h}\vep_{(w,h)}^T(\h\theta_{(3)}-\h\theta_{(4)})
		\eenr
		The expression (\ref{eq:2}) now follows from (\ref{eq:17}) with notational changes. This completes this algebraic part of the proof of Lemma \ref{lem:unif.lb}.
	\end{remark}

	\bc$\rule{3.5in}{0.1mm}$\ec

	\begin{lemma}\label{lem:unif.lb.height} Suppose the model (\ref{model:rvmcp}) and assume that Conditions A, B, C(i)(a) and C(ii)(a,b)  hold. Let $0\le v_{T_h}\le u_{T_h}\le 1,$ be any non-negative sequences. Then,
		\benr\label{eq:3i}
		\inf_{\tau_h\in\cG_h(u_{T_h},v_{T_h})}\cU_h(\h\tau_w,\tau_h,\h\theta)\ge \frac{T_wT_h\xi^2_h}{2}\Big[v_{T_h}-c_u\log(p\vee T_wT_h)\frac{\si}{\xi_h}\Big\{\frac{u_{T_h}s}{T_wT_h}\Big\}^{\frac{1}{2}}\Big]
		\eenr
		with probability at least $1-2\exp\{-c_1\log(p\vee T)\}-\pi_T,$ for some constant $c_1>0$ that does not depend on any model parameters.
	\end{lemma}

	\begin{proof}[{\bf Proof of Lemma \ref{lem:unif.lb.height}}] The proof of this result is analogous to  Lemma \ref{lem:unif.lb} above, and is thus omitted. All assumptions are identical to Lemma \ref{lem:unif.lb}, the only distinction being the desired uniform bound is over the collection $\cG_h$ of the change parameter in the vertical direction, instead of $\cG_w.$
	\end{proof}

	
	\bc$\rule{3.5in}{0.1mm}$\ec

	\begin{proof}[{\bf Proof of Theorem \ref{thm:cp.nearoptimal}}] We begin with Part (i) of this result.
		The proof shall rely on a recursive argument using Lemma \ref{lem:unif.lb}, where the desired rate of convergence is obtained by a series of recursions, with this rate being sharpened at each step.
		
		Consider any $v_{T_w}>0$ and apply Lemma \ref{lem:unif.lb} on the set $\cG_w(1,v_{T_w})$ to obtain,
		\benrr
		\inf_{\tau_w\in\cG_w(1,v_{T_w})}\cU_w(\tau_w,\h\tau_h,\h\theta)\ge \frac{T_wT_h\xi^2_w}{2}\Big[v_{T_w}-c_u\log(p\vee T_wT_h)\frac{\si}{\xi_w}\Big\{\frac{s}{T_wT_h}\Big\}^{\frac{1}{2}}\Big]
		\eenrr
		with probability at least $1-2\exp\{-c_{1}\log (p\vee T_wT_h)\}-\pi_T.$ Now
		upon choosing any,
		\benr
		v_{T_w}>v_{T_w}^*=c_u\log (p\vee T_wT_h)\frac{\si}{\xi_w}\Big\{\frac{s}{T_wT_h}\Big\}^{\frac{1}{2}},\nn
		\eenr
		we obtain $\inf_{\tau_w\in\cG(1,v_{T_w})}\cU_w(\tau_w,\h\tau_h,\h\theta)>0,$ thus implying that $\tilde\tau_w\notin\cG(1,v_{T_w}^*),$ i.e., $|\tilde\tau_w-\tau^0_w|\le  T_wv_{T_w}^*,$ with probability at least  $1-2\exp\{-c_{1}\log (p\vee T_wT_h)\}-\pi_T.$ \footnote{Since by construction of $\tilde\tau_w,$ we have $\cU_w(\tilde\tau_w,\h\tau_h,\h\theta)\le 0.$} Now reset $u_{T_w}=v_{T_w}^*$ and reapply Lemma \ref{lem:unif.lb} for any $v_{T_w}>0$ to obtain,
		\benrr
		\inf_{\tau_w\in\cG_w(u_{T_w},v_{T_w})}\cU_w(\tau_w,\h\tau_h,\h\theta)\ge \frac{T_wT_h\xi^2_w}{2}\Big[v_{T_w}-\Big\{c_u\log(p\vee T_wT_h)\frac{\si}{\xi_w}\Big\}^{1+\frac{1}{2}}\Big\{\frac{s}{T_wT_h}\Big\}^{\frac{1}{2}+\frac{1}{4}}\Big],\nn
		\eenrr
		with probability at least $1-2\exp\{-c_{1}\log (p\vee T_wT_h)\}-\pi_T.$. Again choosing any,
		\benr
		v_T>v_T^*= \Big\{c_u\log(p\vee T_wT_h)\frac{\si}{\xi_w}\Big\}^{1+\frac{1}{2}}\Big\{\frac{s}{T_wT_h}\Big\}^{\frac{1}{2}+\frac{1}{4}},\nn
		\eenr
		we obtain $\inf_{\tau_w\in\cG(u_{T_w},v_{T_w})}\cU_w(\tau_w,\h\tau_h,\h\theta)>0,$ thus yielding $\tilde\tau_w\notin\cG(u_{T_w},v_{T_w}^*),$ i.e.,
		\benr
		|\tilde\tau_w-\tau^0_w|\le T_w\Big\{c_u\log(p\vee T_wT_h)\frac{\si}{\xi_w}\Big\}^{a_2}\Big\{\frac{s}{T_wT_h}\Big\}^{b_2},
		\eenr
		with probability at least $1-2\exp\{-c_{1}\log (p\vee T_wT_h)\}-\pi_T.$ Where,
		\benr
		a_2=1+\frac{1}{2}=\sum_{j=0}^1\frac{1}{2^j},\quad{\rm and}\quad b_2=\frac{1}{2}+\frac{1}{4}=\sum_{j=1}^2\frac{1}{2^j}.\nn
		\eenr
		Note that the rate of convergence of $\tilde\tau_w$ has been sharpened at the second recursion in comparison to the first. Continuing these recursions by resetting $u_T$ to the bound of the previous recursion, and applying Lemma \ref{lem:unif.lb}, we obtain for the $m^{th}$ recursion,
		\benr
		|\tilde\tau_w-\tau^0_w|\le T_w\Big\{c_u\log(p\vee T_wT_h)\frac{\si}{\xi_w}\Big\}^{a_m}\Big\{\frac{s}{T_wT_h}\Big\}^{b_m},
		\eenr
		with probability at least $1-2\exp\{-c_{1}\log (p\vee T_wT_h)\}-\pi_T.$ Repeating these recursions an infinite number of times and noting that $a_\iny=\sum_{j=0}^{\iny}(1/2^j)=2,$ and $b_{\iny}=\sum_{j=1}^{\iny}(1/2^j)=1$ we obtain,
		\benr
		|\tilde\tau_w-\tau^0_w|\le T_w\Big\{c_u\log(p\vee T_wT_h)\frac{\si}{\xi_w}\Big\}^{2}\Big\{\frac{s}{T_wT_h}\Big\}\nn
		\eenr
		with probability at least $1-2\exp\{-c_{1}\log (p\vee T_wT_h)\}-\pi_T.$ Finally, note that despite the recursions in the above argument, the probability of the bound after every recursion is maintained to be at least $1-2\exp\{-c_{1}\log (p\vee T_wT_h)\}-\pi_T.$ This follows since the probability statement of Lemma \ref{lem:unif.lb} arises from stochastic upper bounds of Lemma \ref{lem:nearoptimalcross.subE.special} applied recursively with a tighter bound at each recursion. This yields a sequence of events such that the event at each recursion is a proper subset of the event at the previous recursion. This completes the proof of Part (i) of this theorem. The proof of Part (ii) follows an analogous recursive argument applied on Lemma \ref{lem:unif.lb.height} instead of Lemma \ref{lem:unif.lb}.
	\end{proof}
	
	\bc$\rule{4in}{0.1mm}$\ec

	\begin{lemma}\label{lem:unif.lb.optimal} Suppose the model (\ref{model:rvmcp}) and assume that conditions A, B, C(i)(b) and C(ii)(a,c)  holds. Let $0\le v_{T_w}\le u_{T_w}\le 1,$ be any non-negative sequences. Then, for any $0<a<1,$ choosing $c_{a}\ge \surd{(1/a)},$ we have the following uniform lower bound.
		\benr\label{eq:3}
		\inf_{\tau_w\in\cG_w(u_{T_w},v_{T_w})}\cU_w(\tau_w,\h\tau_h,\h\theta)\ge \frac{T_wT_h\xi^2_w}{2}\Big[v_{T_w}-c_uc_a\frac{\si}{\xi_w}\Big\{\frac{u_{T_w}}{T_wT_h}\Big\}^{\frac{1}{2}}\Big]
		\eenr
		with probability at least $1-a-o(1)-\pi_T.$
	\end{lemma}

	\begin{proof}[{\bf Proof of Lemma \ref{lem:unif.lb.optimal}}] The broader structure of this proof is similar to that of Lemma \ref{lem:unif.lb}. The distinction being that the availability of a sharper assumption of Condition C(ii)(c) on the preliminary mean estimates together with a more delicate analysis of the bounds in Step 1 of the proof of Lemma \ref{lem:unif.lb} shall yields this result.
		
		Recall the inequalities of {\bf Step 1} of Lemma \ref{lem:unif.lb} and note that analogous to (\ref{eq:4}), one may obtain that under Condition C(ii)(c) that,
		\benr\label{eq:25}
		\|\h\eta_{(1)}-\eta^0_{(1)}\|_1\le \frac{8c_{u1}\xi_{\min}}{\log(p\vee T_wT_h)},\quad{\rm and}\quad
		\|\h\eta_{(3)}-\eta^0_{(3)}\|_1\le \frac{8c_{u1}\xi_{\min}}{\log(p\vee T_wT_h)},
		\eenr
		with probability at least $1-\pi_T.$ Moreover, also observe here that since Condition C(ii)(c) assumed here is sharper than C(ii)(b) assumed in Lemma \ref{lem:unif.lb} consequently, the bounds  (\ref{eq:5}), (\ref{eq:7}), (\ref{eq:8}) and (\ref{eq:9}) remain valid here as well, with the same probability.
		
		Next we examine the stochastic terms considered in (\ref{eq:10}) and (\ref{eq:11}) more closely. Applying Lemma \ref{lem:optimalcross}, for any $0<a<1,$ choosing $c_a\ge 4\surd(1/a),$ we obtain,
		\benr\label{eq:24}
		\sup_{\substack{\tau_w\in\cG_w(u_{T_w},v_{T_w});\\ \tau_w\ge\tau^0_w}}\Big|\sum_{w=\tau^0_w+1}^{\tau_w}\Big(\sum_{h=\tau^0_h+1}^{T_h}\vep_{(w,h)}^T\eta^0_{(1)}+\sum_{h=1}^{\tau^0_h}\vep_{w,h}^T\eta^0_{(3)}\Big)\Big|\le
		c_{a}\phi\xi_w\surd(T_wT_hu_{T_w}),
		\eenr
		with probability at least $1-a.$ Additionally, we also have that,
		\benr\label{eq:29}
		\sup_{\substack{\tau_w\in\cG_w(u_{T_w},v_{T_w});\\ \tau_w\ge\tau^0_w}}\Big|\sum_{w=\tau^0_w+1}^{\tau_w}\sum_{h=\tau^0_h+1}^{T_h}\vep_{(w,h)}^T(\h\eta_{(1)}-\eta^0_{(1)})\Big|\hspace{3cm}\nn\\
		\le \sup_{\substack{\tau_w\in\cG_w(u_{T_w},v_{T_w});\\ \tau_w\ge\tau^0_w}}\Big\|\sum_{w=\tau^0_w+1}^{\tau_w}\sum_{h=\tau^0_h+1}^{T_h}\vep_{(w,h)}\Big\|_{\iny}\big\|\h\eta_{(1)}-\eta^0_{(1)}\big\|_1\nn\\
		\le c_uc_{u1}\xi_{\min}\si\surd{(T_wT_hu_{T_w})}\hspace{1.85in},
		\eenr	
		with probability at least $1-o(1)-\pi_T,$ here we have utilized Lemma \ref{lem:nearoptimalcross.subE.special} as well as (\ref{eq:25}) to obtain the final inequality. Similarly, we can also obtain the following bounds,
		\benr\label{eq:26}
		&&\sup_{\substack{\tau_w\in\cG_w(u_{T_w},v_{T_w});\\ \tau_w\ge\tau^0_w}}\Big|\sum_{w=\tau^0_w+1}^{\tau_w}\sum_{h=1}^{\tau^0_h}\vep_{(w,h)}^T(\h\eta_{(3)}-\eta^0_{(3)})\Big|
		\le c_uc_{u1}\xi_{\min}\si\surd{(T_wT_hu_{T_w})},\nn\\
		&&\sup_{\substack{\tau_w\in\cG_w(u_{T_w},v_{T_w});\\ \tau_w\ge\tau^0_w}}\Big|\sum_{w=\tau^0_w+1}^{\tau_w}\sum_{h=\tau_h^0+1}^{\h\tau_h}\vep_{(w,h)}^T\h\eta_{(1)}\Big| \le  c_uc_{u1}\xi_{\min}\si\surd{(T_wT_hu_{T_w})},\nn\\
		&&\sup_{\substack{\tau_w\in\cG_w(u_{T_w},v_{T_w});\\ \tau_w\ge\tau^0_w}}\Big|\sum_{w=\tau^0_w+1}^{\tau_w}\sum_{h=\tau_h^0+1}^{\h\tau_h}\vep_{(w,h)}^T\h\eta_{(3)}\Big| \le  c_uc_{u1}\xi_{\min}\si\surd{(T_wT_hu_{T_w})},
		\eenr	
		with probability at least $1-o(1)-\pi_T.$ In order to obtain the second and third inequalities of (\ref{eq:26}), we have utilized the $\ell_1$ bounds of  (\ref{eq:5}), (\ref{eq:8}). Additionally, towards these bounds we have also utilized Condition C(i)(b) together with Lemma \ref{lem:nearoptimalcross.subE.special}, in particular from Condition C(i)(b) we have that $|\h\tau_h-\tau_h^0|\le T_hu_{T_h},$ where $u_{T_h}=c_{u1}\big/\{s\log^2(p\vee T_wT_h)\},$ with probability $1-\pi_T,$ Lemma \ref{lem:nearoptimalcross.subE.special} is then applied with this choice of $u_{T_h}.$
		
		Next we consider {\bf Step 2} as described in the proof of Lemma \ref{lem:unif.lb}. Consider the decomposition (\ref{eq:2}) and note that it can be further manipulated as the following,
		\benr\label{eq:27}
		\cU_w(\tau_w,\h\tau_h,\h\theta)&=&(\tau_w-\tau_w^0)\Bigg[(T_h-\tau_h^0)\Big\{\|\h\eta_{(1)}\|_2^2+2(\h\theta_{(1)}-\theta_{(1)}^0)^T\h\eta_{(1)}\Big\}\nn\\
		&&+\tau_h^0\Big\{\|\h\eta_{(3)}\|_2^2+2(\h\theta_{(4)}-\theta_{(4)}^0)^T\h\eta_{(3)}\Big\}\Bigg]\nn\\
		&&+(\tau_w-\tau_w^0)(\h\tau_h-\tau_h^0)\Bigg[\|\h\eta_{(3)}\|_2^2-\|\h\eta_{(1)}\|_2^2\nn\\
		&&-2(\h\theta_{(1)}-\theta_{(1)}^0)^T\h\eta_{(1)}+2(\h\theta_{(4)}-\theta_{(1)}^0)^T\h\eta_{(3)}\Bigg]\nn\\
		&&-2\sum_{w=\tau^0_w+1}^{\tau_w}\Big(\sum_{h=\tau^0_h+1}^{T_h}\vep_{(w,h)}^T\eta^0_{(1)}+\sum_{h=1}^{\tau^0_h}\vep_{w,h}^T\eta^0_{(3)}\Big)\nn\\
		&&-2\sum_{w=\tau_w^0+1}^{\tau_w}\sum_{h=\tau_h^0+1}^{T_h}\vep_{(w,h)}^T(\h\eta_{(1)}-\eta_{(1)}^0)\nn\\
		&&-2\sum_{w=\tau_w^0+1}^{\tau_w}\sum_{h=1}^{\tau_h^0}\vep_{(w,h)}^T(\h\eta_{(3)}-\eta^0_{(3)})\nn\\
		&&+2\sum_{w=\tau_w^0+1}^{\tau_w}\sum_{h=\tau_h^0+1}^{\h\tau_h}\vep_{(w,h)}^T\h\eta_{(1)}-2\sum_{w=\tau_w^0+1}^{\tau_w}\sum_{h=\tau_h^0+1}^{\h\tau_h}\vep_{(w,h)}^T\h\eta_{(3)}
		\eenr
		
		We are now ready to proceed to {\bf Step 3} to obtain the desired lower bound. Utilizing the $\ell_2$ lower bounds of  (\ref{eq:7}) and (\ref{eq:8}), and the upper bounds of (\ref{eq:9}) (\ref{eq:24}), (\ref{eq:29}) and (\ref{eq:26}) to the expression (\ref{eq:27}) yields,
		\benr
		\inf_{\substack{\tau_w\in\cG_w(u_{T_w},v_{T_w});\\ \tau_w\ge\tau_w^0}}\cU_w(\tau_w,\h\tau_h,\h\theta)&\ge& T_wT_hv_{T_w}\Big[\om_h\xi_1^2+(1-\om_h)\xi_3^2-c_{u1}\xi_{\min}\overline\xi-c_{u1}\xi_{\min}^2\Big]\nn\\
		&&-c_uc_a\xi_{\min}\si\surd(T_wT_hu_{T_w})\nn\\
		&\ge& \frac{T_wT_h\xi_{w}^2}{2}\Big[v_{T_w}-c_uc_a\frac{\si}{\xi_w}\Big(\frac{u_{T_w}}{T_wT_h}\Big)^{\frac{1}{2}}\Big]
		\eenr
		with probability at least $1-a-o(1)-\pi_T.$ To obtain the second inequality  we have used that by definition $\om_h\xi_1^2+(1-\om_h)\xi_3^2=\xi_w^2,$ and $\xi_{\min}\le \xi_{w},$ additionally from Condition B(ii) we have $\overline\xi\le c_u\xi_{\min}$ and that the constant $c_{u1}$ arises from Condition C where it is chosen to be suitable small enough. Repeating symmetrical arguments for the mirroring permutations of the ordering of $\tau_w,\tau_h$ with respect to $\tau_w^0,\h\tau_h$ shall yield the same uniform lower bound (\ref{eq:12}). This completes the proof of this lemma.
	\end{proof}

	\bc$\rule{3.5in}{0.1mm}$\ec
	
	\begin{lemma}\label{lem:unif.lb.optimal.height} Suppose the model (\ref{model:rvmcp}) and assume that conditions A, B, C(i)(b) and C(ii)(a,c)  holds. Let $0\le v_{T_h}\le u_{T_h}\le 1,$ be any non-negative sequences. Then, for any $0<a<1,$ choosing $c_{a}\ge \surd{(1/a)},$ we have the following uniform lower bound.
		\benr
		\inf_{\tau_h\in\cG_h(u_{T_h},v_{T_h})}\cU_h(\h\tau_w,\tau_h,\h\theta)\ge \frac{T_wT_h\xi^2_h}{2}\Big[v_{T_h}-c_a\frac{\si}{\xi_h}\Big\{\frac{u_{T_h}}{T_wT_h}\Big\}^{\frac{1}{2}}\Big]\nn
		\eenr
		with probability at least $1-a-o(1)-\pi_T.$
	\end{lemma}
	
	\begin{proof}[{\bf Proof of Lemma \ref{lem:unif.lb.optimal.height}}] The proof of this result is analogous to  Lemma \ref{lem:unif.lb.optimal} above, and is thus omitted. All assumptions are identical to Lemma \ref{lem:unif.lb.optimal}, the only distinction being the desired uniform bound is over the collection $\cG_h$ of the change parameter in the vertical direction, instead of $\cG_w.$
	\end{proof}

	\bc$\rule{3.5in}{0.1mm}$\ec
	
	\begin{proof}[{\bf Proof of Theorem \ref{thm:cpoptimal}}] The proof of this result follows a recursive argument similar to that of Theorem \ref{thm:cp.nearoptimal}, the distinction being that these recursions are made on Lemma \ref{lem:unif.lb.optimal} instead of Lemma \ref{lem:unif.lb}. We begin by considering any $v_{T_w}>0$ and apply Lemma \ref{lem:unif.lb} on the set $\cG_w(1,v_{T_w})$ to obtain,
		\benrr
		\inf_{\tau_w\in\cG_w(1,v_{T_w})}\cU_w(\tau_w,\h\tau_h,\h\theta)\ge \frac{T_wT_h\xi^2_w}{2}\Big[v_{T_w}-c_uc_a\frac{\si}{\xi_w}\Big\{\frac{1}{T_wT_h}\Big\}^{\frac{1}{2}}\Big]
		\eenrr
		with probability at least $1-a-o(1)-\pi_T.$ Upon choosing any,
		\benr
		v_{T_w}>v_{T_w}^*=c_uc_a\frac{\si}{\xi_w}\Big\{\frac{1}{T_wT_h}\Big\}^{\frac{1}{2}},\nn
		\eenr
		we obtain $\inf_{\tau_w\in\cG(1,v_{T_w})}\cU_w(\tau_w,\h\tau_h,\h\theta)>0,$ thus implying that $\tilde\tau_w\notin\cG(1,v_{T_w}^*),$  with probability at least  $1-a-o(1)-\pi_T.$ Now reset $u_{T_w}=v_{T_w}^*$ and reapply Lemma \ref{lem:unif.lb} for any $v_{T_w}>0$ to obtain,
		\benrr
		\inf_{\tau_w\in\cG_w(u_{T_w},v_{T_w})}\cU_w(\tau_w,\h\tau_h,\h\theta)\ge \frac{T_wT_h\xi^2_w}{2}\Big[v_{T_w}-\Big\{c_uc_a\frac{\si}{\xi_w}\Big\}^{1+\frac{1}{2}}\Big\{\frac{1}{T_wT_h}\Big\}^{\frac{1}{2}+\frac{1}{4}}\Big],\nn
		\eenrr
		with probability at least $1-a-o(1)-\pi_T.$. Again choosing,
		\benr
		v_{T_w}>v_{T_w}^*= \Big\{c_uc_a\frac{\si}{\xi_w}\Big\}^{1+\frac{1}{2}}\Big\{\frac{1}{T_wT_h}\Big\}^{\frac{1}{2}+\frac{1}{4}},\nn
		\eenr
		we obtain $\inf_{\tau_w\in\cG(u_{T_w},v_{T_w})}\cU_w(\tau_w,\h\tau_h,\h\theta)>0,$ thus yielding,
		\benr
		|\tilde\tau_w-\tau^0_w|\le T_w\Big\{c_uc_a\frac{\si}{\xi_w}\Big\}^{a_2}\Big\{\frac{1}{T_wT_h}\Big\}^{b_2},
		\eenr
		with probability at least $1-a-o(1)-\pi_T.$ Where,
		\benr
		a_2=1+\frac{1}{2}=\sum_{j=0}^1\frac{1}{2^j},\quad{\rm and}\quad b_2=\frac{1}{2}+\frac{1}{4}=\sum_{j=1}^2\frac{1}{2^j}.\nn
		\eenr
		Continuing these recursions by resetting $u_{T_w}$ to the bound of the previous recursion, and applying Lemma \ref{lem:unif.lb}, we obtain for the $m^{th}$ recursion,
		\benr
		|\tilde\tau_w-\tau^0_w|\le T_w\Big\{c_uc_a\frac{\si}{\xi_w}\Big\}^{a_m}\Big\{\frac{1}{T_wT_h}\Big\}^{b_m},
		\eenr
		with probability at least $1-a-o(1)-\pi_T.$ Repeating these recursions an infinite number of times and noting that $a_\iny=\sum_{j=0}^{\iny}(1/2^j)=2,$ and $b_{\iny}=\sum_{j=1}^{\iny}(1/2^j)=1$ we obtain,
		\benr
		|\tilde\tau_w-\tau^0_w|\le T_w\Big\{c_uc_a\frac{\si}{\xi_w}\Big\}^{2}\Big\{\frac{1}{T_wT_h}\Big\}\nn
		\eenr
		with probability at least $1-a-o(1)-\pi_T.$ As earlier for Theorem \ref{thm:cp.nearoptimal}, despite the recursions in the above argument, the probability of the bound after every recursion is maintained to be at least $1-a-o(1)-\pi_T$ since each recursion holds on an event that is a proper subset of the event at the previous recursion. This completes the proof of Part (i) of this theorem. The proof of Part (ii) follows an analogous recursive argument applied on Lemma \ref{lem:unif.lb.optimal.height} instead of Lemma \ref{lem:unif.lb.optimal}.
	\end{proof}

	\bc$\rule{3.5in}{0.1mm}$\ec

	As the reader may have observed, a change of notation has been carried out for the results of Theorem \ref{thm:wc.vanishing} and Theorem \ref{thm:wc.non.vanishing}. These results are presented in more conventional {\it argmax} notation instead of the {\it argmin} notation of the problem setup in Section \ref{sec:intro}. This is purely a notational change and all results can equivalently be stated in the {\it argmin} language. Accordingly we define the following versions. Let $\cU_w(\tau_w,\tau_h,\theta)$ and $\cU_h(\tau_w,\tau_h,\theta)$ be as in (\ref{def:cU}) and consider,
	\benr\label{def:cC}
	\cC_w(\tau_w,\tau_h,\theta)=-\cU_w(\tau_w,\tau_h,\theta)\quad{\rm and,}\quad\cC_h(\tau_w,\tau_h,\theta)=-\cU_h(\tau_w,\tau_h,\theta)
	\eenr
	Then, we can re-express the change point estimators $\tilde\tau_w(\tau_h,\theta)$ and $\tilde\tau_h(\tau_w,\theta)$ as,
	\benr
	\tilde\tau_w(\tau_h,\theta)=\argmax_{1\le \tau_w< T_w}\cC_w(\tau_w,\tau_h,\theta),\quad{\rm and}\quad \tilde\tau_h(\tau_w,\theta)=\argmax_{1\le \tau_h< T_h}\cC_h(\tau_w,\tau_h,\theta)\nn
	\eenr
	
	The proofs of Theorem \ref{thm:wc.vanishing} and Theorem \ref{thm:wc.non.vanishing} below are applications of the Argmax Theorem (reproduced as Theorem \ref{thm:argmax} in Appendix \ref{app:auxiliary}). The arguments here are largely an exercise in verification of requirements of this theorem.
	
	\bc$\rule{4in}{0.1mm}$\ec
	
	\begin{proof}[{\bf Proof of Theorem \ref{thm:wc.vanishing}}] We begin by proving the first part of this result. Here we shall examine the limiting distribution of the sequence $T_h\xi_w^{2}(\tilde\tau_w-\tau^0_w),$ consequently the underlying indexing metric space here is $\R$\footnote{Although $\tilde\tau_w$ is a discrete r.v., however $T_h\xi_w^{2}\tilde\tau_w\in\R$}. Now consider the two cases of known and unknown plug-in parameters.
		
		\vspace{1.5mm}	
		\noi{\bf Case I \big($\tau^0_h$ and $\theta^0$ known\big):} Following is list of requirement of the Argmax theorem that require verification for this case (see, page 288 of \cite{vaart1996weak}).
		\begin{enumerate}
			\item The sequence $T_h\xi_w^{2}(\tilde\tau_w^*-\tau^0_w)$ is uniformly tight (see, Definition \ref{def:utight}).
			\item $\big\{2\si_{(w,\iny)}W_w(\z)-|\z|\}$ satisfies suitable regularity conditions\footnotemark.
			\item For any $\z\in [-c_u,c_u]$ we have
			\benr
			\cC_w(\tau_w^0+\z T_h^{-1}\xi^{-2}_w,\tau_h^0,\theta^0)
			\Rightarrow \big\{2\si_{(w,\iny)}W_w(\z)-|\z|\}.\nn
			\eenr
		\end{enumerate}
		\footnotetext{Almost all sample paths $\z\to \big\{2\si_{(w,\iny)}W_w(\z)-|\z|\}$ are upper semicontinuous and posses a unique maximum at a (random) point $\argmax_{\z\in\R}\big\{2\si_{(w,\iny)}W_w(\z)-|\z|\},$ which as a random map in the indexing metric space is tight.}
		
		Note that by setting $\h\theta_{(j)}=\theta_{(j)}^0,$ $j=1,2,3,4$ and $\h\tau_h=\tau_h^0,$  Condition C(i)(b) and C(ii)(a,c) are trivially satisfied. Now using Theorem \ref{thm:cpoptimal} we have that $T_h\xi^{2}_w(\tilde\tau^*_w-\tau^0_w)=O_p(1).$ This directly yields requirement (1). The second requirement follows from well known properties of Brownian motion's. The only remaining requirement is (3), which is provided below.

		We begin with a couple of observations that shall be useful in the subsequent argument. For any given $w=1,...,T_w,$ define r.v.'s,
		\benr\label{eq:34a}
		\psi_w=\frac{1}{\xi_w\surd(T_h)}\Big[\sum_{h=\tau^0_h+1}^{T_h}\vep_{(w,h)}^T\eta^0_{(1)}+\sum_{h=1}^{\tau_h^0}\vep_{(w,h)}^T\eta^0_{(3)}\Big],
		\eenr
		and note that we have,
		\benr
		{\rm var}(\psi_w)=\frac{1}{\xi_w^2T_h}\Big[T_h\om_h\eta^0_{(1)}\Si\eta^0_{(1)}+T_h(1-\om_h)\eta^0_{(3)}\Si\eta^0_{(3)}\Big]\to \si^2_{(w,\iny)},\nn
		\eenr
		where the convergence follows from Condition D. Now let $\z> 0,$ w.l.o.g. assume $\z T_h^{-1}\xi^{-2}_w$ is integer valued and let $\tau^*_w=\tau^0_w+\z T_h^{-1}\xi^{-2}_w>\tau^0_w$ and consider,
		\benr\label{eq:39}
		\cC_w(\tau_w^*,\tau_h^0,\theta^0)
		&=&\sum_{w=\tau_w^0+1}^{T_w}\sum_{h=\tau^0_h+1}^{T_h}\|x_{(w,h)}-\theta_{(1)}^0\|_2^2+\sum_{w=1}^{\tau_w^0}\sum_{h=\tau_h^0+1}^{T_h}\|x_{(w,h)}-\theta_{(2)}^0\|_2^2\nn\\
		&&+\sum_{w=1}^{\tau_w^0}\sum_{h=1}^{\tau_h^0}\|x_{(w,h)}-\theta_{(3)}^0\|_2^2+\sum_{w=\tau_w^0+1}^{T_w}\sum_{h=1}^{\tau_h^0}\|x_{(w,h)}-\theta_{(4)}^0\|_2^2\nn\\
		&&-\sum_{w=\tau_w^*+1}^{T_w}\sum_{h=\tau_h^0+1}^{T_h}\|x_{(w,h)}-\theta_{(1)}^0\|_2^2-\sum_{w=1}^{\tau_w^*}\sum_{h=\tau_h^0+1}^{T_h}\|x_{(w,h)}-\theta_{(2)}^0\|_2^2\nn\\
		&&-\sum_{w=1}^{\tau_w^*}\sum_{h=1}^{\tau_h^0}\|x_{(w,h)}-\theta_{(3)}^0\|_2^2-\sum_{w=\tau_w^*+1}^{T_w}\sum_{h=1}^{\tau_h^0}\|x_{(w,h)}-\theta_{(4)}^0\|_2^2\nn\\
		&=& \sum_{w=\tau_w^0+1}^{\tau_w^*}\sum_{h=\tau^0_h+1}^{T_h}\Big[\|x_{(w,h)}-\theta_{(1)}^0\|_2^2-\|x_{(w,h)}-\theta_{(2)}^0\|_2^2\Big]\nn\\
		&&+\sum_{w=\tau_w^0+1}^{\tau_w^*}\sum_{h=1}^{\tau_h^0}\Big[\|x_{(w,h)}-\theta_{(4)}^0\|_2^2-\|x_{(w,h)}-\theta_{(3)}^0\|_2^2\Big]\nn\\
		&=&2\sum_{w=\tau_w^0+1}^{\tau_w^*}\Big[\sum_{h=\tau^0_h+1}^{T_h}\vep_{(w,h)}^T\eta^0_{(1)}+\sum_{h=1}^{\tau_h^0}\vep_{(w,h)}^T\eta^0_{(3)}\Big]-(\tau^*_w-\tau^0_w)T_h\xi_w^2\nn\\
		&=&2\xi_w \surd(T_h)\sum_{w=\tau_w^0+1}^{\tau_w^0+\z T_h^{-1}\xi_w^{-2}} \psi_w -\z\Rightarrow 2\si_{(w,\iny)}W_{w1}(\z)-\z
		\eenr
		The final equality obtained by substituting the defining expressions of $\tau^*_w=\tau^0_w+\z T_h^{-1}\xi^{-2}_w>\tau^0_w$ as well as that of $\psi_w$ from (\ref{eq:34a}). The weak convergence here now follows from the functional central limit theorem. Repeating the same argument with $\z\in[-c_u,0),$ yields  $\cC(\tau^0+\z T_h^{-1}\xi^{-2}_w,\tau_h^0,\theta^0)\Rightarrow 2\si_{(w,\iny)}W_{w2}(-\z)-|\z|.$ This completes the proof of requirement (3) for the Argmax theorem and consequently an application of its results yields $T_h^{-1}\xi^{2}_w(\tilde\tau^*-\tau^0)\Rightarrow \argmax_{\z\in\R}\big\{2\si_{(w,\iny)}W_w(\z)-|\z|\},$ which completes the proof of this case of known plug-in parameters.
		
		\vspace{1.5mm}		
		\noi{\bf Case II \big($\tau^0_h$ and $\theta^0$ unknown\big):} In this case the applicability of the argmax theorem requires verification of the following conditions.
		\begin{enumerate}
			\item The sequence $T_h\xi_w^{2}(\tilde\tau_w-\tau^0_w)$ is uniformly tight.
			\item $\big\{2\si_{(w,\iny)}W_w(\z)-|\z|\}$ satisfies suitable regularity conditions.
			\item For any $\z\in [-c_u,c_u]$ we have
			\benr
			\cC_w(\tau_w^0+\z T_h^{-1}\xi^{-2}_w,\h\tau_h,\h\theta)
			\Rightarrow \big\{2\si_{(w,\iny)}W_w(\z)-|\z|\}.\nn
			\eenr
		\end{enumerate}
		Part (i) again follows from the result of Theorem \ref{thm:cpoptimal} under the assumed Condition C(i)(b) and C(ii)(a,c) on the nuisance estimates $\h\tau_h$ and $\h\theta.$  Part (2) is identical to the corresponding requirement of Case I. Finally to prove part (3) note that from Lemma \ref{lem:Capprox} we have that,
		\benr\label{eq:36}	\sup_{\tau_w\in\cG_w(c_uT_w^{-1}T^{-1}_h\xi^{-2}_w,0)}|\cC_w(\tau_w,\h\tau_h,\h\theta)-\cC_w(\tau_w,\tau_h^0,\theta^0)|=o_p(1).
		\eenr
		The approximation (\ref{eq:36}) and Part (3) of Case I together imply Part (3) for this case. This completes the verification of all requirements for this case. The statement of first limiting distribution of the theorem now follows by an application of the Argmax theorem. The second limiting distribution result can be proved by proceeding with symmetrical arguments.
	\end{proof}

	\bc$\rule{4in}{0.1mm}$\ec

	\begin{proof}[{\bf Proof of Theorem \ref{thm:wc.non.vanishing}}] The proof of this theorem follows a similar structure as that of Theorem \ref{thm:wc.vanishing} in that it is also an application of the Argmax theorem. The distinction here is in the limiting distributional structure that is induced by the change of regime of the jump size.
		
		We begin by proving the first part of this result, for which consider the sequence $(\tilde\tau_w-\tau^0_w).$ Consequently, the underlying indexing metric space here is $\Z.$ Now consider the two cases of known and unknown plug-in parameters.
		
		\vspace{1.5mm}	
		\noi{\bf Case I \big($\tau_h^0$ and $\theta^0$ known\big):} The requirements to be verified here are as follows.
		\begin{enumerate}
			\item The sequence $(\tilde\tau_w^*-\tau^0_w)$ is uniformly tight.
			\item $\cC_{(w,\iny)}(\z)$ satisfies suitable regularity conditions.
			\item For any $\z\in \{-c_u,-c_u+1,...,-1,0,1,...c_u\},$ we have
			\benr
			\cC_w(\tau_w^0+\z,\tau_h^0,\theta^0)
			\Rightarrow \cC_{(w,\iny)}(\z).\nn
			\eenr
		\end{enumerate}
		As in the proof of Theorem \ref{thm:wc.vanishing}, requirement (1) follows directly from the result of Theorem \ref{thm:cpoptimal}.  Requirement (2) of regularity of the {\it argmax} of two sided negative drift random walk $\cC_{(w,\iny)}(\z)$ has been proved earlier in Lemma A.3 of the supplement of \cite{kaul2020inference}. The requirement (3) is verified in the following. 
		
		Let $\psi_w$ be as defined in (\ref{eq:34a}), then we begin by noting that under this non-vanishing regime $\surd(T_h)\xi_w\to\xi_{(w,\iny)},$ we have,
		\benr
		{\rm var}(\psi_w)=\frac{1}{\xi_w^2T_h}\Big[T_h\om_h\eta^0_{(1)}\Si\eta^0_{(1)}+T_h(1-\om_h)\eta^0_{(3)}\Si\eta^0_{(3)}\Big]\to \xi_{(w,\iny)}\si^2_{(w,\iny)},\nn
		\eenr
		where the convergence follows from Condition D and the regime under consideration. Now for any $\z\in\{1,2,...,c_u\},$  let $\tau^*_w=\tau^0_w+\z T_h^{-1}\xi^{-2}_w>\tau^0_w$ and note that,
		\benr
		\cC_w(\tau_w^*,\tau_h^0,\theta^0)
		&=&2\sum_{w=\tau_w^0+1}^{\tau_w^*}\Big[\sum_{h=\tau^0_h+1}^{T_h}\vep_{(w,h)}^T\eta^0_{(1)}+\sum_{h=1}^{\tau_h^0}\vep_{(w,h)}^T\eta^0_{(3)}\Big]-(\tau^*_w-\tau^0_w)T_h\xi_w^2\nn\\
		&=&2\sum_{w=\tau_w^0+1}^{\tau_w^0+\z} \psi_w -\z\xi_w^2\Rightarrow \sum_{w=1}^{\z}\cP\big(-\xi_{(w,\iny)}^2,\,\,4\xi_{(w,\iny)}^2\si^2_{(w,\iny)}\big).\nn
		\eenr
		The equalities here follow by performing a algebraic decomposition as provided in (\ref{eq:39}). The weak convergence follows from Condition A$'$. Repeating the same argument with $\z\in\{-c_u,-c_u+1,...,-1\},$ yields  	$	\cC_w(\tau_w^0+\z,\tau_h^0,\theta^0)\Rightarrow \sum_{t=1}^{-\z}\cP(-\xi_{(w,\iny)}^2,4\xi_{(w,\iny)}^2\si^2_{(w,\iny)}).$ An application the Argmax theorem now yields $(\tilde\tau^*_w-\tau^0_w)\Rightarrow \argmax_{\z\in\Z}\cC_{(w,\iny)}(\z),$ which completes the proof of this case.
		
		\vspace{1.5mm}		
		\noi{\bf Case II \big($\tau_h^0$ and $\theta^0$ unknown\big):} In this case, the applicability of the argmax theorem requires verification of the following.
		\begin{enumerate}
			\item The sequence $(\tilde\tau_w-\tau^0_w)$ is uniformly tight.
			\item $\cC_{(w,\iny)}(\z)$ satisfies suitable regularity conditions.
			\item For any $\z\in \{-c_u,-c_u+1,...,-1,0,1,...c_u\},$ we have
			\benr
			\cC_w(\tau_w^0+\z,\h\tau_h,\h\theta)
			\Rightarrow \cC_{(w,\iny)}(\z).\nn
			\eenr
		\end{enumerate}
		Part (1) follows from Theorem \ref{thm:cpoptimal} under the assumed Condition C(i)(b) and C(ii)(a,c) on the nuisance estimates $\h\tau_h$ and $\h\theta_{(j)},$ $j=1,2,3,4.$ Part (2) is identical to the corresponding requirement of Case I. Finally to prove part (3) note that from Lemma \ref{lem:Capprox} where we have under the regime $\surd{T_h}\xi_w\to\xi_{(2,\iny)}$ that,
		\benr\label{eq:38}
		\sup_{\tau\in\cG(c_uT_w^{-1},0)}|\cC_w(\tau_w,\h\tau_h,\h\theta)-\cC_w(\tau_w,\tau_h^0,\theta^0)|=o_p(1).\nn
		\eenr
		The approximation (\ref{eq:38}) and Part (3) of Case I together imply Part (3) for this case. This completes the verification of all requirements for this case. The statement of the first limiting distribution of the theorem now follows by an application of the Argmax theorem. The second limiting distribution can be proved by symmetrical arguments.
	\end{proof}

	\bc$\rule{4in}{0.1mm}$\ec

	\begin{lemma}\label{lem:Capprox} Let $\cC_w(\tau_w,\tau_h,\theta)$ and $\cC_h(\tau_w,\tau_h,\theta)$ be as defined in (\ref{def:cC}) and suppose Condition A and B hold. Additionally assume that Condition C(i)(b) and Condition C(ii)(a,c) holds with the sequence $r_T=\big\{o(1)\big/s^{1/2}\log(p\vee T_wT_h)\big\}.$ Then, for any $c_u>0,$ we have,
		\benr
		&(i)&\,\,\sup_{\tau_w\in\cG_w\big(c_uT_w^{-1}T^{-1}_h\xi^{-2}_w,\,0\big)}\big|\cC_w(\tau_w,\h\tau_h,\h\theta)-\cC_w(\tau_w,\tau_h^0,\theta^0)\big|=o_p(1),\quad{\rm and}\nn\\
		&(ii)&\,\,\sup_{\tau_h\in\cG_h\big(c_uT^{-1}_wT_h^{-1}\xi^{-2}_h,\,0\big)}\big|\cC_h(\h\tau_w,\tau_h,\h\theta)-\cC_w(\tau_w^0,\tau_h,\theta^0)\big|=o_p(1),\nn
		\eenr
		where the orders of (i) and (ii) are w.r.t. $T_w$ and $T_h,$ respectively.
	\end{lemma}
	
	\begin{proof}[Proof of Lemma \ref{lem:Capprox}] We only prove Part (i) below, the proof of Part (ii) follows symmetrically. This proof relies on a complete algebraic decomposition of the difference of interest and an examination of the rates of residual terms. To this end, we begin with a few bounds and definitions of residual terms that shall be required for the said decomposition.
		
		Observing that from Condition C(i)(b) with the assumed choice of $r_T,$ we have that,
		\benr\label{eq:43}
		|\h\tau_h-\tau_h^0|\le T_hr_T^2,\quad{\rm with}\quad r_T=\frac{o(1)}{s^{1/2}\log(p\vee T_wT_h)},
		\eenr
		with probability at least $1-\pi_T.$ Also, by proceeding similar to (\ref{eq:4}) and (\ref{eq:8}), we have under Condition C(ii)(a,c) with the assumed choice of $r_T,$ that,
		\benr\label{eq:44}
		\max_{1\le j\le 4}\|\h\theta_{(j)}-\theta_{(j)}^0\|_1\le \max_{1\le j\le 4}\surd{s}\|\h\theta_{(j)}-\theta_{(j)}^0\|_2\le \frac{o(1)\xi_{\min}}{\log (p\vee T)}.
		\eenr
		with probability at least $1-\pi_T.$ Consequently, we also have,
		\benr\label{eq:47}
		\|\h\eta_{(1)}-\eta^0_{(1)}\|_1\le \frac{o(1)\xi_{\min}}{\log (p\vee T)}\,\,\,{\rm and}\,\,\, \|\h\eta_{(3)}-\eta^0_{(3)}\|_1\le \frac{o(1)\xi_{\min}}{\log (p\vee T_wT_h)}.
		\eenr
		with probability at least $1-\pi_T.$ Finally, following (\ref{eq:5}) and (\ref{eq:8}) we have that,
		\benr\label{eq:48}
		\|\h\eta_{(1)}\|_1\le c_u\surd{s}\xi_{\min}\quad{\rm and} \|\h\eta_{(3)}\|_1\le c_u\surd{s}\xi_{\min}
		\eenr
		with probability at least $1-\pi_T.$  Now consider the case where $\tau_w\ge \tau^0_w$ and  $\h\tau_h\ge\tau_h^0,$ and define the following residual terms,
		\benr
		R1&=&2\sum_{w=\tau_w^0+1}^{\tau_w}\sum_{h=\tau_h^0+1}^{T_h}\vep_{(w,h)}^T(\h\eta_{(1)}-\eta_{(1)}^0)\nn\\
		R2&=&2\sum_{w=\tau_w^0+1}^{\tau_w}\sum_{h=1}^{\tau_h^0}\vep_{(w,h)}^T(\h\eta_{(3)}-\eta^0_{(3)})\nn\\
		R3&=&2\sum_{w=\tau_w^0+1}^{\tau_w}\sum_{h=\tau_h^0+1}^{\h\tau_h}\vep_{(w,h)}^T\h\eta_{(1)}-2\sum_{w=\tau_w^0+1}^{\tau_w}\sum_{h=\tau_h^0+1}^{\h\tau_h}\vep_{(w,h)}^T\h\eta_{(3)}\nn\\
		R4&=&\om_h\big(\|\h\eta_{(1)}\|_2^2-\|\eta_{(1)}^0\|_2^2\big)+(1-\om_h)\big(\|\h\eta_{(3)}\|_2^2-\|\eta_{(3)}^0\|_2^2\big)\nn\\
		R5&=&2\om_h(\h\theta_{(1)}-\theta_{(1)}^0)^T\h\eta_{(1)}+2(1-\om_h)(\h\theta_{(4)}-\theta_{(4)}^0)^T\h\eta_{(3)}\nn\\
		&&-2\frac{(\h\tau_h-\tau_h^0)}{T_h}(\h\theta_{(1)}-\theta_{(1)}^0)^T\h\eta_{(1)}
		+2\frac{(\h\tau_h-\tau_h^0)}{T_h}(\h\theta_{(4)}-\theta_{(1)}^0)^T\h\eta_{(3)}\nn\\
		&&+\frac{(\h\tau_h-\tau_h^0)}{T_h}\|\h\eta_{(3)}\|_2^2-\frac{(\h\tau_h-\tau_h^0)}{T_h}\|\h\eta_{(1)}\|_2^2\nn
		\eenr
		Then under the considered orientation $\tau_w\ge \tau^0_w$ and  $\h\tau_h\ge\tau_h^0,$ we have the following algebraic expansion,	
		\benr\label{eq:46}
		\cC_w(\tau_w,\h\tau_h,\h\theta)-\cC_w(\tau_w,\tau_h^0,\theta^0)&=&\cU_w(\tau_w,\tau_h^0,\theta^0)-\cU_w(\tau_w,\h\tau_h,\h\theta)\nn\\
		&=&-R1+R2-R3-(\tau_w-\tau_w^0)T_h (R4+R5)
		\eenr
		We now examine each of the residual terms in (\ref{eq:46}) individually. Applying Lemma \ref{lem:nearoptimalcross.subE.special} we have that,
		\benr
		\sup_{\substack{\tau_w\in\cG_w\big(c_uT_w^{-1}T^{-1}_h\xi^{-2}_w,\,0\big);\\\tau_w\ge\tau_w^0}}|R1|\le \frac{c_u\si}{\xi_w}\log(p\vee T_wT_h)\|\h\eta_{(1)}-\eta^0_{(1)}\|_1=o(1)\nn
		\eenr
		w.p. $1-o(1).$ Here the equality follows from (\ref{eq:47}). An analogous argument yields,
		\benr
		\sup_{\substack{\tau_w\in\cG_w\big(c_uT_w^{-1}T^{-1}_h\xi^{-2}_w,\,0\big);\\\tau_w\ge\tau_w^0}}|R2|=o_p(1).\nn
		\eenr
		Applying Lemma \ref{lem:nearoptimalcross.subE.special} together with the bounds of (\ref{eq:43}) and (\ref{eq:47}) yields,
		\benr
		\sup_{\substack{\tau_w\in\cG_w\big(c_uT_w^{-1}T^{-1}_h\xi^{-2}_w,\,0\big);\\\tau_w\ge\tau_w^0}}|R3|\le \frac{c_u\si\xi_{\min}\surd{s}}{\xi_w}r_T\log(p\vee T_wT_h)=o(1)
		\eenr
		with probability $1-o(1).$ To bound the remaining two terms $R4$ and $R5,$ recall from the construction (\ref{def:setG}) of the set $\cG_w\big(c_uT_w^{-1}T^{-1}_h\xi^{-2}_w,\,0\big),$ that any $\tau_w$ must satisfy, $|\tau_w-\tau_w^0|\le c_uT_h^{-1}\xi_w^{-2}.$ Now consider,
		\benr
		\sup_{\substack{\tau_w\in\cG_w\big(c_uT_w^{-1}T^{-1}_h\xi^{-2}_w,\,0\big);\\\tau_w\ge\tau_w^0}}|(\tau_w-\tau_w^0)T_hR4|\le c_u\xi_{w}^{-2}|\big(\|\h\eta_{(1)}\|_2^2-\|\eta_{(1)}^0\|_2^2|+\|\h\eta_{(3)}\|_2^2-\|\eta_{(3)}^0\|_2^2\big)\nn\\
		\le  c_u\xi^{-2}_w\big|\|\h\eta_{(1)}-\eta^0_{(1)}\|_2^2+2(\h\eta_{(1)}-\eta^0_{(1)})^T\eta^0_{(1)}\big|\hspace{2cm}\nn\\
		+c_u\xi^{-2}_w\big|\|\h\eta_{(3)}-\eta^0_{(3)}\|_2^2+2(\h\eta_{(3)}-\eta^0_{(3)})^T\eta^0_{(3)}\big|\hspace{1.75cm}\nn\\
		\le  c_u\xi^{-2}_w \|\h\eta_{(1)}-\eta^0_{(1)}\|_2^2+ c_u\xi^{-1}_w \|\h\eta_{(1)}-\eta^0_{(1)}\|_2\hspace{2cm}\nn\\
		+ c_u\xi^{-2}_w \|\h\eta_{(3)}-\eta^0_{(3)}\|_2^2+ c_u\xi^{-1}_w \|\h\eta_{(3)}-\eta^0_{(3)}\|_2
		= o_p(1).\hspace{0.35cm}\nn
		\eenr
		Here the third inequality follows from applications of the Cauchy Schwarz inequality and the equality follows from the bounds in (\ref{eq:47}). The only remaining residual term now is $R5$ which is examined below.
		
		\benr
		\sup_{\substack{\tau_w\in\cG_w\big(c_uT_w^{-1}T^{-1}_h\xi^{-2}_w,\,0\big);\\\tau_w\ge\tau_w^0}}|(\tau_w-\tau_w^0)T_hR5|\le  c_u\xi_w^{-2}\Big[\|\h\theta_{(1)}-\theta_{(1)}^0\|_2\|\h\eta_{(1)}\|_2+\|\h\theta_{(4)}-\theta_{(4)}^0\|_2\|\h\eta_{(3)}\|_2\Big]\nn\\
		+c_u\xi_w^{-2}r_T^2\|\h\theta_{(1)}-\theta_{(1)}^0\|_2\|\h\eta_{(1)}\|_2
		+c_u\xi_w^{-2}r_T^2\|\h\theta_{(4)}-\theta_{(1)}^0\|_2\|\h\eta_{(3)}\|_2\nn\\
		+c_ur_T^2\|\h\eta_{(3)}\|_2^2+c_ur_T^2\|\h\eta_{(1)}\|_2^2=o_p(1)\hspace{4.15cm}\nn
		\eenr
		The inequality here follows from several applications of the Cauchy Schwarz inequality and the equality follows by substituting the choice of $r_T$ from (\ref{eq:43}), as well as the available bounds for the mean estimates. Substituting the uniform bounds for $R1,R2,R3,R4$ and $R5$ obtained above into the expression (\ref{eq:46}) yields,
		\benr
		\sup_{\substack{\tau_w\in\cG_w\big(c_uT_w^{-1}T^{-1}_h\xi^{-2}_w,\,0\big);\\\tau_w\ge\tau_w^0}}\Big|\cC_w(\tau_w,\h\tau_h,\h\theta)-\cC_w(\tau_w,\tau_h^0,\theta^0)\Big|=o_p(1)
		\eenr
		Repeating symmetrical arguments on the remaining three orientations of the ordering of $(\tau_w,\h\tau_h)$ w.r.t $(\tau_w^0,\tau_h^0),$ in particular, $\tau_w\le\tau_w^0,$ $\h\tau_h\ge\tau_h^0,$ and $\tau_w\le\tau_w^0,$ $\h\tau_h\le\tau_h^0,$ and $\tau_w\ge\tau_w^0,$ $\h\tau_h\le\tau_h^0,$ shall yield the same $o_p(1)$ approximation. This completes the proof of Part (i) and the statement of the lemma.
	\end{proof}

	\bc$\rule{4in}{0.1mm}$\ec
	
The proof of Corollary \ref{cor:alg1.validity} requires some preliminary work, in particular we first need to examine the behavior of the estimates $\tilde\theta_{(j)}(\tau),$ $j=1,2,3,4,$ and uniformly over a collection of values of $\tau.$ This is provided in the following theorem.

	\begin{theorem}\label{thm:unifmean} Let $0\le u_{T_w},\, u_{T_w}\le 1,$ be non-negative sequences and let $\psi=\max_j\|\eta^0_{(j)}\|_{\iny}.$ Additionally, for any constants $c_u,c_{u1}>0,$ and each $j=1,2,3,4,$ let,
		\benr\label{eq:la}
		\la:=\la_j=8\max\Big\{\si\Big\{\frac{2c_{u1}\log(p\vee T_wT_h)}{c_uT_wT_h\underline\om}\Big\}^{\frac{1}{2}},\,\,\frac{3\psi}{c_u\underline\om}(u_{T_w}\vee u_{T_h})\Big\}.
		\eenr
		Suppose Condition A and B holds and that $c_uT_wT_h\underline\om\ge \log (p\vee T_wT_h).$ Then, $\h\theta_{(j)}(\tau),$ $j=1,2,3,4$ of (\ref{est:softthresh}) satisfy the following two results with probability at least $1-\pi_T.$\\~
		(i) For any $j=1,2,3,4,$ and any $\tau=(\tau_w,\tau_h)^T\in \cG_w(u_{T_w},0)\times\cG_h(u_{T_h},0),$ with $|Q_j(\tau)|\ge c_uT_wT_h\underline\om,$ we have $\big\|\big(\h\theta_{(j)}(\tau)\big)_{S_j^c}\big\|_1\le 3\big\|\big(\h\theta_{(j)}(\tau)-\theta_{(j)}^0\big)_{S_j}\big\|_1.$\\~
		(ii) The following bound is satisfied,
		\benr
		\max_{1\le j\le 4}\,\,\sup_{\substack{\tau\in\cG_w(u_{T_w},0)\times \cG_h(u_{T_h},0)\\ |Q_j(\tau)|\ge c_uT_wT_h\underline\om}} \|\h\theta_{(j)}(\tau)-\theta_{(j)}^0\|_2\le 6\surd{s}\la.\nn
		\eenr
		Here  $\pi_T=8\exp\big\{-\big(c_{u2}-2\big)\log (p\vee T_wT_h)\big\},$ where $c_{u2}=c_{u1}\wedge \surd(c_uc_{u1}/2).$	
	\end{theorem}

	\begin{proof}[Proof of Theorem \ref{thm:unifmean}]
		Consider $j=1$ and any $(\tau_w,\tau_h)^T\in\cG_w(u_{T_w},0)\times\cG_w(u_{T_h},0),$ such that $\tau=(\tau_w,\tau_h)^T$ satisfies $|Q_1(\tau)|\ge c_uT_wT_h\underline\om.$ Without loss of generality assume $\tau_w\le\tau^0_w,$ $\tau_h\le \tau_h^0$ The remaining permutations of the ordering of $\tau$ w.r.t. $\tau^0$ can be proved using symmetrical arguments.
		
		An algebraic rearrangement of the elementary inequality $\big\|\bar x_{(1)}(\tau)-\h\theta_{(1)}(\tau)\big\|^2_2+\la_1\|\h\theta_{(1)}(\tau)\|_1\le \big\|\bar x_{(1)}(\tau)-\theta_{(1)}^0\big\|^2_2+\la_1\|\theta_{(1)}^0\|_1$ yields,
		\benr\label{eq:20}
		\big\|\h\theta_{(1)}(\tau)-\theta_{(1)}^0\big\|_2^2+\la_1\big\|\tilde\theta_{(1)}(\tau)\big\|_1&\le& \la_1\big\|\theta^0_{(1)}\big\|_1\nn\\
		&&+ \frac{2}{|Q_1(\tau)|}\sum_{w=\tau_w+1}^{T_w}\sum_{h=\tau_h+1}^{T_h}\tilde\vep_{(w,h)}^T(\h\theta_{(1)}(\tau)-\theta_{(1)}^0).\nn\\
		&=&\la_1\big\|\theta^0_{(1)}\big\|_1+\frac{2}{|Q_1(\tau)|}\sum_{w=\tau_w+1}^{T_w}\sum_{h=\tau_h+1}^{T_h}\vep_{(w,h)}^T(\h\theta_{(1)}(\tau)-\theta_{(1)}^0)\nn\\
		&&-\frac{2}{|Q_1(\tau)|}(\tau_w^0-\tau_w)(\tau_h^0-\tau_h)(\theta_{(1)}^0-\theta_{(3)}^0)^T(\h\theta_{(1)}(\tau)-\theta_{(1)}^0)\nn\\
		&&-\frac{2}{|Q_1(\tau)|}(\tau_w^0-\tau_w)(T_h-\tau_h^0)(\theta_{(1)}^0-\theta_{(2)}^0)^T(\h\theta_{(1)}(\tau)-\theta_{(1)}^0)\nn\\
		&&-\frac{2}{|Q_1(\tau)|}(T_w-\tau_w^0)(\tau_h^0-\tau_h)(\theta_{(1)}^0-\theta_{(3)}^0)^T(\h\theta_{(1)}(\tau)-\theta_{(1)}^0).\nn\\
		&=&\la_1\big\|\theta^0_{(1)}\big\|_1+2R1-2R2-2R3-2R4
		\eenr
		Here $\tilde\vep_{(w,h)}=\big(x_{(w,h)}-\h\theta_{(1)}(\tau)\big).$ Next we consider the residual terms $R1,R2,R3,R4$ on the r.h.s of (\ref{eq:20}). For this purpose, first note from Lemma \ref{lem:1.to.tau.bound} we have,
		\benr
		\frac{2}{|Q_1(\tau)|}\Big\|\sum_{w=\tau_w+1}^{T_w}\sum_{h=\tau_h+1}^{T_h}\vep_{(w,h)}\Big\|_{\iny}\le 2\si\Big\{\frac{2c_{u1}\log(p\vee T_wT_h)}{c_uT_wT_h\underline\om}\Big\}^{\frac{1}{2}},
		\eenr
		with probability at least $1-8\exp\{-(c_{u2}-2)\log(p\vee T_wT_h)\}.$ Additionally recall we have $\psi=\max_j\|\eta^0_{(j)}\|,$ and $|Q_1(\tau)|\ge c_uT_wT_h\underline\om$ thus,
		\benr
		2|R2+R3+R4|\le \frac{6\psi}{c_u\underline\om}(u_{T_w}\vee u_{T_h})\|\h\theta_{(1)}(\tau)-\theta_{(1)}^0\|_1.
		\eenr
		Consequently, upon choosing,
		\benr
		\la^*=\max\Big\{4\si\Big\{\frac{2c_{u1}\log(p\vee T_wT_h)}{c_uT_wT_h\underline\om}\Big\}^{\frac{1}{2}},\,\,\frac{12\psi}{c_u\underline\om}(u_{T_w}\vee u_{T_h})\Big\},\nn
		\eenr
		and substituting these bounds in (\ref{eq:20}) we obtain,
		\benr\label{eq:21}
		\big\|\h\theta_{(1)}(\tau)-\theta_{(1)}^0\big\|_2^2+\la_1\big\|\h\theta_{(1)}(\tau)\big\|_1\le\la_1\big\|\theta^0_{(1)}\big\|_1+\la^*\big\|\h\theta_{(1)}(\tau)-\theta_{(1)}^0\big\|_1,
		\eenr
		with probability at least $1-8\exp\{-(c_{u2}-2)\log(p\vee T_wT_h)\}.$  Now choosing $\la_1= 2\la^*,$ leads to $\|\big(\h\theta_{(1)}(\tau)\big)_{S_1^c}\|_1\le 3\|\big(\h\theta_{(1)}(\tau)-\theta_{(1)}^0\big)_{S_1}\|_1,$ which proves part (i) of this theorem for $j=1.$ From inequality (\ref{eq:21}) we also have that,
		\benr
		\|\h\theta_{(1)}(\tau)-\theta_{(1)}^0\|_2^2\le \frac{3}{2}\la_1\|\h\theta_{(1)}(\tau)-\theta_{(1)}^0\|_1\le 6\la_1\surd{s}\|\h\theta_{(1)}(\tau)-\theta_{(1)}^0\|_2
		\eenr
		This directly implies that  $\|\h\theta_{(1)}(\tau)-\theta_{(1)}^0\|_2\le 6\la_1\surd{s},$ where we have used $\|\h\theta_{(1)}(\tau)-\theta_{(1)}^0\|_1\le 4\sqrt{s}\|\h\theta_{(1)}(\tau)-\theta_{(1)}^0\|_2,$ which follows in turn from $\|\big(\h\theta_{(1)}(\tau)\big)_{S_1^c}\|_1\le 3\|\big(\h\theta_{(1)}(\tau)-\theta_{(1)}^0\big)_{S_1}\|_1.$ To supply uniformity over $\tau,$ recall that the only stochastic bound used here is Lemma \ref{lem:1.to.tau.bound} which holds uniformly over $\tau,$ consequently the final bound also holds uniformly over the given collection. A symmetrical argument can be replicated for each $j=2,3,4$ and recalling that Lemma \ref{lem:1.to.tau.bound} also holds uniformly over these $j$'s. This finishes the proof of the Theorem. This result can alternatively be proved using the properties of the soft-thresholding operator $k_{\la}(\cdotp),$ by building uniform versions of arguments such as those in \cite{kaul2017structural}.
	\end{proof}

	\bc$\rule{3.5in}{0.1mm}$\ec

	Following is another preliminary result required to prove Corollary \ref{cor:alg1.validity}. This results uses Theorem \ref{thm:unifmean} to provide the rate of convergence of Step 1 mean estimates of Algorithm \ref{alg:single}.
	
	\begin{lemma}\label{lem:step1mean} Assume Condition A, B and F holds and that the model dimensions together with the least jump size are restricted by the following condition,
		\benr\label{eq:23}
		\frac{c_u\si}{\xi_{\min}}\Big\{\frac{s\log (p\vee T_wT_h)}{T_wT_h\underline\om}\Big\}^{\frac{1}{2}}\le c_{u1},
		\eenr	
		for an appropriately chosen small enough constant $c_{u1}>0.$ Additionally assume that  $c_uT_wT_h\underline\om\ge \log (p\vee T_wT_h).$ Then with a suitably chosen regularizer $\la,$ the Step 1 mean estimates of Algorithm 1, $\check\theta_{(j)}=\h\theta_{(j)}(\check\tau),$ $j=1,2,3,4,$ satisfy the following, w.p. $1-o(1).$\\~
		(i)  $\big\|\big(\check\theta_{(j)}\big)_{S_j^c}\big\|_1\le 3\big\|\big(\check\theta_{(j)}-\theta_{(j)}^0\big)_{S_j}\big\|_1,$ for any $j=1,2,3,4.$ \\~
		(ii) The following bound is satisfied,
		\benr
		\max_{1\le j\le 4}\|\check\theta_{(1)}-\theta_{(1)}^0\|_2\le c_{u1}\xi_{\min}.\nn
		\eenr
		Consequently, the mean estimates $\check\theta_{(j)},$ $j=1,2,3,4$ satisfy Condition C(ii)(a,b).
	\end{lemma}
	
	\begin{proof}[Proof of Lemma \ref{lem:step1mean}] From Condition F, the initializer  $\check\tau=(\check\tau_w,\check\tau_h)^T$ 	is assumed to satisfy, $(i)\,\,|\check\tau_w-\tau^0_w|\le T_wu_{T_w},$ $(ii)\,\,|\check\tau_h-\tau_w^0|\le T_hu_{T_h}$ and $(iii)\,\,\min_{1\le j\le 4}|Q_j(\check\tau)|\ge c_{u}T_wT_h\underline\om,$ where,
		\benr\label{eq:28}
		u_{T_w}=u_{T_h}=c_{u1}\underline\om\xi_{\min}\big/(\surd s\psi),
		\eenr	
		i.e., $\check\tau\in\cG_w(u_T,0)\times\cG_h(u_T,0).$ Now applying Theorem \ref{thm:unifmean} while choosing,
		\benr\label{eq:la.step1.choice}
		\la\,\,{\textrm {as prescribed in (\ref{eq:la}) with}} \,\,u_{T_w},\,u_{T_h}\,\,{\textrm {as given in (\ref{eq:28})}},
		\eenr
		we obtain the following two results that hold with probability $1-o(1).$  First,  $\big\|\big(\check\theta_{(j)}\big)_{S_j^c}\big\|_1\le 3\big\|\big(\check\theta_{(j)}-\theta_{(j)}^0\big)_{S_j}\big\|_1,$ for any $j=1,2,3,4.$ Second,
		\benr\label{eq:22}
		\max_{1\le j\le 4}\|\check\theta_{(j)}-\theta_{(j)}^0\|_2&\le& \max\Big[c_u\si\Big\{\frac{s\log(p\vee T_wT_h)}{T_wT_h\underline\om}\Big\}^{\frac{1}{2}},\,\,c_u\frac{(u_{T_w}\vee u_{T_h})\surd{s}\psi}{\underline\om}\Big],\nn\\
		&=& \xi_{\min}\max\Big[\frac{c_u\si}{\xi_{\min}}\Big\{\frac{s\log(p\vee T_wT_h)}{T_wT_h\underline\om}\Big\}^{\frac{1}{2}},\,\,c_u\frac{(u_{T_w}\vee u_{T_h})\surd{s}\psi}{\xi_{\min}\underline\om}\Big]\nn\\
		&=& \xi_{\min}\Big[R_1,R_2\Big]
		\eenr	
		Here the first equality is simply an algebraic manipulation. Now for a suitable chosen $c_{u1}>0,$ we have from assumption (\ref{eq:23}) that,
		\benr
		\frac{c_u\si}{\xi_{\min}}\Big\{\frac{s\log (p\vee T_wT_h)}{T_wT_h\underline\om}\Big\}^{\frac{1}{2}}\le c_{u1},\nn
		\eenr
		which provides a bound for term $R_1$ on the RHS of (\ref{eq:22}). Next we bound term $R_2$ of the same expression. Substituting the choice of $u_T$ from (\ref{eq:28}) in term $R_2$, together with the earlier bound for $R_1,$ we obtain,
		\benr\max_{1\le j\le 4}\|\check\theta_{(1)}-\theta_{(1)}^0\|_2\le
		\xi_{\min}\Big[R_1,R_2\Big]\le c_{u1}\xi_{\min},\nn
		\eenr
		with probability $1-o(1).$ Thereby the requirement Condition C(ii)(a,b) are met and this completes the proof of the lemma.
	\end{proof}
	
	\bc$\rule{3.5in}{0.1mm}$\ec

	\begin{proof}[{\bf Proof of Corollary \ref{cor:alg1.validity}}] The logical progression of the argument to follow is as described in Figure \ref{fig:schematic}, effectively, we show that once Algorithm \ref{alg:single} is initialized under Condition F, then, under the assumed rate conditions on model parameters all remaining conditions fall in line for Step 1 and Step 2, thereby allowing applicability of the main results of Sub-section \ref{subsec:rate.limiting.dist}.
		
		We begin by noting that Lemma \ref{lem:step1mean} establishes that $\check\theta_{(j)}=\tilde\theta_{(j)}(\check\tau),$ $j=1,2,3,4,$ of Step 1 of Algorithm \ref{alg:single} satisfies Condition C(ii)(a,c), under the dimensional rate assumption (\ref{eq:23}), which is weaker than Condition E, and therefore this result continues to hold. Also observe that Condition C(i)(a) is weaker than assumed Condition F on initializer $\check\tau.$ Thus, a direct application of Theorem \ref{thm:cp.nearoptimal} yields the rate of convergence of $\h\tau$ of Step 1 of Algorithm \ref{alg:single} as,
		\benr\label{eq:49}
		&&|\h\tau_w-\tau^0_w|\le c_{u}\si^2T_h^{-1}\xi^{-2}_ws\log^2(p\vee T_wT_h),\quad{\rm and}\nn\\
		&&|\h\tau_h-\tau^0_h|\le c_{u}\si^2T_w^{-1}\xi^{-2}_hs\log^2(p\vee T_wT_h),
		\eenr	
		with probability at least $1-o(1).$ This completes the proof of Part (a) of this theorem.

		Moving onto Part (b), first observe that the bounds (\ref{eq:49}) together with rate assumption of  Condition E implies that $\h\tau=(\h\tau_w,\h\tau_h)^T$ of Step 1 satisfies the stronger Condition C(i)(b). Next we show that the updated mean estimates $\h\theta_{(j)},$ $j=1,2,3,4,$ satisfy Condition C(ii)(a,c). For this purpose, note that from (\ref{eq:49}) we have that $\h\tau_w\in\cG_w(u_{T_w},0),$ and $\h\tau_h\in\cG_h(u_{T_h},0)$ with probability $1-o(1),$ where,
		\benr\label{eq:ut}
		(u_{T_w}\vee u_{T_h})\le c_{u}\si^2T_{w}^{-1}T_h^{-1}\xi^{-2}_{\min}s\log^2(p\vee T_wT_h),
		\eenr
		Moreover, (\ref{eq:49}) and Condition E also imply that with the same probability as above, we have $|Q_j(\h\tau)|\ge c_uT_wT_h\underline\om.$ Now applying Theorem \ref{thm:unifmean} with,
		\benr\label{eq:la.step2.choice}
		\la\,\,{\rm as\, prescribed\,\, in\,\, (\ref{eq:la})\,\, with}\,\,(u_{T_w}\vee u_{T_h})\,\,{\rm as\,\,in\,\,(\ref{eq:ut})},
		\eenr
		we obtain  $\h\theta_{(j)}=\tilde\theta_{(j)}(\h\tau),$ $j=1,2,3,4,$  of Step 2 of Algorithm \ref{alg:single} satisfies Condition C(ii)(a). Furthermore,
		\benr\label{eq:25a}
		\max_{1\le j\le 4}\big\|\h\theta_{(j)}-\theta_{(j)}^0\big\|_2&\le& \max\Big[\si\Big\{\frac{c_{u}s\log(p\vee T_wT_h)}{T_wT_h\underline\om}\Big\}^{\frac{1}{2}},\,\,\frac{c_u\surd{s}\psi}{\underline\om}(u_{T_w}\vee u_{T_h})\Big].\nn\\
		&=&\frac{\xi_{\min}}{s^{1/2}\log (p\vee T_wT_h)} \max\Big[\si\Big\{\frac{c_{u}s\log^{3/2}(p\vee T_wT_h)}{\xi_{\min}\surd{(T_wT_h\underline\om)}}\Big\},\,\nn\\
		&&\hspace{4cm}\,\frac{c_us\log (p\vee T_wT_h)\psi}{\underline\om\xi_{\min}}(u_{T_w}\vee u_{T_h})\Big]\nn\\
		&=&\frac{\xi_{\min}}{s^{1/2}\log (p\vee T_wT_h)}\max\big[R1,R2\big]
		\eenr
		with probability at $1-o(1).$ Here the first equality is simply an algebraic manipulation. From Condition E we have that $R1\le c_{u1},$ where $c_{u1}>0,$ is an appropriately chosen small enough constant. Next consider term $R2$ of (\ref{eq:25a}). Substituting $(u_{T_w}\vee u_{T_h})$ from (\ref{eq:ut}) in term $R2$ we obtain,
		\benr
		c_us\log (p\vee T_wT_h)\frac{\psi}{\underline\om\xi_{\min}}(u_{T_w}\vee u_{T_h})&\le &c_{u}\si^2\Big(\frac{\psi}{\xi_{\min}}\Big)\Big\{\frac{s^2\log^3(p\vee T_wT_h)}{\xi^{2}_{\min}T_{w}T_h\underline\om}\Big\},\nn\\
		&\le& c_u\Big\{\Big(\frac{\si}{\xi_{\min}}\Big)\frac{s\log^{2} (p\vee T_wT_h)}{\surd(T_wT_h\om)}\Big\}^2
		\le c_{u1}.\nn
		\eenr
		Here the second inequality follows from the assumption $(\psi\big/\xi)\le \log(p\vee T_wT_h).$ The third inequality follows from Condition E. Substituting the bounds for terms $R1$ and $R2$ back in (\ref{eq:25a}) yields,
		\benr\label{eq:25aa}
		\max_{1\le j\le 4}\big\|\h\theta_{(j)}-\theta_{(j)}^0\big\|_2\le
		\frac{c_{u1}\xi_{\min}}{s^{1/2}\log (p\vee T_wT_h)}
		\eenr
		with probability at $1-o(1).$ Thus, the estimates $\h\theta_1$ and $\h\theta_2$ of Step 2 of Algorithm 1 satisfy all requirement of  Condition C(ii)(a,c). We now appeal to Theorem \ref{thm:cpoptimal} which yields Part (b) of this theorem.
		
		Nearly all ingredients for Part (c) are already available above. The only observation required here is that
		repeating the above arguments under the tightened rate assumption $\big[\{s\log^2(p\vee T_wT_h)\}\big/\surd(T_wT_h\underline\om)\big]=o(1),$ yields that Step 2 estimates $\h\tau$ and $\h\theta$ satisfy Condition C(i)(b) and C(ii)(a,c) with the additional requirement of Theorem \ref{thm:wc.vanishing} and Theorem \ref{thm:wc.non.vanishing}. The statement of Part (c) is now a direct consequence of these theorem's under respective jump size regimes. This completes the proof of this corollary.
	\end{proof}

	\bc$\rule{3.5in}{0.1mm}$\ec

	\section{Deviation bounds}\label{app:deviation}

	\begin{lemma}\label{lem:nearoptimalcross.subE.special} Suppose Condition A and B(i) holds and let $0\le v_{T_w}\le u_{T_w}\le 1,$ and $0\le v_{T_h}\le u_{T_h}\le 1,$ be any non-negative sequences. Then for any $c_u\ge 1,$ we have,
		\benr
		\sup_{\substack{\tau_w\in\cG_w(u_{T_w},v_{T_w});\\\tau_w\ge\tau^0_w}}\sup_{\substack{\tau_h\in\cG_h(u_{T_h},v_{T_h});\\\tau_h\ge\tau^0_h}}\Big\|\sum_{w=\tau^0_w+1}^{\tau_w}\sum_{h=\tau_h^0+1}^{\tau_h}\vep_{(w,h)}\Big\|_{\iny}\le 2c_u\si\log (p\vee T_wT_h)\surd \big(T_wu_{T_w}T_hu_{T_h}\big)\nn
		\eenr
		with probability at least $1-2\exp\{-(c_u-2)\log(p\vee T_wT_h)\}.$ 	
	\end{lemma}
	
	\begin{proof}[Proof of Lemma \ref{lem:nearoptimalcross.subE.special}] Without loss of generality assume $v_{T_w}\ge (1/T_w),$ and $v_{T_h}\ge (1/T_h),$ else, the sum of interest is over an empty set and is thus trivially zero. Now consider any $k\in\{1,2,...,p\}$ and any $\tau_w>\tau^0_w,$ $\tau_h>\tau_h^0$ and apply the Bernstein's inequality (Lemma \ref{lem:bernstein}) for any $d>0$ to obtain,
		\benr\label{eq:6}
		pr\Big(\big|\sum_{w=\tau^0_w+1}^{\tau_w}\sum_{h=\tau_h^0+1}^{\tau_h}\vep_{(w,h,k)}\big|>d(\tau_w-\tau^0_w)(\tau_h-\tau^0_h)\Big)\le\hspace{2cm}\nn\\
		2\exp\Big\{-\frac{1}{2}(\tau_w-\tau^0_w)(\tau_h-\tau^0_h)\Big(\frac{d^2}{\si^2}\wedge\frac{d}{\si}\Big)\Big\}.
		\eenr	
		Choose $d=2c_u\si\{\log^2 (p\vee T_wT_h)\big/\big((\tau_w-\tau^0_w)(\tau_h-\tau_h^0)\big)\}^{1/2},$ and note that,
		\benr\label{eq:1}
		(\tau_w-\tau^0_w)(\tau_h-\tau_h^0)\frac{d^2}{2\si^2}&=&2c_u^2\log^2(p\vee T_wT_h),\quad {\rm and},\nn\\
		(\tau_w-\tau^0_w)(\tau_h-\tau_h^0)\frac{d}{2\si}&\ge& c_u\log(p\vee T_wT_h),
		\eenr
		where we have used $(\tau_w-\tau^0_w)\ge T_wv_{T_w}\ge 1,$ $(\tau_h-\tau^0_h)\ge T_hv_{T_h}\ge 1,$  to obtain the first inequality. Thus, substituting this choice of $d$ in (\ref{eq:6}) and recalling that by choice $c_u\ge 1$, we obtain,
		\benr
		\Big|\sum_{w=\tau^0_w+1}^{\tau_w}\sum_{h=\tau_h^0+1}^{\tau_h}\vep_{(w,h,k)}\Big|&\le& 2c_u\si(\tau_w-\tau^0_w)^{1/2}(\tau_h-\tau^0_h)^{1/2}\{\log^2(p\vee T_wT_h)\}^{1/2}\nn\\
		&\le& 2c_u\si\{T_wu_{T_w}T_hu_{T_h}\log^2 (p\vee T_wT_h)\}^{1/2},\nn
		\eenr
		with probability at least $1-2\exp\{-c_u\log (p\vee T_wT_h)\}.$ The statement of this lemma follows by applying a union bound over $k=1,...,p,$ $\tau_w=1,...,T_w,$ and $\tau_h=1,...T_h.$
	\end{proof}

	\begin{remark} Lemma \ref{lem:nearoptimalcross.subE.special} provides a uniform bound over a collection with the restriction $\tau_w\ge\tau_w^0$ and $\tau_h\ge\tau_h^0.$ This restriction is considered for clarity of presentation of the proof. A more general result without this restriction can be obtained following identical arguments, specifically,
		\benr
		\sup_{\tau_w\in\cG_w(u_{T_w},v_{T_w})}\sup_{\tau_h\in\cG_h(u_{T_h},v_{T_h})}\Big\|\underset{(w,h)\in Q(\tau,\tau^0)}{\sum\sum}\vep_{(w,h)}\Big\|_{\iny}\le 2c_u\si\log (p\vee T_wT_h)\surd \big(T_wu_{T_w}T_hu_{T_h}\big)\nn
		\eenr
		with probability at least $1-2\exp\{-(c_u-2)\log(p\vee T_wT_h)\}.$ Here $Q(\tau,\tau^0)$ represents the quadrant of observations bounded by $\tau=(\tau_w,\tau_h)^T$ and $\tau^0=(\tau_w^0,\tau^0_h)^T,$ i.e.,
		\benr
		Q(\tau,\tau^0)=\big\{(w,h)\in\{\tau_w\wedge\tau_w^0,...,\tau_w\vee\tau_w^0\}\times\{\tau_h\wedge\tau_h^0,...,\tau_h\vee\tau_h^0\}\big\}.
		\eenr
		To prove this more general result one may proceed case by case based on the permutations of the orientation between $\tau=(\tau_w,\tau_h)^T$ and $\tau^0=(\tau_w^0,\tau^0_h)^T,$ and follow the same argument as the proof of Lemma \ref{lem:nearoptimalcross.subE.special}.
	\end{remark}
	
	\bc$\rule{3.5in}{0.1mm}$\ec

	\begin{lemma}\label{lem:optimalcross} Suppose Condition A and B(i) hold and let $u_{T_w},v_{T_w}$ be any non-negative sequences satisfying $0\le v_{T_w}\le u_{T_w}\le 1.$ Then for any $0<a<1,$ choosing $c_a\ge 4\surd(1/a),$ we have,
		\benr
		\sup_{\substack{\tau_w\in\cG_w(u_{T_w},v_{T_w});\\ \tau_w\ge\tau^0_w}}\Big|\sum_{w=\tau^0_w+1}^{\tau_w}\Big(\sum_{h=\tau^0_h+1}^{T_h}\vep_{(w,h)}^T\eta^0_{(1)}+\sum_{h=1}^{\tau^0_h}\vep_{w,h}^T\eta^0_{(3)}\Big)\Big|\le
		c_{a}\phi\xi_w\surd(T_wT_hu_{T_w}),\nn
		\eenr
		with probability at least $1-a.$	
	\end{lemma}

	\begin{proof} Begin by defining for any $w$ the r.v.,
		\benr
		\psi_{w}=\sum_{h=\tau^0_h+1}^{T_h}\vep_{(w,h)}^T\eta^0_{(1)}+\sum_{h=1}^{\tau^0_h}\vep_{(w,h)}^T\eta^0_{(3)}\nn
		\eenr
		Then from Lemma \ref{lem:lcsubE} we have $\psi_{w}\sim{\rm subE}(\la^2),$ where $\la^2=\phi^2\big\{(T_h-\tau^0_h)\xi_1^2+\tau^0_h\xi_3^2\}=\phi^2T_h\xi_w^2.$ Consequently from Lemma \ref{lem:momentprop} we also have ${\rm var}\psi_{w}\le 16\la^2.$ Next, we note that there are at most $T_wu_{T_w}$ values of $\tau_w$ in the set $\cG_w(u_{T_w},v_{T_w}),$ and then apply the Kolmogorov's inequality (Theorem \ref{thm:kolmogorov}) for any $d>0$ to obtain,
		\benr
		pr\Big(\sup_{\tau_w\in\cG_w(u_{T_w},v_{T_w})}\Big|\sum_{w=\tau_w^0+1}^{\tau_w}\psi_{w}\Big|>d\Big)\le 16\frac{T_wu_{T_w}}{d^2}\la^2.\nn
		\eenr
		choosing $d=c_a\phi\xi_w\surd(T_wT_hu_{T_w}),$ with $c_a\ge 4\surd(1/a)$ yields the result of the lemma.
	\end{proof}


	\bc$\rule{3.5in}{0.1mm}$\ec

	\begin{lemma}\label{lem:1.to.tau.bound} Assume Condition A and B(i) holds. 	Additionally assume for $c_u> 0$ that $c_uT_wT_h\underline\om\ge\log (p\vee T_wT_h).$ Then for any $c_{u1}>0,$ we have,
		\benr\label{eq:18}
		\max_{1\le j\le 4}\sup_{\substack{\tau_w\in\{1,.....,T\};\\ \substack{\tau_h\in\{1,.....,T\};\\
					|Q_j(\tau)|\ge c_uT_wT_h\underline\om}}}\Big\|\frac{1}{|Q_j(\tau)|}\underset{(w,h)\in Q_j(\tau)}{\sum\sum}\vep_{(w,h)}\Big\|_{\iny}\le  \si\Big\{\frac{2c_{u1}\log(p\vee T_wT_h)}{c_uT_wT_h\underline\om}\Big\}^{\frac{1}{2}}\nn
		\eenr
		with probability at least $1-8\exp\big\{-(c_{u2}-2)\log (p\vee T_wT_h)\big\},$ where $c_{u2}=c_{u1}\wedge \surd(c_uc_{u1}/2).$
	\end{lemma}

	\begin{proof}[Proof of Lemma \ref{lem:1.to.tau.bound}]
		First consider the case of $j=1,$ where $Q_1(\tau)=\big\{(w,h)\in \{\tau_w+1,...T_w\}\times\{\tau_h+1,...,T_h\}\big\},$ then we have $\sum_{w=\tau_w+1}^{T_w}\sum_{h=\tau_h+1}^{T_h}\vep_{(w,h,k)}\sim{\rm subE}\big(|Q_1(\tau)|\si^2\big).$  Now, applying Bernstein's inequality (Lemma \ref{lem:bernstein}) for any $d>0,$ we have,
		\benr\label{eq:19}
		pr\Big(\Big|\sum_{w=\tau_w+1}^{T_w}\sum_{h=\tau_h+1}^{T_h}\vep_{(w,h,k)}\Big|>d|Q_1(\tau)|\Big)\le 2\exp\Big\{-\frac{|Q_1(\tau)|}{2}\Big(\frac{d^2}{\si^2}\wedge\frac{d}{\si}\Big)\Big\}.
		\eenr	
		Choose $d=\si\{2c_{u1}\log (p\vee T_wT_h)\big/|Q_1(\tau)|\}^{1/2},$ and due to the assumption $|Q_1(\tau)|\ge c_uT_wT_h\underline\om\ge \log (p\vee T_wT_h),$ we have,
		\benr
		|Q_1(\tau)|\frac{d^2}{2\si^2}&=&c_{u1}\log(p\vee T_wT_h),\quad {\rm and},\nn\\
		|Q_1(\tau)|\frac{d}{2\si}&\ge&\surd(c_{u1}/2)(c_uT_wT_h\underline\om)^{1/2}\{\log(p\vee T_wT_h)\}^{1/2}\ge \surd(c_uc_{u1}/2)\log(p\vee T_wT_h).\nn
		\eenr	
		Now substituting this choice of $d$ in (\ref{eq:19}),  we obtain,	
		\benr
		\frac{1}{|Q_1(\tau)|}\Big|\sum_{w=\tau_w+1}^{T_w}\sum_{h=\tau_h+1}^{T_h}\vep_{(w,h,k)}\Big|\le \si\{2c_{u1}\log (p\vee T_wT_h)\big/|Q_1(\tau)|\}^{1/2}\le\si\Big\{\frac{2c_{u1}\log(p\vee T_wT_h)}{c_uT_wT_h\underline\om}\Big\}^{1/2}\nn
		\eenr
		with probability at least $1-2\exp\{-c_{u2}\log (p\vee T_wT_h)\},$ where $c_{u2}=c_{u1}\wedge\surd(c_uc_{u1}/2).$ Uniformity of the inner collection in the lemma follows by applying union bounds over all values of $\tau_w,\tau_h$ and $k.$ Uniformity over $j=1,2,3,4$ can be obtained by proceeding with identical arguments as above for each respective quadrant to obtain the same upper bound and finally applying a union bound to obtain the statement of the lemma.
	\end{proof}
	\bc$\rule{3.5in}{0.1mm}$\ec

	\section{Definitions and auxiliary results}\label{app:auxiliary}
	
	The following definition's and results provide basic properties of subexponential distributions. These are largely reproduced from \cite{vershynin2019high} and \cite{rigollet201518}. Theorem \ref{thm:kolmogorov} and \ref{thm:argmax} below reproduce the Kolmogorov's inequality and the argmax theorem. We also refer to Appendix B and Appendix F of \cite{kaul2020inference} and \cite{kaul2021graphical}, respectively, where these results and some additional proofs have been compiled.

	\begin{definition}\label{def:sube}[Subexponential r.v.] A random variable $X\in\R$ is said to be sub-exponential with parameter $\si^2>0$ \big(denoted by $X\sim{\rm subE(\si^2)}$\big) if $E(X)=0$ and its moment generating function
		\benr
		E(\e^{tX})\le \e^{t^2\si^2/2},\qquad \forall\,\, |t|\le \frac{1}{\si}\nn
		\eenr
	\end{definition}
	
	\begin{definition}\label{def:submult} A random vector $X\in\R^p$ shall said to be subexponential with parameter $\si^2,$ if the inner product $\langle X, v\rangle\sim {\rm subE}(\si^2),$ respectively, for any $v\in\R^p$ with $\|v\|_2 = 1.$
	\end{definition}
	
	Following is the elementary definition of uniform tightness of a sequence of random variables reproduced from Page 166, Chapter 2 of \cite{durrett2010probability}.
	\begin{definition}\label{def:utight} A sequence of random variables $X_n$ is said to be uniformly tight if for every $\ep>0,$ there is a compact set $K$ such that $pr(X_n\in K)>1-\ep.$
	\end{definition}

	\begin{lemma}\label{lem:tailb}[Tail bounds] If $X\sim {\rm subE}(\si^2),$ then
		\benr
		pr(|X|\ge \la)\le 2\exp\Big\{-\frac{1}{2}\Big(\frac{\la^2}{\si^2}\wedge\frac{\la}{\si}\Big) \Big\}.\nn
		\eenr
	\end{lemma}

	\begin{lemma}[Moment bounds]\label{lem:momentprop} If $X\sim {\rm subE}(\si^2),$ then
		\benr
		E|X|^k\le 4\si^k k^k, \qquad k> 0.\nn
		\eenr	
	\end{lemma}

	\begin{lemma}\label{lem:lcsubE} Assume that $X\sim{\rm subE(\si^2)},$ and that $\al\in\R,$ then $\alpha X\sim{\rm subE}(\alpha^2\si^2).$ Moreover, assume that $X_1\sim{\rm subE(\si_1^2)}$ and $X_2\sim{\rm subE(\si_2^2)},$ then $X_1+X_2\sim{\rm subE((\si_1+\si_2)^2)},$ additionally, if $X_1$ and $X_2$ are independent, then $X_1+X_2\sim{\rm subE(\si_1^2+\si_2^2)}.$
	\end{lemma}

	\begin{lemma}[Bernstein's inequality]\label{lem:bernstein} Let $X_1,X_2,...,X_T$ be independent random variables such that $X_t\sim {\rm subE}(\la^2).$ Then for any $d>0$ we have,
		\benr
		pr(|\bar X|>d)\le 2\exp\Big\{-\frac{T}{2}\Big(\frac{d^2}{\la^2}\wedge \frac{d}{\la}\Big)\Big\}\nn
		\eenr
	\end{lemma}

	The next result is the Kolmogorov's inequality reproduced from \cite{hajek1955generalization}
	\begin{theorem}[Kolmogorov's inequality]\label{thm:kolmogorov} If $\xi_1,\xi_2,...$ is a sequence of mutually independent random variables with mean values $E(\xi_k)=0$ and finite variance ${\rm var}(\xi_k)=D_k^2$ $(k=1,2,...),$ we have, for any $\vep>0,$
		\benr
		pr\Big(\max_{1\le k\le m}\big|\xi_1+\xi_2+...+\xi_k\big|>\vep\Big)\le \frac{1}{\vep^2}\sum_{k=1}^mD_k^2\nn
		\eenr	
	\end{theorem}

	Following is the well known `Argmax' theorem reproduced from Theorem 3.2.2 of \cite{vaart1996weak}.
	\begin{theorem}[Argmax Theorem]\label{thm:argmax} Let $\cM_n,\cM$ be stochastic processes indexed by a metric space $H$ such that $\cM_n\Rightarrow\cM$ in $\ell^{\iny}(K)$ for every compact set $K\subseteq H$. Suppose that almost all sample paths $h\to \cM(h)$ are upper semicontinuous and posses a unique maximum at a (random) point $\h h,$ which as a random map in $H$ is tight. If the sequence $\h h_n$ is uniformly tight and satisfies $\cM_n(\h h_n)\ge \sup_h \cM_n(h)-o_p(1),$ then $\h h_n\Rightarrow \h h$ in $H.$
	\end{theorem}

\section{Additional details and results}\label{app:numerical.supplement}
	
This section contains is divided in three subsections. Sub-section \ref{subsec:app:application} provides additional results of image denoising carried out in Section \ref{sec:application}. Sub-section \ref{subsec:app:simulations} provides remaining simulation results of Section \ref{sec:numerical}. Finally, sub-section \ref{subsec:app:par.est} provides details pertaining to estimation of additional parameters such as drifts and asymptotic variances, as well as on computation of quantiles which are in turn necessary for computation of confidence intervals presented in Section \ref{sec:application} and Section \ref{sec:numerical}.

\subsection{Additional results of Section \ref{sec:application}}\label{subsec:app:application}		
	
\begin{figure}[H]
		\includegraphics[width=.2\textwidth]{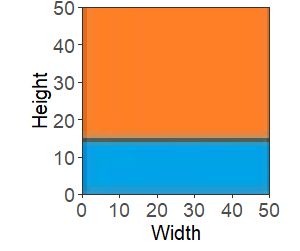}\qquad
		\includegraphics[width=.2\textwidth]{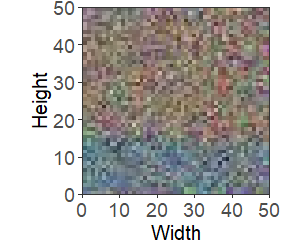}\qquad
		\includegraphics[width=.2\textwidth]{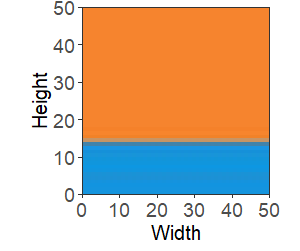}
		\\[\smallskipamount]
		\includegraphics[width=.2\textwidth]{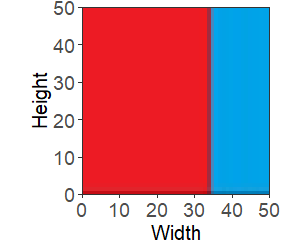}\qquad
		\includegraphics[width=.2\textwidth]{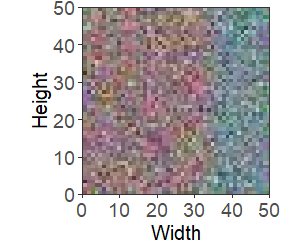}\qquad
		\includegraphics[width=.2\textwidth]{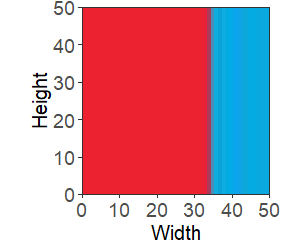}
		\\[\smallskipamount]
		\includegraphics[width=.2\textwidth]{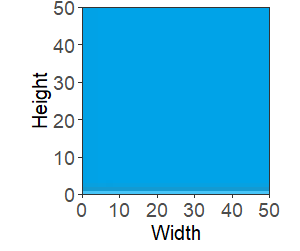}\qquad
		\includegraphics[width=.2\textwidth]{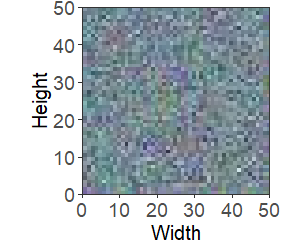}\qquad
		\includegraphics[width=.2\textwidth]{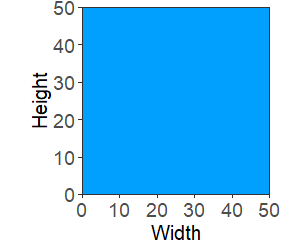}
		\caption{\footnotesize{Image denoising with Algorithm \ref{alg:quarter.seg}. Images are $50\times 50$ pixels with at most one $2d$-change point. {\it Left panels:} True images (unobserved), {\it Center panels:} Noisy images (observed), {\it Right panels:} Recovered images. Noise set to $\Si=I_{3\times 3},$ and tuning constant $c_{bic}=1.$}}
		\label{fig:ex1.single.supp}
	\end{figure}

\begin{figure}[H]
		\includegraphics[width=.25\textwidth]{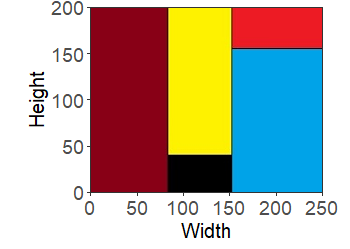}\qquad
		\includegraphics[width=.25\textwidth]{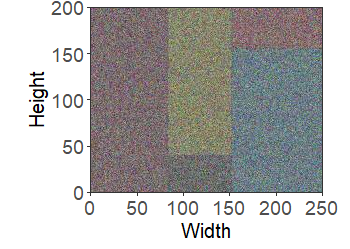}\qquad
		\includegraphics[width=.25\textwidth]{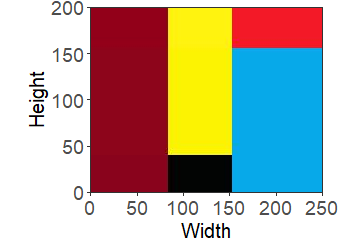}
		\caption{\footnotesize{Image denoising with Algorithm \ref{alg:quarter.seg}. Image is $250\times 200$ pixels with one $2d$-change point. {\it Left panel:} True image (unobserved), {\it Center panel:} Noisy image (observed), {\it Right panel:} Recovered image. Estimated model recovered with $\ell=2$ hierarchical change points inducing a total of $8$ disjoint partitions of the sampling space. Noise set to $\Si=I_{3\times 3},$ and tuning constant $c_{bic}=1.$}}
		\label{fig:ex1.multiple.supp}
	\end{figure}

\begin{figure}[H]
		\includegraphics[width=.33\textwidth]{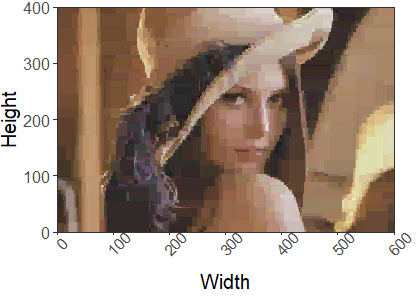}\qquad
		\includegraphics[width=.33\textwidth]{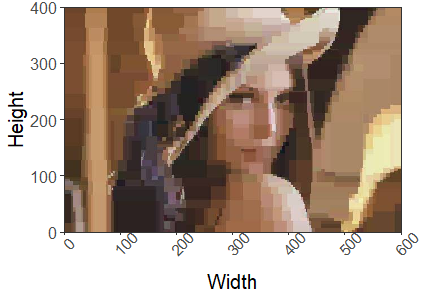}
		\caption{\footnotesize{Image denoising with Algorithm \ref{alg:quarter.seg}. Image is $600\times 400$ pixels. {\it Left panel:} Recovered image with $c_{bic}=0.5,$ estimated model recovered with $\ell=11$ hierarchical change points inducing a total of $2569$ disjoint partitions of the sampling space. {\it Right panel:}  Recovered image with $c_{bic}=1,$ estimated model recovered with $\ell=11$ hierarchical change points inducing a total of $1278$ disjoint partitions of the sampling space.  Noise set to $\Si=0.05\cdotp I_{3\times 3},$ see, Figure \ref{fig:ex2.lena} for true and noisy images.}}
		\label{fig:ex2.lena.supp}
	\end{figure}

\begin{figure}[H]
		\includegraphics[width=.33\textwidth]{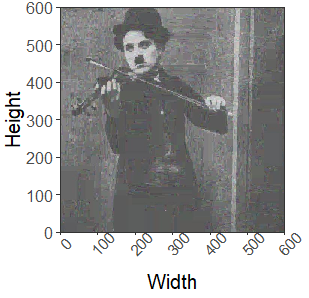}\qquad
		\includegraphics[width=.33\textwidth]{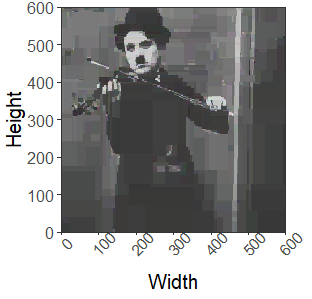}
		\caption{\footnotesize{Image denoising with Algorithm \ref{alg:quarter.seg}. Image is $600\times 600$ pixels. {\it Left panel:} Recovered image with $c_{bic}=0.25,$ estimated model recovered with $\ell=15$ hierarchical change points inducing a total of $15952$ disjoint partitions of the sampling space. {\it Right panel:}  Recovered image with $c_{bic}=1,$ estimated model recovered with $\ell=13$ hierarchical change points inducing a total of $1401$ disjoint partitions of the sampling space. Noise set to $\Si=0.05\cdotp I_{3\times 3},$ see, Figure \ref{fig:ex2.chaplin} for true and noisy images.}}
		\label{fig:ex2.chaplin.supp}
	\end{figure}

Here we provide the remaining results of image denoising application described in sub-Section \ref{subsec:image}. These are provided in Figure \ref{fig:ex1.single.supp} (single change point synthetic examples), Figure \ref{fig:ex1.multiple.supp} (multiple change point synthetic examples), Figure \ref{fig:ex2.lena.supp} (lena image) and Figure \ref{fig:ex2.chaplin.supp} (Chaplin image).	
	
	

	

\subsection{Additional simulation results of Section \ref{sec:numerical}}\label{subsec:app:simulations}

Below are the additional results of the simulation designs described in Section \ref{sec:numerical}. Table \ref{tab:hres1} -Table \ref{tab:hres4} provide results in context of estimation of the height change parameter $\tau_h,$ these results are under symmetrical designs to those provided in context of the width change parameter $\tau_w$ in Section \ref{sec:numerical}. A large number of other cases were also evaluated the results of which were in accordance to the discussion of Section \ref{sec:numerical} and these are omitted to avoid redundancy.	
	\begin{table}[H]	
		\centering
		\resizebox{0.8\textwidth}{!}{
			\begin{tabular}{llllllll}
				\toprule
				\multicolumn{2}{c}{\multirow{2}{*}{\begin{tabular}[c]{@{}c@{}}$T_w=30,$\\ $\tau^0_w/T_w=0.2$\end{tabular}}} & \multicolumn{3}{c}{$p=10$}                                                                                           & \multicolumn{3}{c}{$p=50$}                                                                                           \\ \cmidrule{3-8}
				\multicolumn{2}{c}{}                                                                                        & \multicolumn{1}{c}{\multirow{2}{*}{bias (rmse)}} & \multicolumn{2}{c}{coverage (av. ME)}                             & \multicolumn{1}{c}{\multirow{2}{*}{bias (rmse)}} & \multicolumn{2}{c}{coverage (av. ME)}                             \\ \cmidrule{1-2} \cmidrule{4-5} \cmidrule{7-8}
				\multicolumn{1}{c}{$\tau^0_h/T_h$}                        & \multicolumn{1}{c}{$T_h$}                       & \multicolumn{1}{c}{}                             & \multicolumn{1}{c}{Vanishing} & \multicolumn{1}{c}{Non-Vanishing} & \multicolumn{1}{c}{}                             & \multicolumn{1}{c}{Vanishing} & \multicolumn{1}{c}{Non-Vanishing} \\ \midrule
				0.2                                                       & 30                                              & 0.122 (0.574)                                    & 0.912 (0.527)                 & 0.926 (0.07)                      & 0.142 (0.581)                                    & 0.9 (0.465)                   & 0.904 (0.018)                     \\
				0.2                                                       & 35                                              & 0.03 (0.205)                                     & 0.958 (0.514)                 & 0.968 (0.05)                      & 0.168 (1.251)                                    & 0.916 (0.483)                 & 0.922 (0.018)                     \\
				0.2                                                       & 40                                              & 0.012 (0.126)                                    & 0.984 (0.515)                 & 0.984 (0.028)                     & 0.116 (0.429)                                    & 0.912 (0.486)                 & 0.916 (0.018)                     \\
				0.2                                                       & 45                                              & 0.032 (0.268)                                    & 0.952 (0.529)                 & 0.956 (0.038)                     & 0.078 (0.366)                                    & 0.936 (0.498)                 & 0.936 (0.016)                     \\ \midrule
				0.4                                                       & 30                                              & 0.042 (0.293)                                    & 0.95 (0.5)                    & 0.954 (0.03)                      & 0.138 (0.76)                                     & 0.928 (0.505)                 & 0.932 (0.012)                     \\
				0.4                                                       & 35                                              & 0.018 (0.205)                                    & 0.964 (0.518)                 & 0.966 (0.014)                     & 0.076 (0.663)                                    & 0.95 (0.502)                  & 0.954 (0.01)                      \\
				0.4                                                       & 40                                              & 0.004 (0.179)                                    & 0.968 (0.532)                 & 0.968 (0.02)                      & 0.04 (0.228)                                     & 0.966 (0.511)                 & 0.966 (0.008)                     \\
				0.4                                                       & 45                                              & 0.008 (0.219)                                    & 0.958 (0.529)                 & 0.962 (0.022)                     & 0.036 (0.2)                                      & 0.96 (0.506)                  & 0.962 (0.006)                     \\ \midrule
				0.6                                                       & 30                                              & 0.012 (0.245)                                    & 0.956 (0.487)                 & 0.958 (0.024)                     & 0.052 (0.329)                                    & 0.946 (0.458)                 & 0.946 (0)                         \\
				0.6                                                       & 35                                              & 0.018 (0.241)                                    & 0.96 (0.522)                  & 0.96 (0.022)                      & 0.048 (0.261)                                    & 0.956 (0.463)                 & 0.956 (0)                         \\
				0.6                                                       & 40                                              & 0 (0.253)                                        & 0.958 (0.526)                 & 0.96 (0.012)                      & 0.05 (0.3)                                       & 0.95 (0.478)                  & 0.954 (0.004)                     \\
				0.6                                                       & 45                                              & 0.022 (0.338)                                    & 0.94 (0.53)                   & 0.948 (0.022)                     & 0.028 (0.237)                                    & 0.95 (0.482)                  & 0.952 (0.002)                     \\ \midrule
				0.8                                                       & 30                                              & 0.016 (0.502)                                    & 0.988 (0.483)                 & 0.988 (0.036)                     & 0.022 (0.784)                                    & 0.968 (0.384)                 & 0.968 (0.004)                     \\
				0.8                                                       & 35                                              & 0.042 (0.241)                                    & 0.96 (0.516)                  & 0.962 (0.022)                     & 0.034 (0.911)                                    & 0.968 (0.408)                 & 0.968 (0.002)                     \\
				0.8                                                       & 40                                              & 0.04 (0.303)                                     & 0.942 (0.517)                 & 0.956 (0.036)                     & 0.014 (0.148)                                    & 0.978 (0.415)                 & 0.978 (0)                         \\
				0.8                                                       & 45                                              & 0.03 (0.249)                                     & 0.966 (0.522)                 & 0.97 (0.032)                      & 0 (0.19)                                         & 0.97 (0.422)                  & 0.97 (0)                          \\ \bottomrule
		\end{tabular}}
		\vspace{1mm}
		\caption{\footnotesize{Simulation results for estimation of $\tau^0_h$ based on 500 replications. All reported metrics rounded to three decimals. Other data generating parameters: $T_w=30,$ $\tau^0_w=\lfloor 0.2\cdotp T_w\rfloor$ and $p\in\{10,50\}.$}}
		\label{tab:hres1}
	\end{table}
	
	\begin{table}[H]
		\centering
		\resizebox{0.8\textwidth}{!}{
			\begin{tabular}{llllllll}
				\toprule
				\multicolumn{2}{c}{\multirow{2}{*}{\begin{tabular}[c]{@{}c@{}}$T_w=30,$\\ $\tau^0_w/T_w=0.2$\end{tabular}}} & \multicolumn{3}{c}{$p=100$}                                                                                          & \multicolumn{3}{c}{$p=250$}                                                                                          \\ \cmidrule{3-8}
				\multicolumn{2}{c}{}                                                                                        & \multicolumn{1}{c}{\multirow{2}{*}{bias (rmse)}} & \multicolumn{2}{c}{coverage (av. ME)}                             & \multicolumn{1}{c}{\multirow{2}{*}{bias (rmse)}} & \multicolumn{2}{c}{coverage (av. ME)}                             \\ \cmidrule{1-2} \cmidrule{4-5} \cmidrule{7-8}
				\multicolumn{1}{c}{$\tau^0_h/T_h$}                        & \multicolumn{1}{c}{$T_h$}                       & \multicolumn{1}{c}{}                             & \multicolumn{1}{c}{Vanishing} & \multicolumn{1}{c}{Non-Vanishing} & \multicolumn{1}{c}{}                             & \multicolumn{1}{c}{Vanishing} & \multicolumn{1}{c}{Non-Vanishing} \\ \midrule
				0.2                                                       & 30                                              & 0.276 (1.346)                                    & 0.854 (0.444)                 & 0.854 (0.018)                     & 0.514 (2.32)                                     & 0.808 (0.422)                 & 0.81 (0.03)                       \\
				0.2                                                       & 35                                              & 0.212 (1.397)                                    & 0.896 (0.457)                 & 0.904 (0.02)                      & 0.17 (0.828)                                     & 0.892 (0.415)                 & 0.892 (0.01)                      \\
				0.2                                                       & 40                                              & 0.158 (0.553)                                    & 0.884 (0.464)                 & 0.884 (0.01)                      & 0.28 (1.114)                                     & 0.86 (0.443)                  & 0.86 (0.022)                      \\
				0.2                                                       & 45                                              & 0.124 (1.043)                                    & 0.926 (0.473)                 & 0.928 (0.012)                     & 0.144 (0.678)                                    & 0.898 (0.441)                 & 0.898 (0.006)                     \\ \midrule
				0.4                                                       & 30                                              & 0.134 (0.677)                                    & 0.918 (0.492)                 & 0.92 (0.016)                      & 0.216 (0.984)                                    & 0.894 (0.469)                 & 0.896 (0.012)                     \\
				0.4                                                       & 35                                              & 0.074 (0.564)                                    & 0.94 (0.489)                  & 0.94 (0.006)                      & 0.078 (0.387)                                    & 0.938 (0.474)                 & 0.938 (0)                         \\
				0.4                                                       & 40                                              & 0.05 (0.319)                                     & 0.952 (0.498)                 & 0.952 (0)                         & 0.108 (0.756)                                    & 0.94 (0.483)                  & 0.94 (0.004)                      \\
				0.4                                                       & 45                                              & 0.052 (0.261)                                    & 0.954 (0.509)                 & 0.954 (0.002)                     & 0.09 (0.332)                                     & 0.908 (0.486)                 & 0.908 (0)                         \\ \midrule
				0.6                                                       & 30                                              & 0.05 (0.307)                                     & 0.948 (0.455)                 & 0.948 (0.002)                     & 0.048 (0.303)                                    & 0.966 (0.438)                 & 0.966 (0)                         \\
				0.6                                                       & 35                                              & 0.09 (0.553)                                     & 0.94 (0.465)                  & 0.94 (0.006)                      & 0.06 (0.385)                                     & 0.948 (0.449)                 & 0.948 (0)                         \\
				0.6                                                       & 40                                              & 0.038 (0.214)                                    & 0.966 (0.469)                 & 0.966 (0)                         & 0.048 (0.31)                                     & 0.948 (0.46)                  & 0.948 (0)                         \\
				0.6                                                       & 45                                              & 0.034 (0.286)                                    & 0.954 (0.475)                 & 0.954 (0.002)                     & 0.056 (0.322)                                    & 0.95 (0.467)                  & 0.95 (0)                          \\ \midrule
				0.8                                                       & 30                                              & 0.032 (0.93)                                     & 0.958 (0.353)                 & 0.958 (0.006)                     & 0.002 (0.504)                                    & 0.972 (0.303)                 & 0.972 (0.002)                     \\
				0.8                                                       & 35                                              & 0.02 (0.932)                                     & 0.948 (0.362)                 & 0.948 (0.002)                     & 0.028 (0.268)                                    & 0.946 (0.317)                 & 0.946 (0)                         \\
				0.8                                                       & 40                                              & 0.006 (0.118)                                    & 0.986 (0.377)                 & 0.986 (0)                         & 0.02 (0.2)                                       & 0.96 (0.334)                  & 0.96 (0)                          \\
				0.8                                                       & 45                                              & 0.016 (0.21)                                     & 0.968 (0.387)                 & 0.968 (0.002)                     & 0.082 (1.477)                                    & 0.952 (0.354)                 & 0.952 (0.006)                     \\ \bottomrule
		\end{tabular}}
		\vspace{1mm}
		\caption{\footnotesize{Simulation results for estimation of $\tau^0_h$ based on 500 replications. All reported metrics rounded to three decimals. Other data generating parameters: $T_w=30,$ $\tau^0_w=\lfloor 0.2\cdotp T_w\rfloor$ and $p\in\{100,250\}.$}}
		\label{tab:hres2}
	\end{table}

	\begin{table}[H]
		\centering
		\resizebox{0.8\textwidth}{!}{
			\begin{tabular}{llllllll}
				\toprule
				\multicolumn{2}{c}{\multirow{2}{*}{\begin{tabular}[c]{@{}c@{}}$T_w=30,$\\ $\tau^0_w/T_w=0.4$\end{tabular}}} & \multicolumn{3}{c}{$p=10$}                                                                                           & \multicolumn{3}{c}{$p=50$}                                                                                           \\ \cmidrule{3-8}
				\multicolumn{2}{c}{}                                                                                        & \multicolumn{1}{c}{\multirow{2}{*}{bias (rmse)}} & \multicolumn{2}{c}{coverage (av. ME)}                             & \multicolumn{1}{c}{\multirow{2}{*}{bias (rmse)}} & \multicolumn{2}{c}{coverage (av. ME)}                             \\ \cmidrule{1-2} \cmidrule{4-5} \cmidrule{7-8}
				\multicolumn{1}{c}{$\tau^0_h/T_h$}                        & \multicolumn{1}{c}{$T_h$}                       & \multicolumn{1}{c}{}                             & \multicolumn{1}{c}{Vanishing} & \multicolumn{1}{c}{Non-Vanishing} & \multicolumn{1}{c}{}                             & \multicolumn{1}{c}{Vanishing} & \multicolumn{1}{c}{Non-Vanishing} \\ \midrule
				0.2                                                       & 30                                              & 0.122 (0.574)                                    & 0.912 (0.527)                 & 0.926 (0.07)                      & 0.142 (0.581)                                    & 0.9 (0.465)                   & 0.904 (0.018)                     \\
				0.2                                                       & 35                                              & 0.03 (0.205)                                     & 0.958 (0.514)                 & 0.968 (0.05)                      & 0.168 (1.251)                                    & 0.916 (0.483)                 & 0.922 (0.018)                     \\
				0.2                                                       & 40                                              & 0.012 (0.126)                                    & 0.984 (0.515)                 & 0.984 (0.028)                     & 0.116 (0.429)                                    & 0.912 (0.486)                 & 0.916 (0.018)                     \\
				0.2                                                       & 45                                              & 0.032 (0.268)                                    & 0.952 (0.529)                 & 0.956 (0.038)                     & 0.078 (0.366)                                    & 0.936 (0.498)                 & 0.936 (0.016)                     \\ \midrule
				0.4                                                       & 30                                              & 0.042 (0.293)                                    & 0.95 (0.5)                    & 0.954 (0.03)                      & 0.138 (0.76)                                     & 0.928 (0.505)                 & 0.932 (0.012)                     \\
				0.4                                                       & 35                                              & 0.018 (0.205)                                    & 0.964 (0.518)                 & 0.966 (0.014)                     & 0.076 (0.663)                                    & 0.95 (0.502)                  & 0.954 (0.01)                      \\
				0.4                                                       & 40                                              & 0.004 (0.179)                                    & 0.968 (0.532)                 & 0.968 (0.02)                      & 0.04 (0.228)                                     & 0.966 (0.511)                 & 0.966 (0.008)                     \\
				0.4                                                       & 45                                              & 0.008 (0.219)                                    & 0.958 (0.529)                 & 0.962 (0.022)                     & 0.036 (0.2)                                      & 0.96 (0.506)                  & 0.962 (0.006)                     \\ \midrule
				0.6                                                       & 30                                              & 0.012 (0.245)                                    & 0.956 (0.487)                 & 0.958 (0.024)                     & 0.052 (0.329)                                    & 0.946 (0.458)                 & 0.946 (0)                         \\
				0.6                                                       & 35                                              & 0.018 (0.241)                                    & 0.96 (0.522)                  & 0.96 (0.022)                      & 0.048 (0.261)                                    & 0.956 (0.463)                 & 0.956 (0)                         \\
				0.6                                                       & 40                                              & 0 (0.253)                                        & 0.958 (0.526)                 & 0.96 (0.012)                      & 0.05 (0.3)                                       & 0.95 (0.478)                  & 0.954 (0.004)                     \\
				0.6                                                       & 45                                              & 0.022 (0.338)                                    & 0.94 (0.53)                   & 0.948 (0.022)                     & 0.028 (0.237)                                    & 0.95 (0.482)                  & 0.952 (0.002)                     \\ \midrule
				0.8                                                       & 30                                              & 0.016 (0.502)                                    & 0.988 (0.483)                 & 0.988 (0.036)                     & 0.022 (0.784)                                    & 0.968 (0.384)                 & 0.968 (0.004)                     \\
				0.8                                                       & 35                                              & 0.042 (0.241)                                    & 0.96 (0.516)                  & 0.962 (0.022)                     & 0.034 (0.911)                                    & 0.968 (0.408)                 & 0.968 (0.002)                     \\
				0.8                                                       & 40                                              & 0.04 (0.303)                                     & 0.942 (0.517)                 & 0.956 (0.036)                     & 0.014 (0.148)                                    & 0.978 (0.415)                 & 0.978 (0)                         \\
				0.8                                                       & 45                                              & 0.03 (0.249)                                     & 0.966 (0.522)                 & 0.97 (0.032)                      & 0 (0.19)                                         & 0.97 (0.422)                  & 0.97 (0)                          \\ \bottomrule
		\end{tabular}}
		\vspace{1mm}
		\caption{\footnotesize{Simulation results for estimation of $\tau^0_h$ based on 500 replications. All reported metrics rounded to three decimals. Other data generating parameters: $T_w=30,$ $\tau^0_w=\lfloor 0.4\cdotp T_w\rfloor$ and $p\in\{10,50\}.$}}
		\label{tab:hres3}
	\end{table}

	\begin{table}[H]
		\centering
		\resizebox{0.8\textwidth}{!}{
			\begin{tabular}{llllllll}
				\toprule
				\multicolumn{2}{c}{\multirow{2}{*}{\begin{tabular}[c]{@{}c@{}}$T_w=30,$\\ $\tau^0_w/T_w=0.4$\end{tabular}}} & \multicolumn{3}{c}{$p=100$}                                                                                          & \multicolumn{3}{c}{$p=250$}                                                                                          \\ \cmidrule{3-8}
				\multicolumn{2}{c}{}                                                                                        & \multicolumn{1}{c}{\multirow{2}{*}{bias (rmse)}} & \multicolumn{2}{c}{coverage (av. ME)}                             & \multicolumn{1}{c}{\multirow{2}{*}{bias (rmse)}} & \multicolumn{2}{c}{coverage (av. ME)}                             \\ \cmidrule{1-2} \cmidrule{4-5} \cmidrule{7-8}
				\multicolumn{1}{c}{$\tau^0_h/T_h$}                        & \multicolumn{1}{c}{$T_h$}                       & \multicolumn{1}{c}{}                             & \multicolumn{1}{c}{Vanishing} & \multicolumn{1}{c}{Non-Vanishing} & \multicolumn{1}{c}{}                             & \multicolumn{1}{c}{Vanishing} & \multicolumn{1}{c}{Non-Vanishing} \\ \midrule
				0.2                                                       & 30                                              & 0.276 (1.346)                                    & 0.854 (0.444)                 & 0.854 (0.018)                     & 0.514 (2.32)                                     & 0.808 (0.422)                 & 0.81 (0.03)                       \\
				0.2                                                       & 35                                              & 0.212 (1.397)                                    & 0.896 (0.457)                 & 0.904 (0.02)                      & 0.17 (0.828)                                     & 0.892 (0.415)                 & 0.892 (0.01)                      \\
				0.2                                                       & 40                                              & 0.158 (0.553)                                    & 0.884 (0.464)                 & 0.884 (0.01)                      & 0.28 (1.114)                                     & 0.86 (0.443)                  & 0.86 (0.022)                      \\
				0.2                                                       & 45                                              & 0.124 (1.043)                                    & 0.926 (0.473)                 & 0.928 (0.012)                     & 0.144 (0.678)                                    & 0.898 (0.441)                 & 0.898 (0.006)                     \\ \midrule
				0.4                                                       & 30                                              & 0.134 (0.677)                                    & 0.918 (0.492)                 & 0.92 (0.016)                      & 0.216 (0.984)                                    & 0.894 (0.469)                 & 0.896 (0.012)                     \\
				0.4                                                       & 35                                              & 0.074 (0.564)                                    & 0.94 (0.489)                  & 0.94 (0.006)                      & 0.078 (0.387)                                    & 0.938 (0.474)                 & 0.938 (0)                         \\
				0.4                                                       & 40                                              & 0.05 (0.319)                                     & 0.952 (0.498)                 & 0.952 (0)                         & 0.108 (0.756)                                    & 0.94 (0.483)                  & 0.94 (0.004)                      \\
				0.4                                                       & 45                                              & 0.052 (0.261)                                    & 0.954 (0.509)                 & 0.954 (0.002)                     & 0.09 (0.332)                                     & 0.908 (0.486)                 & 0.908 (0)                         \\ \midrule
				0.6                                                       & 30                                              & 0.05 (0.307)                                     & 0.948 (0.455)                 & 0.948 (0.002)                     & 0.048 (0.303)                                    & 0.966 (0.438)                 & 0.966 (0)                         \\
				0.6                                                       & 35                                              & 0.09 (0.553)                                     & 0.94 (0.465)                  & 0.94 (0.006)                      & 0.06 (0.385)                                     & 0.948 (0.449)                 & 0.948 (0)                         \\
				0.6                                                       & 40                                              & 0.038 (0.214)                                    & 0.966 (0.469)                 & 0.966 (0)                         & 0.048 (0.31)                                     & 0.948 (0.46)                  & 0.948 (0)                         \\
				0.6                                                       & 45                                              & 0.034 (0.286)                                    & 0.954 (0.475)                 & 0.954 (0.002)                     & 0.056 (0.322)                                    & 0.95 (0.467)                  & 0.95 (0)                          \\ \midrule
				0.8                                                       & 30                                              & 0.032 (0.93)                                     & 0.958 (0.353)                 & 0.958 (0.006)                     & 0.002 (0.504)                                    & 0.972 (0.303)                 & 0.972 (0.002)                     \\
				0.8                                                       & 35                                              & 0.02 (0.932)                                     & 0.948 (0.362)                 & 0.948 (0.002)                     & 0.028 (0.268)                                    & 0.946 (0.317)                 & 0.946 (0)                         \\
				0.8                                                       & 40                                              & 0.006 (0.118)                                    & 0.986 (0.377)                 & 0.986 (0)                         & 0.02 (0.2)                                       & 0.96 (0.334)                  & 0.96 (0)                          \\
				0.8                                                       & 45                                              & 0.016 (0.21)                                     & 0.968 (0.387)                 & 0.968 (0.002)                     & 0.082 (1.477)                                    & 0.952 (0.354)                 & 0.952 (0.006)                     \\ \bottomrule
		\end{tabular}}
		\vspace{1mm}
		\caption{\footnotesize{Simulation results for estimation of $\tau^0_h$ based on 500 replications. All reported metrics rounded to three decimals. Other data generating parameters: $T_w=30,$ $\tau^0_w=\lfloor 0.4\cdotp T_w\rfloor$ and $p\in\{100,250\}.$}}
		\label{tab:hres4}
	\end{table}

\subsection{Estimation of drifts, asymptotic variances and quantiles}\label{subsec:app:par.est}

We begin with a discussion on the estimation of $\xi_{w},$ $\xi_{h},$ and $\si^2_{(w,\iny)},\si^2_{(h,\iny)}$ which  utilized for the computation of confidence intervals for $\tau^0=(\tau^0_w,\tau^0_h)^T$ using the result of Theorem \ref{thm:wc.vanishing} and Theorem \ref{thm:wc.non.vanishing}. To avoid redundancy, we only describe part of this estimation process in context of the width change parameter $\tau^0_w,$ the procedure for the height change parameter is symmetrical.

First, in order to alleviate finite sample regularization biases we utilize refitted mean estimates computed as $\breve\theta_{(j)}=\big[\bar x_{(j)}(\tilde\tau)\big]_{\h S_j}$ $j=1,2,3,4,$ where $\tilde\tau=(\tilde\tau_w,\tilde\tau_h)^T$ is the change point estimate of Algorithm 1. Here $\h S_j=\{k\,\,\h\theta_{(1k)}\ne 0\},$ $j=1,2,3,4,$ are the estimated sparsity sets, where $\h\theta_{(j)},$ $j=1,2,3,4$ are the Step 2 mean estimates of Algorithm 1. It is well known in the literature that refitted mean estimates preserve the rate of convergence of the regularized version while reducing finite sample biases, see, e.g. \cite{belloni2011square} and \cite{belloni2017pivotal}. The jump vector and individual jump size are then evaluated as $\breve\eta_{(j)}$ and $\breve\xi_j,$ $j=1,2,3,4,$ as plug-in estimates as per the relations (\ref{def:jump.size.quad}). The width and height-wise proportions are estimated as the observed versions, i.e., $\breve\om_w=(T_w-\tilde\tau^0_w)/T_w$ and $\breve\om_h=(T_h-\tilde\tau_h)/T_h.$ The directional jump sizes are obtained as $\breve\xi_w=\breve\om_h\breve\xi_1^2+(1-\breve\om_h)\breve\xi_4^2$ and symmetrically for $\breve\xi_h.$

Next consider the asymptotic variance $\si^2_{(w,\iny)}$ of Condition D. Note the finite sample representation of this parameter, $\xi_{w}^{-2}\big[\om_h\eta^0_{(1)}\Si\eta^0_{(1)}+(1-\om_h)\eta^0_{(3)}\Si\eta^0_{(3)}\big].$  A plug in version $\breve\si^2_{w,\iny}$ is computed by utilizing the above described estimated parameters. The covariance matrix $\Si$ is estimated as the sample covariance $\breve\Si$ computed by utilizing the entire data set with centering done in correspondence with the estimated mean parameters over quadrants. We note that since we are not interested in the estimation of the covariance itself but instead the quadratic form described above, thus utilizing the sample covariance here is effectively identical to utilizing refitted covariance on the adjacency matrix estimated by the jump vectors $\breve\eta_{(1)},$ and $\breve\eta_{(3)},$ in turn making this shortcut valid despite potential high dimensionality.

Finally, regarding quantiles of limiting distributions characterized in Theorem \ref{thm:wc.vanishing} and Theorem \ref{thm:wc.non.vanishing} in the vanishing and non-vanishing regimes, respectively. For the quantiles of the former case, we utilize the cdf of the underlying distributions which was first presented in \cite{yao1987approximating}. For the latter case, we assume in all calculations that underlying distribution is Gaussian and consequently the distribution of the increments $\cP$ of Condition A$'$ is also Gaussian. The above estimated parameters are then utilized to produce realizations of this incremental distribution, which are then used to produce realizations of the two-sided random walk and in turn those of its {\it argmax}. The quantiles are then estimated as a monte-carlo approximation.
\end{appendix}

\bibliographystyle{imsart-number} 
\bibliography{meanchange}       

\end{document}